\documentclass[12pt]{article}
\usepackage{bbm} 
\usepackage{dsfont}
\usepackage{amsmath,amssymb,amsfonts,color,graphicx,cite,color,feynarts,soul}
\usepackage{epsfig}
\usepackage{float}
\input paperdef
\graphicspath{{figs/}}

\evensidemargin \oddsidemargin
\allowdisplaybreaks

\newcommand{\DR}{\overline{\text{DR}}}
\newcommand{\mDR}{\text{m}\overline{\text{DR}}}
\newcommand{\one}{\id}

\begin{document}
\thispagestyle{empty}

\def\thefootnote{\fnsymbol{footnote}}

\begin{flushright}
IFT-UAM/CSIC-14-054\\
FTUAM-14-21
\end{flushright}

\vspace{0.5cm}

\begin{center}

{\large\sc {\bf Radiative corrections to $\mathbf{M_h}$ from three generations}}

\vspace{0.4cm}

{\large\sc {\bf  of Majorana neutrinos and sneutrinos}}

\vspace{1cm}

{\sc
S.~Heinemeyer$^{1}$%
\footnote{email: Sven.Heinemeyer@cern.ch}%
, J.~Hernandez-Garcia$^{2}$%
\footnote{email: josu.hernandez@uam.es}%
, M.J.~Herrero$^{2}$%
\footnote{email: maria.herrero@uam.es}%
, X.~Marcano$^{2}$%
\footnote{email: xabier.marcano@uam.es}%

\vspace{0.4em}

~and A.M.~Rodriguez-Sanchez$^{2}$%
\footnote{email: anam.uam@gmail.com}%
}

\vspace*{.7cm}

{\sl
$^1$Instituto de F\'isica de Cantabria (CSIC-UC), Santander, Spain

\vspace*{0.1cm}

$^2$Departamento de F\'isica Te\'orica and Instituto de F\'isica Te\'orica,
UAM/CSIC\\
Universidad Aut\'onoma de Madrid, Cantoblanco, Madrid, Spain

}

\end{center}

\vspace*{0.1cm}

\begin{abstract}
\noindent
In this work we study the radiative corrections to the mass of the
lightest Higgs boson of the MSSM from three generations of Majorana
neutrinos and sneutrinos. The spectrum of the MSSM is augmented by three
right handed neutrinos and their supersymmetric partners. A seesaw
mechanism of type I is used to generate the physical neutrino masses and
oscillations that we require to be in agreement with present neutrino
data. We present a full one-loop computation of these Higgs mass
corrections, and analyze in full detail their  numerical size in terms
of both the MSSM and the new (s)neutrino parameters. A  critical
discussion on the different possible renormalization schemes and their
implications, in particular concerning decoupling, is included.    
\end{abstract}

\def\thefootnote{\arabic{footnote}}
\setcounter{page}{0}
\setcounter{footnote}{0}

\newpage


\section{Introduction}

In order to account for the impressive experimental data on neutrino
mass differences and  neutrino mixing angles~\cite{pdg} physics beyond
the Standard Model (SM) is needed. 
On the other hand, after the discovery of a Higgs boson at the
Large Hadron Collider (LHC)~\cite{ATLASdiscovery,CMSdiscovery}, the problem of stabilizing the
Higgs mass at the electroweak scale within the SM became even more
relevant. Similarly, the existence of Cold Dark Matter (CDM)~\cite{Planck} has
to be accounted for by an extension of the SM.
Consequently, in order to incorporate neutrino masses into the
SM, to stabilize the Higgs-boson mass scale and to provide a viable CDM
we choose here one of the most popular extensions of the SM: 
the simplest version of a supersymmetric extension of the SM, the Minimal
Supersymmetric Standard Model (MSSM)~\cite{mssm1,mssm2,mssm3}, with the addition of heavy 
right-handed Majorana neutrinos, and      
where the well known seesaw mechanism of type I~\cite{seesaw1,seesaw2,seesaw3,seesaw4,seesaw5,seesaw6} is implemented 
to generate the  observed small neutrino masses. From now on we will
denote this model by ``MSSM-seesaw''. 
The lightest Higgs boson in this model can be interpreted as the Higgs
particle discovered at the LHC~\cite{Mh125}.

In this MSSM-seesaw context, the smallness of the light
neutrino masses, $m_{\nu} \sim \mD^2/\mM$, appears naturally due to the 
induced large suppression by the ratio of the two very distant mass scales.
Namely, the  Majorana 
neutrino mass $\mM$, that represents the new physics scale, and the Dirac
neutrino mass $\mD$, which is related to the electroweak scale via the 
neutrino Yukawa couplings $\Ynu$, by $\mD=\Ynu v \sin \beta$. The Higgs 
sector content in the MSSM-seesaw is as in the MSSM,
i.e.\ composed of two Higgs doublets. $\tb$ is the ratio of the two 
vacuum expectation values, $\tb = v_2/v_1$, and 
$v^2 = v_1^2 + v_2^2 = (174 \gev)^2$. 
Small neutrino masses of the order of $\mnu \sim\mathcal O(0.1)$ eV can be easily 
accommodated with large Yukawa couplings, $\Ynu \sim \order{1}$, if the new
physics scale is very large, within the range $\mM \sim 10^{13}-10^{15} \gev$.
This is to be compared with the Dirac neutrino 
case where, in order to get similar small neutrino masses, extremely tiny,
hence irrelevant, Yukawa couplings of the order of 
$\Ynu \sim 10^{-12}-10^{-13}$ are required.

As for all SM fermions, the neutrinos in the MSSM are accompanied by
their respective super partners, the scalar neutrinos.    
The hypothesis of Majorana massive (s)neutrinos is 
very appealing for various reasons, including the interesting possibility of
generating satisfactorily baryogenesis via
leptogenesis~\cite{Fukugita:1986hr}. 
Furthermore, they can produce
an interesting phenomenology due to their potentially large Yukawa
couplings to the Higgs sector of the MSSM, such as corrections to the
light $\cp$-even Higgs-boson mass, $\Mh$~\cite{Ana,Haber}
(see also ~\cite{Cao:2004hs,Farzan:2004cm,Kang1,Kang2} for previous
evaluations). 
Further striking phenomenological implications~\cite{Raidal:2008jk} of
the MSSM-seesaw scenario are the prediction of sizeable rates for lepton 
flavor violating processes (within the present experimental reach
for specific areas of the model parameters~\cite{LFV1,LFV2,LFV3,LFV4,LFV5,LFV6,LFV7,LFV8,LFV9}),
non-negligible contributions to electric dipole moments of charged
leptons~\cite{EDM1,EDM2,EDM3}, and also the occurrence of sneutrino-antisneutrino 
oscillations~\cite{Grossman:1997is} as well as sneutrino
flavor-oscillations~\cite{Dedes:2007ef}.

It is worth recalling that the seesaw mechanism is not the only way to generate neutrino masses in the context of supersymmetry (see, for instance, \cite{Valle:2009tt,Munoz:2007fa}). In fact there are many well known extensions of the MSSM that can generate small neutrino masses besides the various types of high and low scale Seesaw models (see e.g,\cite{Porod:2010ax} for a review and references therein).
One possible alternative to the addition of right-handed neutrinos is the incorporation of R-parity violating interactions to the MSSM, which can introduce the lepton number violating terms that are needed for the small neutrino mass generation. Indeed, R-parity violation can be produced in many ways: spontaneously, explicitly, by bi-linear terms, by trilinear terms, etc., see, e.g, ref.\cite{Hirsch:2000ef,Barbier:2004ez}.
Another popular extension of the MSSM is the Next-to-Minimal-Supersymmetric-Standard-Model (NMSSM) (see, for instance, the review in \cite{Ellwanger:2009dp}), which includes an extra chiral singlet superfield with zero lepton number, offering a solution to the so-called $\mu$-problem of the MSSM and providing an extra tree level mass term to the SM-like Higgs boson which raises its mass above that of the lightest Higgs boson of the MSSM. In this NMSSM, as in the MSSM, the small neutrino masses can be generated either by allowing for R-parity violating terms or by adding extra chiral singlet superfields carrying non-vanishing lepton number (like, for instance, right-handed neutrinos). The $\mu\nu$SSM \cite{LopezFogliani:2005yw} can also solve the $\mu$ problem and generate masses for the neutrinos by adding to the MSSM right-handed neutrino superfields and R-parity breaking terms. 

It should be noted that
each of the above mentioned extensions of the MSSM leads to different phenomenological implications, including those in the neutrino and in the Higgs boson sectors. Our preference for the particular choice of extended MSSM with three generations of right handed neutrinos and sneutrinos, and with a seesaw mechanism of type I, is mainly because, as we have said above, it is the simplest extension of the MSSM compatible with neutrino data that naturally allows for large neutrino Yukawa couplings. It is precisely this interesting possibility of large neutrino Yukawa couplings what can induce large radiative corrections to the lightest Higgs boson mass, and thus the (s)neutrino sector phenomenology is directly linked to the Higgs sector.
Other extensions of the MSSM could also induce relevant corrections to the Higgs boson mass from the additional superfields and the new input parameters associated to the neutrino mass generation. For instance, within the NMSSM, in addition to the tree level enhanced Higgs boson mass, one may generate relevant mass corrections from the TeV-scale right-handed neutrinos via their interactions with the zero-lepton-number chiral singlet superfield while having small neutrino Yukawa couplings\cite{Wang:2013jya}. Alternatively, one may also generate relevant corrections to the Higgs boson mass from TeV-scale right-handed neutrinos, within the context of the Inverse Seesaw Models, that allow for large Yukawa couplings but introduce in addition a small lepton number violating parameter\cite{Chun:2014tfa}.

We are interested here in the 
indirect effects of Majorana neutrinos and sneutrinos via their radiative corrections to
the MSSM Higgs boson masses within the MSSM-seesaw framework. 
While the initial evaluations and analyses of corrections to
$\Mh$  concentrated on the one-generation case to analyze the
general analytic behavior of this type of contributions, in this paper we
investigate the Majorana neutrino and sneutrino sectors with three generations which can accommodate the present neutrino data.
We will focus here on the corrections to the lightest $\Mh$ and will present the full one-loop contributions from  
the complete three generations of neutrinos and sneutrinos and without using any approximation. 
It should be noted that the extrapolation from the one generation to the three generations case cannot be trivially done due to the relevant generation mixing in the latter and, therefore, the corresponding radiative corrections must be explicitly and separately computed.
A crucial issue of interest in relation with the present computation is the question of decoupling of the heavy Majorana mass scales.
While it was shown for the one generation case~\cite{Ana,Haber} that this
strongly depends on the choice of the renormalization scheme, no such scheme 
could be identified being superior to the other in all respects. 
Consequently, we will also comment comparatively the advantages and
disadvantages of the various renormalization schemes in the present case of three generations where 
there are several mass scales involved.
On the one hand it will not be possible to obtain
information from a precise $\Mh$ measurement on the Majorana mass
scale. On the other hand, however, the precise prediction of $\Mh$ in the
presence of Majorana (s)neutrinos and the understanding of these
corrections in the different schemes (and their respective
decoupling behavior) used in the $\Mh$ calculations, is
desirable.

For the estimates of the total corrections to $\Mh$ in the MSSM-seesaw,  
obviously, the one-loop corrections from the neutrino/sneutrino sector
that we are interested here have to be added to the existing MSSM
corrections. The status of radiative corrections to $\Mh$ in the
non-$\nu/\Snu$~sector, i.e.\ in the MSSM {\em without} massive
neutrinos, can be summarized as follows. Full one-loop
calculations~\cite{mhiggsf1l,mhiggsf1l-2,mhiggsf1l-3} have been supplemented by the leading and
subleading two-loop corrections, see~\cite{mhiggsAEC} and references
therein. Together with leading three-loop corrections~\cite{mhiggs3l,mhiggs3l-2,mhiggs3l-3}
and the recently added resummation of logarithmic
contributions~\cite{Mh-logresum}, 
the current precision in $\Mh$ is estimated to be 
$\sim 2 - 3 \gev$~\cite{mhiggsAEC,Mh-logresum,ehowp}.

A summary and discussion of the previous estimates of neutrino/sneutrino
radiative corrections to the Higgs mass parameters can be found in
\cite{Ana}, where (as discussed above) the one-generation case was
calculated and analyzed. 
In this work, we will consider the more general three generation MSSM-seesaw
scenarios with no universality conditions imposed, and explore the full 
parameter space, without restricting 
ourselves just to large or small values of any of the relevant
neutrino/sneutrino parameters. 
In principle, since the right handed Majorana neutrinos and 
their SUSY partners are $SU(2) \times U(1)$ singlets, there is no a priori 
reason why the size of their associated parameters should be related to the 
size of the other sector parameters. 
In the numerical estimates, we will therefore explore
a wide interval for all the involved 
neutrino/sneutrino relevant input parameters.

The paper is organized as follows. In \refse{sec:nN}, we summarize the
most important 
ingredients of the MSSM-seesaw scenario that are needed for the present 
computation of the Higgs mass loop corrections. These include, the setting of
the model parameters and the complete list of the Lagrangian relevant terms. A
complete set of the corresponding relevant Feynman rules in the physical basis 
is also provided here. They are collected in \refapp{sec:fmr} (to 
our knowledge, they  are not available in the previous literature).
In \refse{sec:corrections} we discuss the renormalization procedure and
emphasize the differences between the selected renormalization schemes.
The corresponding analytic analysis can be found in
\refse{Analytic}. A numerical evaluation and in particular the
dependence on the (hierarchical) Majorana mass scales is given in
\refse{Numerical}. 
Finally, our conclusions can be found in \refse{Conclusions}.


\section{The MSSM-seesaw model}
\label{sec:nN}

\noindent
In order to include the proper neutrino masses and oscillations in
agreement with present neutrino data (see, for
  instance,~\cite{Tortola:2012te,GonzalezGarcia:2012sz,Capozzi:2013csa}),
we employ an extended version of the MSSM, where three right handed
neutrinos and their supersymmetric partners are included, in addition to
the usual MSSM spectra. A seesaw mechanism of  
type~I~\cite{seesaw1,seesaw2,seesaw3,seesaw4,seesaw5,seesaw6} is implemented which requires in addition to the
Dirac neutrino mass matrix, $m_D$, the introduction of a new $3\times 3$
so-called Majorana mass matrix, $m_M$.
This matrix
$m_M$ is the responsible for the Majorana character of the physical
neutrinos in this MSSM-seesaw model.

The terms of the superpotential within the MSSM-seesaw that are relevant for neutrino and Higgs related physics are described by \cite{Ana,Grossman:1997is,Dedes:2007ef}:
\begin{equation}
\label{RH_superpotential}
W\,=\,\epsilon_{ab}\left[Y_\nu^{ij} \hat{\mathcal H}_2^a\, \hat L_i^b \hat N_j \,-\, 
Y_l^{ij} \hat{\mathcal H}_1^a\,\hat L_i^b\, \hat R_j \,+\,\mu \hat{\mathcal H}_1^a \hat{\mathcal H}_2^b\right]\,+\,
\dfrac{1}{2}\, m_M^{ij}\,\hat N_i \,\hat N_j \,.
\end{equation}
$Y_\nu$ is a $3\times3$ complex Yukawa matrix, while $m_M$ is a complex
symmetric $3\times3$ mass matrix. The indices $i,\,j$ represent
generations (with $i,\,j=1,2,3$), the indices $a,\,b$ refer to $SU(2)$
doublets components, and $\epsilon_{12}=-1$. Omitting the generation
indexes, for brevity, the involved superfields are as follows:  $\hat N =
\{\tilde\nu_R^*, (\nu_R)^c\}$ is the new superfield that contains the
right-handed neutrinos $\nu_{R_i}$ and their partners $\tilde\nu_{R_i}$,
while the other superfields are as in the MSSM, i.e.,  
$\hat L$ contains the $SU(2)$ lepton doublet, $(\nu_L,l_L)$ and its
superpartner $(\tilde\nu_L,\tilde l_L)$, $\hat R$ contains the $SU(2)$
sfermion and fermion singlets $\{\tilde l_R,(l_R)^c\}$, and the
$\mathcal{\hat H}_1$ and $\hat{\mathcal H}_2$ are the Higgs superfields
that give masses to the down and up-type (s)fermions, respectively. Here
and in the following, $f^c$ refers to the particle-antiparticle
conjugate of a fermion $f$ defined as follows: 
\begin{eqnarray}
\label{charge_conjugate}
\hat{C}:f \rightarrow f^c &= C \left.\bar{f}^T\right.\,,
\end{eqnarray}
where $\hat{C}$ and $C$ are the particle-antiparticle conjugation and charge conjugation respectively.\\

The superfields $\hat L, \hat N $ and $ \hat R$ can be chosen such that $Y_l$ and $m_M$  are real and non-negative diagonal $3\times 3$ matrices, whereas $Y_\nu$, in contrast, is a general complex $3\times3$ matrix.

The additional sneutrinos $\tilde\nu_{R_{i}}$ induce new relevant terms in the soft SUSY-breaking potential. Following \cite{Ana,Grossman:1997is,Dedes:2007ef} it can be written as:
\begin{equation}
\label{V_soft}
V^{\tilde\nu}_{\text{soft}}= \left(m^2_{\tilde L}\right)^{ij} \tilde\nu_{L_i}^* \tilde\nu_{L_j} +\left(m^2_{\tilde R}\right)^{ij} \tilde\nu_{R_i} \tilde\nu_{R_j}^* + \left( A_\nu^{ij} H^2_2 \tilde\nu_{L_i} \tilde\nu_{R_j}^* +  \left(m_B^2\right)^{ij} \tilde\nu_{R_i}^* \tilde\nu_{R_j}^* + {\text{h.c.}}\right),
\end{equation}
where $m^2_{\tilde L}$, $m^2_{\tilde R}$ are $3\times3$ hermitian matrices in the flavor space, $A_\nu$ is a $3\times 3$ generic complex matrix and $m_B^2$ is a complex symmetric matrix.

After the Higgs fields develop a vacuum expectation value, the charged lepton and Dirac neutrino mass matrix elements can be written as:
\begin{eqnarray}
\label{Dirac}
m_l^{ij}\,&=&\,Y_l^{ij}\,\,v_1\,, \nonumber \\
m_D^{ij}\,&=&\,Y_\nu^{ij}\,v_2\,,
\end{eqnarray}
where $v_i$ are the vacuum expectation values (vev) of the $H_i$ fields,
$v_1=v \cos\beta$, $v_2=v \sin\beta$ and
$v^2=\dfrac{2\MW^2}{g^2}=\dfrac{2\MZ^2}{g^2+g^{\prime 2}}=(174\text{ GeV})^2$.
$\MW$ and $\MZ$ denote the masses of the $W$~and $Z$~boson,
  respectively. 
 
Finally, starting with the superpotential of \refeq{RH_superpotential}, 
the Yukawa couplings of the neutrinos and their corresponding mass terms can be derived:
\begin{eqnarray}
 -{\mathcal L}_{\rm {mass}} -{\mathcal L}_{\rm{Yukawa}} = 
\dfrac{1}{2}\sum_{ij}\left[\dfrac{\partial^2 W\left( \phi\right)}{\partial{\phi_i}\partial{\phi_j}}\psi_i\psi_j + \rm {h.c.}\right] \, ,
\end{eqnarray}
where the $\psi_i$ are the two component fermion field superpartners of the corresponding scalar component $\phi_i$ of the super fields.

\subsection{Neutrino mass and interaction Lagrangians}
After the Higgs field develops a vacuum expectation value, the mass Lagrangian of neutrinos in the MSSM-seesaw model with three generations of $\nu_L$ and $\nu_R$ is given by:
\begin{equation}
\label{nu_mass_L}
- \mathcal{L}_{\text{mass}}^{\nu} = \overline{\nu_{R_i}}\,\, m_{D_{ij}}^{\dag} 
\nu_{L_j} + \overline{\nu_{L_i}}\,\, m_{D_{ij}} \nu_{R_j} +
\frac{1}{2} \overline{ \left(\nu_{R_i}\right)^{c}}\,\, m_{M_{ij}} \nu_{R_j} +
\frac{1}{2} \overline{\nu_{R_i}}\, m_{M_{ij}}^{\dag} \left(\nu_{R_j}\right)^{c},
\end{equation} 
where we have used again the notation $i,j$ for
generation indexes, and $m_D$ and $m_M$ are
the Dirac and Majorana mass matrices, respectively,
which have been introduced in the previous
section \refeq{Dirac}. 

Notice that the particle-antiparticle conjugation operator $\hat C$
flips the chirality of a particle and changes all the quantum numbers of
it. Then, it changes a left handed neutrino by a right handed
antineutrino and a right handed neutrino by a left handed
antineutrino. Following \refeq{charge_conjugate}: 
\begin{eqnarray}
&&\hat{C}:\,\, \nu_L \to \left( \nu_L\right)^c= \left(\nu^c\right)_R \,,\nonumber \\
&&\hat{C}:\,\, \nu_R \to \left( \nu_R\right)^c= \left(\nu^c\right)_L\, .
\end{eqnarray}
If a neutrino is a Majorana fermion it is invariant under $\hat{C}$. As a result, $\nu^c=\nu$.\\

The $\mathcal{L}_{\text{mass}}^{\nu}$ of \refeq{nu_mass_L} can be rewritten in a more compact form: \begin{equation}
-\mathcal{L}^{\nu}_{\text{mass}}=\dfrac{1}{2}\left(\overline{\nu_L},\overline{\left(\nu_R\right)^c}\right)_i M^{\nu}_{ij}\left(\begin{array}{c}\left(\nu_L\right)^c\\\nu_R\end{array}\right)_j+h.c.\,,
\end{equation}
where
\begin{equation}
M^\nu=\left(\begin{array}{cc}
0&m_D\\m_D^T&m_M
\end{array}\right)
\end{equation}
is a $6\times 6$ complex symmetric matrix which can be diagonalized by an unitary matrix $U$:
\begin{equation}
\label{neu_diagonalization}
U^TM^{\nu}U=\hat{M}^{\nu}=\text{diag}\left(m_{n_1},m_{n_2},m_{n_3},m_{n_4},m_{n_5},m_{n_6}\right)	\,.
\end{equation}
Here, the diagonal elements of $\hat{M}^{\nu}$, $m_{n_i}$, are the non negative square roots of the eigenvalues of $\left(M^{\nu}\right)^\dagger M^{\nu}$.\\

The interaction eigenstates are the left and right handed components of the neutrino fields, $\nu_{L_i}$ and $\nu_{R_i}$ (with $i=1,2,3$), and are related to the mass eigenstates $n_j$ (with $j=1,...,6$) in the following way:
\begin{eqnarray}\label{rotationR}
\left(\nu_L\right)^c_i&=&U_{ij}P_R n_j\,,\nonumber\\
\nu_{R_i}&=&U_{i+3,j}P_R n_j\,.
\end{eqnarray}
where here and from now on, we shorten the notation to $U_{ij}\equiv U_{i,j}$.
Similarly for the $\hat C$-conjugate relations:
\begin{eqnarray}\label{rotationL}
\nu_{L_i}&=&U_{ij}^*P_L n_j\,,\nonumber\\
\left(\nu_R\right)^c_i&=&U_{i+3,j}^*P_L n_j\,.
\end{eqnarray}
In the seesaw limit, i.e. if $\vert\vert m_D\vert\vert \ll \vert\vert
m_M\vert\vert$ \footnote{The euclidean matrix norm is defined by $\vert\vert
  A\vert\vert =\left[ {\rm tr}\left(
    A^{\dagger}A\right)\right]^{1/2}=\left[\Sigma_{ij}\vert
    a_{ij}\vert^2\right]^{1/2}$ for a matrix $A$ whose elements are given by
  $a_{ij}$}, an analytic perturbative diagonalization in blocks can be
performed by expanding in powers of the dimensionless parameter matrix
$\xi=m_Dm_M^{-1}$. This allows us to separate the light sector from the heavy
sector by the introduction of a $6\times 6$ matrix:
\begin{equation}
\label{rot_matrix}
\hat{U}^\nu\ =\ \left( \begin{array}{cc} (1-\frac{1}{2} \xi^* \xi^T)  & \xi^* (1-\frac{1}{2} \xi^T \xi^*)\\ -\xi^T (1- \frac{1}{2} \xi^* \xi^T)& (1-\frac{1}{2} \xi^T \xi^*) \end{array} \right) + \mathcal{O}(\xi ^4) \,.
\end{equation}

Two independent blocks of $3\times 3$ neutrino mass matrices are obtained once this $\hat{U}^\nu$ matrix is inserted in \refeq{neu_diagonalization}:
\begin{eqnarray}
m_{\nu} &=& -m_D \xi^T + \mathcal{O}(m_D \xi^3) \simeq -m_D m_M^{-1}m_D^T \,,\label{light_seesaw} \\ 
m_N &=& m_M + \mathcal{O}(m_D \xi) \simeq m_M \,.\label{heavy_seesaw}
\end{eqnarray}

The matrix $m_N$ of \refeq{heavy_seesaw} is already diagonal and its  diagonal elements $m_{N_1}$, $m_{N_2}$ and $m_{N_3}$ are approximately the three respective Majorana masses, $m_{M_1}$, $m_{M_2}$, $m_{M_3}$. The diagonalization of the matrix $m_\nu$ of \refeq{light_seesaw} is performed as usual by the Pontecorvo-Maki-Nakagawa-Sakata (PMNS) unitary matrix \cite{PMNS1,PMNS2}, $U_\text{PMNS}$ given by:
\begin{equation}
\label{UPMNS}
U_\text{PMNS}=
\left( 
\begin{array}{ccc} 
c_{12} \,c_{13} & s_{12} \,c_{13} & s_{13} \, e^{-i \delta} \\ 
-s_{12}\, c_{23}\,-\,c_{12}\,s_{23}\,s_{13}\,e^{i \delta} 
& c_{12} \,c_{23}\,-\,s_{12}\,s_{23}\,s_{13}\,e^{i \delta} 
& s_{23}\,c_{13} \\ 
s_{12}\, s_{23}\,-\,c_{12}\,c_{23}\,s_{13}\,e^{i \delta} 
& -c_{12}\, s_{23}\,-\,s_{12}\,c_{23}\,s_{13}\,e^{i \delta} 
& c_{23}\,c_{13}
\end{array} \right) \, \times \, V \,,
\end{equation}
where
\begin{equation}
 V=\text{diag}\,(e^{-i\phi_1/2},e^{-i\phi_2/2},1) \,,
\end{equation}
and the notation $c_{ij} \equiv \cos \theta_{ij}$, $s_{ij} \equiv \sin \theta_{ij}$ has been used. Here, $\theta_{ij}$ are the mixing angles of the light neutrinos, $\delta$ is the Dirac phase and $\phi_{1,2}$ are the two Majorana phases.

As a result, the mass eigenvalues $m_{n_j}$, corresponding to light Majorana neutrinos $\left(\nu\right)$ and heavy Majorana neutrinos $\left(N\right)$ are given respectively by:
\begin{eqnarray}
\label{mass_eigenvalues}
m_{\nu}^\text{diag}&=&U_\text{PMNS}^T m_{\nu} U_\text{PMNS}= \text{diag} \, (m_{\nu_1},m_{\nu_2},m_{\nu_3}) \,,\\ 
m_N^\text{diag} &=& \text{diag} (m_{N_1},m_{N_2},m_{N_3})\simeq 
\text{diag} \, (m_{M_1},m_{M_2},m_{M_3}) \,. 
 \end{eqnarray}

In this work, in order to make contact with the experimental data, we have used the Casas-Ibarra parametrization \cite{Casas:2001sr}, which provides a simple way to reconstruct the Dirac mass matrix by using as inputs the physical light $m_{\nu_i}$ and heavy $m_{N_i}$ neutrino masses, the $U_\text{PMNS}$ matrix, and a general complex and orthogonal matrix $R$:

\begin{equation} 
\label{Casas}
m_D^T =i \sqrt{m_N^\text{diag}} \, R \, \sqrt{m_{\nu}^\text{diag}} \, U_\text{PMNS}^\dag \,,
\end{equation}
where $R^T R = R R^T = \one$ and where we have considered the
following parametrization: 
\begin{equation}
\label{R_Casas}
R = \left( \begin{array}{ccc} c_{2} c_{3} 
& -c_{1} s_{3}-s_1 s_2 c_3& s_{1} s_3- c_1 s_2 c_3\\ c_{2} s_{3} & c_{1} c_{3}-s_{1}s_{2}s_{3} & -s_{1}c_{3}-c_1 s_2 s_3 \\ s_{2}  & s_{1} c_{2} & c_{1}c_{2}\end{array} \right) \,,
\end{equation}
where $c_i\equiv \cos \theta_i$, $s_i\equiv \sin\theta_i$ and $\theta_1$, $\theta_2$ and $\theta_3$ are arbitrary complex angles.

Thus, our set of input values consist of $m_{M_1}$, $m_{M_2}$, $m_{M_3}$
and $\theta_i$, and for $m_{\nu_1}$, $m_{\nu_2}$, $m_{\nu_3}$ and
$U_\text{PMNS}$ we use their suggested values from the experimental data
are used. For the numerical estimates in this work we will use the
following input values for the light neutrino mass squared differences
and the angles in the $U_\text{PMNS}$ matrix: 
\begin{align} 
\label{input_values}
& 
\Delta m^2_{21} \,=7.50\times 10^{-5} {\rm eV}^2 \,,
\quad \quad 
\vert\Delta  m^2_{32}\vert \,= 2.42 \times 10^{-3} {\rm eV}^2 \,,
\nonumber \\
& 
\sin^2{(2\theta_{12})}=0.857 \,, 
\quad 
\sin^2{(2\theta_{23})}=0.95 \,, 
\quad 
\sin^2{(2\theta_{13})}=0.098\,, 
\quad \quad 
\delta=\phi_1=\phi_2=0 \,,
\end{align}
 Notice that $\Delta  m^2_{32}>0$ for light neutrinos with a normal hierarchy and $\Delta  m^2_{32}<0$ for an inverted light neutrino hierarchy. These values are compatible with the present experimental data. Specifically, the recent global fit NuFIT~1.3 (2014)~\cite{GonzalezGarcia:2012sz} sets:
\begin{align}
 \sin^2\theta_{12}&=0.304^{+0.012}_{-0.012}\,, 					 & \Delta m^2_{21}&=7.50^{+0.19}_{-0.17}\times10^{-5}\;\mathrm{eV}^2\,,\nonumber\\
 \sin^2\theta_{23}&=0.451^{+0.001}_{-0.001}\,, 					&  \Delta m^2_{31}&=2.458^{+0.002}_{-0.002}\times10^{-3}\;\mathrm{eV}^2\,({\rm NH}),\\
\sin^2\theta_{13}&=0.0219^{+0.0010}_{-0.0011}\,,				&
\Delta m^2_{32}&=-2.448^{+0.047}_{-0.047}\times10^{-3}\;\mathrm{eV}^2\,({\rm IH}),
\nonumber
\end{align}
where NH and IH refer to the normal hierarchy and inverted hierarchy
cases for the light neutrinos, respectively.

The interaction Lagrangian of the MSSM neutral Higgs bosons  with the three $\nu_L$ and three $\nu_R$ neutrinos is given, in compact form, by:
\begin{eqnarray}
\label{nu_EWL}
\mathcal{L}^{{\rm Higgs}}_{\nu_{L}\nu_{R}}&=& -\dfrac{g}{2 \MW \sin\beta}\left(\overline{\nu_R} m_D^{\dagger} \nu_L + \overline{\nu_L}m_D\nu_R \right)\left(H \sin\alpha +h\cos\alpha\right) \nonumber \\
&&-\dfrac{i g}{2 \MW \sin\beta}\left(\overline{\nu_R} m_D^{\dagger} \nu_L-\overline{\nu_L}m_D\nu_R \right)A \cos\beta.
\end{eqnarray}
Here $\al$ is the angle that diagonalizes the $\cp$-even Higgs sector at
the tree-level.

By using \refeq{rotationR} and \refeq{rotationL} the interaction
Lagrangian in \refeq{nu_EWL} can be expressed in terms of the neutrino
mass eigenstates $n_i=(n_1,\ldots,n_6)$: 
\begin{eqnarray}
\label{nu_PhL}
\mathcal{L}^{{\rm Higgs}}_{n_j n_i} &=& \dfrac{-g}{2 \MW \sin\beta}\, \overline{n}_j \Big[ U_{l+3,j}^*\left(m_D^{\dagger}\right)_{lm}U_{mi}^* P_L + U_{lj} \left(m_D\right)_{lm} U_{m+3,i} P_R\Big] n_i\left(H \sin\alpha+h \cos\alpha\right)  \nonumber \\
&&-\dfrac{i g}{2 \MW \sin\beta}\, \overline{n}_j \Big[ U_{l+3,j}^* \left(m_D^{\dagger}\right)_{lm}U_{mi}^* P_L  
- U_{lj} \left(m_D\right)_{lm} U_{m+3,i} P_R\Big] n_i\, A \cos\beta\,,
\end{eqnarray}
where $j$ and $i$ indexes run from 1 to 6 and $l$ and $m$ indexes run from 1 to 3.\\

The gauge interactions of $\nu_L$ (the $\nu_R$ have no interactions since they are singlets) with the neutral gauge boson $Z$ are given, in compact form, by:
\begin{equation}
\label{nu_Z_EWL}
\mathcal{L}_{\nu_{L} \nu_{L}}^{Z}= 
-\dfrac{g}{2 \cw}\left(\overline{\nu_{L}}\,\gamma^{\mu}\nu_{L}\right)Z_{\mu}\, . 
\end{equation}
When expressed in terms of the physical neutrino basis it gives:
\begin{equation}
\label{nu_Z_PhL}
\mathcal{L}_{n_j n_i }^{Z}=- \dfrac{g}{2 \cw}\left(\overline{n }_jU_{mj}U_{mi}^*\gamma^{\mu}P_{L}n_i\right)Z_{\mu}
\end{equation}
where the indexes $i$ and $j$ run from 1 to 6 and $m$ runs from 1 to 3.


\subsection{Sneutrino mass and interaction Lagrangians}

Following \cite{Dedes:2007ef}, we will express the sneutrino mass terms in a compact $6\times6$ matrix form by defining two six-dimensional vectors $\phi_L =(\tilde{\nu}_L\ \ \tilde{\nu}_L^*)^T$ and $\phi_N =(\tilde{N}\ \ \tilde{N}^*)^T=(\tilde{\nu}_R^*\ \ \tilde{\nu}_R)^T$. 
In this new basis, the mass Lagrangian of the sneutrinos has the form:
\begin{eqnarray}
\label{sneu_mas_matrix_L}
-\mathcal{L}_{\rm mass}\,\, &=& \dfrac{1}{2}\left (\begin{array}{cc}
                        \phi_L^{\dagger} & \phi_N^{\dagger}
                       \end{array}\right) \left(\begin{array}{cc}
M_{LL}^2 & M_{LN}^2 \\
 \left(M_{LN}^2\right)^{\dagger} &  M_{NN}^2
\end{array}
\right)\left(
\begin{array}{c}
\phi_L \\ \phi_N
\end{array}\right)\nonumber\\ 
&=& \dfrac{1}{2}\left(
\begin{array}{cccc}
\tilde{\nu}_L^{*T} & \tilde{\nu}_L^T &\tilde{\nu}_R^T& \tilde{\nu}_R^{*T}
 \end{array}
\right) M_{\tilde{\nu}}^2\left(
\begin{array}{c}
\tilde{\nu}_L \\ \tilde{\nu}_L^* \\ \tilde{\nu}_R^* \\ \tilde{\nu}_R   
\end{array}
\right)\, ,
\end{eqnarray}
here $M_{LL}^2$ and $M_{NN}^2$ are $6\times6$ hermitian matrices while $M_{LN}^2$ is a $6\times6$ complex matrix; and the three of them 
can be expressed in blocks of $3\times3$ matrices as follows:
\begin{eqnarray}
M_{AB}^2= \left( \begin{array}{cc}
                   M_{A^{\dagger}B}^2 &  M_{A^TB}^{2*}\\
M_{A^TB}^2 & M_{A^{\dagger}B}^{2*}
                  \end{array}\right)\, ,
\end{eqnarray}
where the subscripts $A,B$ stand for $L$ and/or $N$. The matrices
$M_{A^{\dagger}B}^2$ and $M_{A^TB}^2$ for $A\neq B$ are general complex
matrices with no restrictions, but the $M_{A^{\dagger}A}^2$ and
$M_{A^TA}^2$, for $A= L, N$, are $3\times3$ hermitian matrices and
complex symmetric matrices, respectively. 

The expressions of the different blocks of matrices that enter in the complete $12\times12$ sneutrino mass matrix $M^2_{\tilde{\nu}}$ are the following:
  \begin{eqnarray}\hspace{-2.cm}
M_{LL}^2&=&\left(\begin{array}{cc}
  m^2_{\tilde L}+ m_D^*m_D^T + \frac{1}{2} \MZ^2 \cos 2\beta   & 0\\
0 &       m^{2*}_{\tilde L}+ m_D m_D^{\dagger} + \frac{1}{2} \MZ^2\cos 2\beta        
                  \end{array}
\right), \\
&&\phantom{M_{LN}^2} \nonumber \\
M_{NN}^2&=&  \left(\begin{array}{cc}
  m^2_{\tilde R}+ m_D^{\dagger}m_D + m_M^{\dagger}m_M   & 2 b_\nu^* m_M^*\\
2 b_\nu m_M & m^{2*}_{\tilde R}+ m_D^Tm_D^* + m_M^{T}m_M^*     
                  \end{array}
\right),\\
&&\phantom{M_{LN}^2} \nonumber \\
M_{LN}^2&=& \left(\begin{array}{cc}
m_D^*m_M  & m_D^*\left(a_\nu^* -\mu^* \cot\beta\right)\\
m_D\left( a_\nu -\mu \cot\beta\right) & m_D m_M^*   
                  \end{array}
\right),
\end{eqnarray}
where we have assumed:
\begin{eqnarray}
m_B^{2}&=&b_\nu m_M,\nonumber\\
A_\nu&=&a_\nu Y_\nu
\end{eqnarray}
with the convention of:
\begin{equation}
\label{Yuk_mat}
Y_{\nu}=\dfrac{g m_D}{\sqrt{2}\MW\sin\beta}\, .
\end{equation}

We have to diagonalize the sneutrino mass matrix in \refeq{sneu_mas_matrix_L} in order to obtain the  twelve mass eigenstates. This matrix is hermitian, so it can be diagonalized by an $12\times12$  unitary matrix $\tilde{U}$ as follows:
\begin{equation}
 \tilde{U}^{\dag} M_{\tilde{\nu}}^2 \tilde{U} = M_{\tilde{n}}^2 = {\rm diag}\left(m_{\tilde{n}_1}^2,...,m_{\tilde{n}_{12}}^2\right)\ .
\end{equation}

The relations between the interaction eigenstates and the mass eigenstates are then given by:
\begin{eqnarray}
\label{snu_rot_matrix}
\tilde{\nu}_{L_i}&=& \tilde{U}_{ij} \tilde{n}_j\,,\nonumber\\
\tilde{\nu}_{L_i}^*&=& \tilde{U}_{i+3,j} \tilde{n}_j\,,\nonumber\\
\tilde{\nu}_{R_i}^*&=& \tilde{U}_{i+6,j} \tilde{n}_j\,,\nonumber\\
\tilde{\nu}_{R_i}&=& \tilde{U}_{i+9,j} \tilde{n}_j\,,
\end{eqnarray}
where $i$ runs from 1 to 3 and $j$ from 1 to 12. Again we shorten the notation to $\tilde U_{ij}\equiv\tilde U_{i,j}$.\\

Finally, the contributions from the $F$-terms, the $D$-terms and the soft SUSY breaking terms to the interactions of  the sneutrinos with the MSSM neutral Higgs bosons are given   by:
\begin{eqnarray}
\label{snu_F_D_Int_intL}
\mathcal{L}_{\text{int}-\tilde{\nu}-\text{Higgs}}^{F-\text{terms}}&=&  \dfrac{g}{2\MW \sin\beta}\left(H\cos\alpha-h\sin\alpha\right) 
\left[\mu^*\,\tilde{\nu}_L^T m_D \tilde{\nu}_R^* +\mu\, \tilde{\nu}_L^{*T}m_D^* \tilde{\nu}_R\right] \nonumber \\
&& - \,\,i \dfrac{g }{2\MW } A \left[\mu^*\,\tilde{\nu}_L^T m_D\tilde{\nu}_R^* -\mu\, \tilde{\nu}_L^{*T}m_D^* \tilde{\nu}_R\right]\nonumber\\
&& -\,\,\dfrac{g}{\MW\sin\beta}\left(H\sin\alpha+h\cos\alpha\right) \left[\tilde{\nu}_R^Tm_D^{\dagger} m_D \tilde{\nu}_R^* +\tilde{\nu}_L^T m_Dm_D^{\dagger} \tilde{\nu}_L^* \right]\nonumber\\
&& -\,\,\dfrac{g^2}{4\MW^2\sin^2\beta}\left(H^2\sin^2\alpha +h^2\cos^2\alpha + 2H h \sin\alpha\cos\alpha +A^2\cos^2\beta\right)\times\nonumber \\
&&\left[\tilde{\nu}_R^T m_D^{\dagger} m_D \tilde{\nu}_R^* +\tilde{\nu}_L^T m_Dm_D^{\dagger} \tilde{\nu}_L^* \right]\nonumber \\
&& -\,\,\dfrac{g}{2\MW \sin\beta}\left(H\sin\alpha+h\cos\alpha\right) \left[\tilde{\nu}_L^{*T} m_D^* m_M \tilde{\nu}_R^*+ \tilde{\nu}_L^T m_D m_M^* \tilde{\nu}_R \right]
\nonumber\\
&& +\,\,i\dfrac{g\cos\beta}{2\MW \sin\beta}A\left[\tilde{\nu}_L^{*T} m_D^* m_M \tilde{\nu}_R^*- \tilde{\nu}_L^T m_D m_M^* \tilde{\nu}_R \right]\, ,\nonumber\\
&&\phantom{\mathcal{L}_{\text{int}-\tilde{\nu}-\text{Higgs}}^{F-\text{terms}}} \nonumber \\
\mathcal{L}_{\text{int}-\tilde{\nu}-\text{Higgs}}^{D-\text{terms}}&=&-\dfrac{g\MZ}{2\cw}\left(H\cos\left(\alpha+\beta\right)-h\sin\left(\alpha+\beta\right)\right)\tilde{\nu}_L^{*T}\tilde{\nu}_L \nonumber \\
&&-\dfrac{g^2}{8\cw^2}\left(H^2\cos2\alpha -h^2\cos2\alpha - 2H h \sin2\alpha-A^2\cos2\beta\right)\tilde{\nu}_L^{*T}\tilde{\nu}_L \label{snu_D_intL} \, , \nonumber\\
&&\phantom{\mathcal{L}_{\text{int}-\tilde{\nu}-\text{Higgs}}^{F-\text{terms}}} \nonumber \\
\mathcal{L}_{\text{int}-\tilde{\nu}-\text{Higgs}}^{\text{soft-terms}}&=&-\dfrac{1}{\sqrt{2}}\left(H\sin\alpha+h\cos\alpha\right)\left[\tilde{\nu}_L^T A_\nu \tilde{\nu}_R^* 
+ \tilde{\nu}_L^{*T}A_\nu^* \tilde{\nu}_R\right]\nonumber \\
&&-\,\,i \dfrac{\cos\beta}{\sqrt{2}} A \left[\tilde{\nu}_L^T A_\nu \tilde{\nu}_R^* - \tilde{\nu}_L^{*T}A_\nu^* \tilde{\nu}_R\right]\label{snu_SUSY_intL}\, . 
\end{eqnarray}

By using the rotations given in \refeq{snu_rot_matrix}, the previous
Lagrangians of \refeq{snu_F_D_Int_intL} to \refeq{snu_SUSY_intL} can be
expressed in terms of the physical sneutrino basis $\tilde{n}_j$,
$(j=1,\ldots,12)$. We have omitted to write them here for brevity. The
derived Feynman Rules for both neutrinos and sneutrinos are collected in
\refapp{sec:fmr}.


\section{Radiative corrections to the Higgs mass}

\label{sec:corrections}

Contrary to the SM, in the MSSM two Higgs doublets are required, $\mathcal{H}_1$ and $\mathcal{H}_{ 2}$, which can be decomposed as:
\begin{eqnarray}
\mathcal{H}_1 &=& \left( \begin{array}{c}
H_1^0 \\ H_1^-
\end{array} \right) = \left( \begin{array}{c}
 v_1 + \dfrac{1}{\sqrt{2}}\left(\phi_1^0 - i\chi_1^0\right) \\ -\phi_1^- \end{array} \right)\, ,
        \nonumber \\
\mathcal{H}_2 &=&  \left( \begin{array}{c}
H_2^+ \\ H_2^0
\end{array} \right) = \left( \begin{array}{c}
\phi_2^+ \\
 v_2 + \dfrac{1}{\sqrt{2}}\left(\phi_2^0 + i\chi_2^0\right) \end{array} \right)\, .
\label{Higgses}
\end{eqnarray}

The Higgs spectrum contains two $\cp$-even neutral bosons $(h,H)$, one $\cp$-odd neutral boson $(A)$, two charged bosons, $(H^\pm)$, and three unphysical Goldstone bosons, $(G,G^\pm)$, and are related to the components of $\mathcal{H}_1$ and $\mathcal{H}_{ 2}$ via the orthogonal transformations:
\begin{eqnarray}
\label{hHdiag}
\left(\begin{array}{c}
 H \\[0.5ex] h \end{array}\right) &=& 
\left(\begin{array}{cc} \cos\alpha & \sin\alpha \\[0.5ex] -\sin\alpha & \cos\alpha \end{array}\right) 
\left(\begin{array}{c} \phi_1^0 \\[0.5ex] \phi_2^0 \end{array}\right) ~, \nonumber\\
\label{AGdiag}
\left(\begin{array}{c}
 G \\[0.5ex] A \end{array}\right) &=& 
\left(\begin{array}{cc} \cos\beta& \sin\beta \\[0.5ex] -\sin\beta & \cos\beta \end{array}\right) 
\left(\begin{array}{c} \chi_1^0 \\[0.5ex] \chi_2^0 \end{array}\right) ~, \nonumber\\
\label{Hpmdiag}
\left(\begin{array}{c}
 G^{\pm} \\[0.5ex] H^{\pm} \end{array}\right) &=& 
\left(\begin{array}{cc} \cos\beta& \sin\beta \\[0.5ex] -\sin\beta & \cos\beta \end{array}\right) 
\left(\begin{array}{c} \phi_1^{\pm}  \\[0.5ex] \phi_2^{\pm}  \end{array}\right) ~,
\end{eqnarray}
where
\begin{equation}
\label{tanbdef}
\tb=\dfrac{v_2}{v_1}\,.
\end{equation}

In the Feynman diagrammatic (FD) approach and assuming $\cp$
conservation, the higher-order corrected  $\cp$-even Higgs
boson masses in the MSSM are derived by finding the poles of the
$(h,H)$-propagator matrix. The inverse of this matrix is given by: 
\begin{equation}
\big(\Delta_{\text{Higgs}}\big)^{-1}=-i\left(\begin{array}{cc} 
p^2-m_{H,\text{tree}}^2+\hat\Sigma_{HH}(p^2) &\hat\Sigma_{hH}(p^2)\\
\hat\Sigma_{hH}(p^2) & p^2-m_{h,\text{tree}}^2+\hat\Sigma_{hh}(p^2)
\end{array}\right)\ ,
\end{equation}
where the tree-level masses of the $\cp$-even Higgs bosons are given by
\begin{equation}
\label{mhH_tree}
m_{{H,h,\text{tree}}}^2=
\dfrac{1}{2}\left(M_A^2+\MZ^2 \pm\sqrt{(M_A^2+\MZ^2)^2-
4\MZ^2M_A^2\cos^2 2\beta}\right) ~,
 \end{equation}
 and $\hat\Sigma$ denotes the renormalized self-energy. 
The poles of the propagator $\Delta_{\text{Higgs}}$ are obtained by solving the equation
\begin{equation}
\Big[p^2-m_{h,\text{tree}}^2+\hat\Sigma_{hh}(p^2)\Big]\Big[p^2-m_{H,\text{tree}}^2+\hat\Sigma_{HH}(p^2)\Big]-\Big[\hat\Sigma_{hH}(p^2)\Big]^2=0\ .
\end{equation}

It has been shown \cite{Ana} that the mixing between these two Higgs
bosons can be neglected in a good approximation for the
neutrino/sneutrino contributions. Morevoer, if the one-loop
contributions due to neutrinos and 
sneutrinos are small in comparison with the pure MSSM contributions,
the correction to the light $\cp$-even Higgs boson mass from
the neutrino/sneutrino sector can be can be approximated by: 
\begin{equation}
\label{delta_mass}
\Delta \Mh \simeq -\dfrac{\hat{\Sigma}_{hh}^{\nu/\tilde{\nu}}\left(M^2_h\right)}{2M_h}\,.
\end{equation}
Here $\hat{\Sigma}_{hh}^{\nu/\tilde{\nu}}$ denotes the one-loop corrections to
the renormalized Higgs-boson self-energy from the neutrinos/sneutrinos sector,
and $M_h$ denotes the higher-order corrected light $\cp$-even Higgs boson
mass, calculated with the help of
\texttt{FeynHiggs}~\cite{feynhiggs1,feynhiggs2, mhiggslong, mhiggsAEC,mhcMSSMlong,Mh-logresum}.  
In this way $\Delta M_h$ approximates the new corrections arising from the new
neutrino/sneutrino sectors with respect to the MSSM corrected Higgs mass, as
shown in \cite{Ana}. 
It should be noted that the two class of mass corrections, the ones from
the MSSM sectors and the ones from the new neutrino/sneutrino sectors are
separately renormalizable.
Therefore, in this paper we will use \refeq{delta_mass} in order to compute the one-loop radiative corrections to the lightest Higgs boson mass. \\

\subsection{Renormalized Higgs boson self-energy}

At one-loop level, the renormalized self-energies can be expressed through the unrenormalized self-energies, $\Sigma(p^2)$, the field renormalization constants, $\delta Z$, and the mass counterterms, $\delta m^2$:
\begin{subequations}
\begin{align}
\hat\Sigma_{hh}(p^2)&=\Sigma_{hh}(p^2)+\delta Z_{hh}(p^2-m_{h,\text{tree}}^2)-\delta m_h^2\ ,\\
\hat\Sigma_{hH}(p^2)&=\Sigma_{hH}(p^2)+\delta Z_{hH}\big(p^2-\frac12(m_{h,\text{tree}}^2+m_{H,\text{tree}}^2)\big)-\delta m_{hH}^2\ ,\\
\hat\Sigma_{HH}(p^2)&=\Sigma_{HH}(p^2)+\delta Z_{HH}(p^2-m_{H,\text{tree}}^2)-\delta m_H^2\ .
\end{align}
\end{subequations}

The mass counterterms arise from the Higgs potential. We introduce the
following counterterms: 
\begin{equation}\label{CT}
\begin{array}{rlcrl}
\MZ^2\longrightarrow&\MZ^2+\delta \MZ^2  &\hskip 2cm&   T_h\longrightarrow & T_h+\delta T_h  \\[.3em]
\MW^2\longrightarrow&\MW^2+\delta \MW^2  &\hskip 2cm&  T_H\longrightarrow & T_H+\delta T_H\\[.3em]
M_{A}^2\longrightarrow &  M_{A}^2+\delta  M_{A}^2   &\hskip 2cm&  \tb\longrightarrow &\tb(1+\delta\tb) \\[.3em]

\end{array}
\end{equation}
$\MA$ denotes the mass of the $\cp$-odd Higgs boson, $T_{h,H}$ are the
tadpoles in the Higgs potential, i.e.\ the terms linear in the fields 
$h, H$, respectively.

Choosing $\delta \MZ^2,\, \delta \MW^2, \, \delta T_h, \, \delta T_H, \, \delta M_A^2$ and $\delta\tb$ as independent counterterms, we can express the Higgs mass counterterms as follows:
\begin{subequations}
\begin{align}\label{deltamh}
\delta m_h^2=&\ \delta M_A^2\cos^2(\alpha-\beta)+\delta \MZ^2\sin^2(\alpha+\beta)\nonumber\\
&+\frac e{2\MW\sw}\Big(\delta T_H\cos(\alpha-\beta)\sin^2(\alpha-\beta)+\delta T_h\sin(\alpha-\beta)\big(1+\cos^2(\alpha-\beta)\big)\Big)\nonumber\\
&+\delta\tb \MZ^2\sin2\beta \sin2(\alpha+\beta)\ , \\[.3em]
\de m_{hH}^2 =& 
\edz \KL \de\MA^2 \sin 2(\al - \be) 
            - \de\MZ^2 \sin 2(\al + \be) \KR \non \\
  &+\frac e{2\MW\sw} \KL \de T_H \sin^3 (\al - \be) 
                       - \de T_h \cos^3 (\al - \be) \KR \non \\
  &-\de\tb \sin\be\cos\be 
    \KL \MA^2 \cos 2 (\al - \be) + \MZ^2 \cos 2 (\al + \be) \KR~, \\[.3em]
\delta m_H^2=&\ \delta M_A^2\sin^2(\alpha-\beta)+\delta \MZ^2\cos^2(\alpha+\beta)\label{deltamH}\nonumber\\
&-\frac e{2\MW\sw}\Big(\delta T_H\cos(\alpha-\beta)\big(1+\sin^2(\alpha-\beta)\big)+\delta T_h\sin(\alpha-\beta)\cos^2(\alpha-\beta)\Big)\nonumber\\
&-\delta\tb \MZ^2\sin2\beta \sin2(\alpha+\beta)\ ,
\end{align}
\end{subequations}
where we have used the tree level relation 
$\MA^2 \sin 2(\al - \be) = \MZ^2 \sin 2 (\al + \be)$.

On the other hand, the field renormalization constants read, 
\begin{equation}
\left(\begin{array}{c} H\\h\end{array}\right)\longrightarrow \left(\begin{array}{cc} 1+\frac12\delta Z_{HH} & \frac12\delta Z_{hH} \\\frac12\delta Z_{hH} &1+\frac12\delta Z_{hh}\end{array}\right) \left(\begin{array}{c}H\\h\end{array}\right)\ .
\end{equation}
If we choose to give one renormalization constant to each Higgs doublet, 
\begin{equation}
 \mathcal H_1\longrightarrow (1+\frac12\delta Z_{\mathcal H_1})\mathcal H_1\qquad\mbox{and} \qquad  \mathcal H_2\longrightarrow (1+\frac12\delta Z_{\mathcal H_2})\mathcal H_2\ , 
 \end{equation}
we obtain the relations,
\begin{subequations}
\label{deltaZ}
\begin{align}
\delta Z_{hh}=&\sin^2\alpha\ \delta Z_{\mathcal H_1}+\cos^2\alpha \ \delta Z_{\mathcal H_2}\ ,\\
\delta Z_{hH}=&\sin\alpha\cos\alpha\  (\delta Z_{\mathcal H_2}-\delta Z_{\mathcal H_1})\ ,\\
\delta Z_{HH}=&\cos^2\alpha\ \delta Z_{\mathcal H_1}+\sin^2\alpha\ \delta Z_{\mathcal H_2}\ .
\end{align}
\end{subequations}
Using the renormalization of the vacuum expectation values $v_i$ of the Higgs doublets,
\begin{equation}
v_1\longrightarrow (1+\frac12\delta Z_{\mathcal H_1})(v_1+\delta v_1)\qquad,\qquad v_2\longrightarrow (1+\frac12\delta Z_{\mathcal H_2})(v_2+\delta v_2)\ ,
\end{equation}
the $\tb$ counterterm can be expressed in terms of the field renormalization constants:
\begin{equation}
\label{dtb}
\delta\tb=\frac12\big(\delta Z_{\mathcal H_2}-\delta Z_{\mathcal H_1}\big)\ .
\end{equation}

This last relation is based on the fact that the divergent parts of $\delta v_1/v_1$ and $\delta v_2/v_2$ are equal, so one can set:
\begin{equation}
\dfrac{\delta v_1}{v_1}-\dfrac{\delta v_2}{v_2}=0 \,.
\end{equation}
The validity of this equation has been discussed in~\cite{Sperling:2013eva}. 


\subsection{Renormalization conditions}

Since there are six independent counterterms, six renormalization conditions are needed. For the masses, we choose an on-shell renormalization condition:
\begin{equation}
\mbox{Re}\hat\Sigma_{ZZ}(\MZ^2)=0\quad,\quad \mbox{Re}\hat\Sigma_{WW}(\MW^2)=0\quad,\quad \mbox{Re}\hat\Sigma_{AA}(M_A^2)=0\ ,
\end{equation}
which sets the mass counterterms to,  
\begin{equation}\label{deltaMA}
\delta \MZ^2=\mbox{Re} \Sigma_{ZZ}(\MZ^2)\quad,\quad \delta \MW^2=\mbox{Re}\Sigma_{WW}(\MW^2)\quad,\quad\delta M_A^2=\mbox{Re}\Sigma(M_A^2)\ ,
\end{equation}
where the gauge bosons self energies are to be understood as the transverse parts of the full self-energies. 

The tadpole condition requires that the tadpole coefficients must vanish in all orders, implying at the one-loop level, 
\begin{equation}
 T_{h,H(1)}+\delta T_{h,H}=0\ ,
 \end{equation}
so we choose the tadpole counterterms as,
\begin{equation} \label{deltaTp}
\delta T_h=-T_{h(1)}\quad,\quad \delta T_H=-T_{H(1)}\ ,
\end{equation}
where $T_{h,H(1)}$ denotes the one loop contributions to the respective Higgs tadpole graph.\\

On the other hand, $\tb$ is just a Lagrangian parameter, it is not a directly measurable quantity. Therefore, there is no obvious relation of this parameter to a specific physical observable which would favor a particular renormalization scheme. 
Furthermore, the choice of one particular renormalization scheme sets the actual definition of $\tb$, its physical meaning and its relation to observables, as it happens within the SM for the weak mixing angle $\theta_W$.
 

\subsection{Renormalization schemes for \boldmath{$\tb$}}
\label{sec:tb-ren} 

There are different possible renormalization schemes for $\tb$, as has
been extensively discussed in the literature, see for instance, the
discussion in \cite{tbren1,tbren2}. Notice that, due to the relation in
\refeq{dtb}, the renormalization scheme for $\tb$ is closely related to
the scheme for the field renormalization constants $\delta Z_{\mathcal
  H_1}$ and $\delta Z_{\mathcal H_2}$. 
Next, we will review some different choices for the renormalization of
$\tb$ that have been considered previously  in the literature,
  and discuss their respective advantages and disadvantages.


\subsubsection{$\mathbf{\DR}$ scheme }

One possibility is to use the field counterterms to remove just the terms proportional to the divergence in dimensional reduction. This defines the most frequently used scheme, the so-called $\DR$ scheme: 
\begin{subequations}
\label{deltaDR}
\begin{align}
&\delta Z_{\mathcal H_1}^{\DR}=-\left[\re \Sigma'_{HH}\right]_{\alpha=0}^{\text{div}}\ ,\label{deltaZ1}\\
&\delta Z_{\mathcal H_2}^{\DR}=-\left[\re \Sigma'_{hh}\right]_{\alpha=0}^{\text{div}}\ ,\label{deltaZ2}
\end{align}
where we have used the notation $\Sigma'\equiv \partial\Sigma/\partial p^2$. Following \refeq{dtb}, the $\tb$ counterterm is then given by:
\begin{equation}
\delta\tb^{\DR}=\frac12\left(\delta Z_{\mathcal H_2}^{\DR}-\delta Z_{\mathcal H_1}^{\DR}\right)\ .
\end{equation}
\end{subequations}
The notation $[\ ]^{\text{div}}$ used here means that one takes just the terms that are proportional to the divergence $\Delta$, which is defined, as it is usual in dimensional regularization/reduction, by:  
\begin{equation}
\Delta \equiv \frac2\epsilon-\gamma_E+\log(4\pi)\ ,
\label{deltaeps}
\end{equation} 
where $\epsilon$ is related to the dimension $d$ by $d=4-\epsilon$ and $\gamma_E$ is the Euler constant.
Notice that we have not specified the particular momentum $p^2$ at which
$\Sigma'$ is evaluated in eqs.(\ref{deltaDR}) because these
$[\ ]^{\text{div}}$ terms are not $p^2$ dependent. 

In this scheme, there is still a remaining dependence of the
renormalized Green functions on the renormalization scale $\mu_{\DR}$,
which has to be fixed to a ``proper'' value.  
This choice will be discussed in more detail below.

The $\DR$ scheme is often used in the literature, because it is process
independent and numerically stable by avoiding threshold effects,
although it induces a gauge
dependence on the $\tb$ parameter already at one-loop  
level\cite{tbren2}. It was also shown in \cite{tbren2} that for the
particular case of $R_\xi$ gauges the $\xi$  
dependence cancels at one-loop resulting in a gauge invariant result.
Nevertheless, this numerical stability could be lost in presence of
large scales, such as the Majorana mass, since large logarithmic
corrections, proportional to $\log(m_M^2/\mu_{\DR}^2)$, could
appear, and in these cases decoupling should be added ``by hand''.


\subsubsection{Modified $\mathbf{\DR}$ scheme ($\mathbf{\mDR}$)}

In models where there is one mass scale much larger than the rest of the mass scales, the remaining dependence on the $\mu_{\DR}$ scale in the $\DR$ scheme is associated to the large scale.
In our case of study, the large scale is the Majorana mass (or
Majorana masses in the case they are  different for each of the three
generations), and this will give rise to new terms in the radiative
corrections involving the neutrino Yukawa coupling that are
proportional to $\log (m_M^2/\mu_{\DR}^2)$ as well as numerically
smaller non-logarithmic terms. These logarithmic terms
can give large contributions for large Majorana masses, worsening the
convergence of the perturbative expansion.  

However, these terms can be absorbed in the $\tb$ and field counterterms 
including not only the terms proportional to the divergence
$\Delta$ but also those large logarithms. 
This choice defines the modified $\DR$ scheme ($\mDR$), which sets the $\tb$ and field counterterms as follows \cite{Ana}:
\begin{subequations}
\begin{align}
\delta Z_{\mathcal H_1}^{\mDR}&=-\left[\re \Sigma'_{HH}\right]_{\alpha=0}^{\text{mdiv}}\ ,\\
\delta Z_{\mathcal H_2}^{\mDR}&=-\left[\re \Sigma'_{hh}\right]_{\alpha=0}^{\text{mdiv}}\ ,\\
\delta\tb^{\mDR}&=\frac12\left(\delta Z_{\mathcal H_2}^{\mDR}-\delta Z_{\mathcal H_1}^{\mDR}\right)\ ,
\end{align}
\end{subequations}
where the notation $[\ ]^{\text{mdiv}}$  means that one now takes only the  
terms proportional to $\Delta_m\equiv\Delta-\log(m_M^2/\mu_{\DR}^2)$.
One can see that if there is only one large scale, this scheme corresponds effectively to the choice $\mu_{\DR}=m_M$ in the $\DR$ scheme, namely:
\begin{equation}
\hat\Sigma_{hh}(p^2)^{\mDR}=\hat\Sigma_{hh}(p^2)^{\DR}\Big|_{\mu_{\DR}=m_M}\ .
\end{equation}
In a general type~I seesaw with three generations, however, there will
be different Majorana masses, $m_{M_1}$,  $m_{M_2}$ and  $m_{M_3}$ , so
the choice of the ``proper'' renormalization scale  
$\mu_{\DR}$ becomes more involved. Besides, there are also new
additional (soft) mass scales from the sneutrino sector, which can be
different for the three generations, and these could also a priori enter
in a non-negligible way into the renormalization procedure. 
This will be discussed in more detail below.

This scheme conserves the good properties that the $\DR$ scheme has, but
is safe from  large logarithmic contributions 
(while leaving the smaller non-logarithmic contributions untouched).
Consequently, this option is often used in the literature
when a large scale is present in the problem. 
One well known example  are the loop
corrections to the beta function in QCD with massive fermions. In fact
such a modified $\DR$ scheme was precisely first proposed in that QCD
context in order to implement properly the matching conditions when
crossing through the various  thresholds, which relate the value of the
strong coupling constant for the case of $n_f+1$ active flavors with the
one with  $n_f$ active flavors. In this QCD case the matching scale is
chosen to be precisely the mass of this fermion ``+1'' that is crossed
by (see, for instance, \cite{NasonDawsonEllis}).

\subsubsection{On-shell scheme}

An on-shell (OS) renormalization requires the derivative of the renormalized self-energy to cancel at the physical mass:
\begin{subequations}
\begin{align}
\re\hat\Sigma_{hh}'(m_h^2)&=0\ ,\\
\re\hat\Sigma_{HH}'(m_H^2)&=0\ .
\end{align}
\label{OSmasses}
\end{subequations}

At one loop level, the physical masses in (\ref{OSmasses}) can be
consistently replaced by the corresponding tree masses, so the field
renormalization constants are set to: 
\begin{subequations}
\begin{align}
\delta Z^{\OS}_{hh}&=-\re\Sigma'(p^2=m_{h,\text{tree}}^2)\ ,\\
\delta Z^{\OS}_{HH}&=-\re\Sigma'(p^2=m_{H,\text{tree}}^2)\ .
\end{align}
\end{subequations}
Using \refeq{deltaZ}, we can write the following relations:
\begin{subequations}
\begin{align}
\delta Z_{\mathcal H_1}^{\OS}&=\frac1{\cos2\alpha}\left(\sin^2\alpha\ \re\hat\Sigma'_{hh}(m_{h,\text{tree}}^2)-\cos^2\alpha \ \re\hat\Sigma'_{HH}(m_{H,\text{tree}}^2)\right)\ ,\\
\delta Z_{\mathcal H_2}^{\OS}&=\frac1{\cos2\alpha}\left(-\cos^2\alpha\ \re\hat\Sigma'_{hh}(m_{h,\text{tree}}^2)+\sin^2\alpha\ \re\hat\Sigma'_{HH}(m_{H,\text{tree}}^2)\right)\ ,
\end{align}
\end{subequations}
which yields for the $\tb$ counterterm, using \refeq{dtb}, 
\begin{equation}
\delta\tb^{\OS}=\frac{-1}{2\cos2\alpha}\left(\re\hat\Sigma'_{hh}(m_{h,\text{tree}}^2)-\re\hat\Sigma'_{HH}(m_{H,\text{tree}}^2)\right)\ .
\end{equation}

Although this OS scheme is interesting due to its intuitive physical
interpretation and its decoupling properties, it can lead to large
corrections to the Higgs boson 
self-energy, which could spoil the convergence of the perturbative
expansion\cite{tbren1,tbren2}.  
Moreover, it  also induces gauge dependence at one-loop level and,
contrary to the $\DR$ scheme, this dependence remains even if one
chooses the class of $R_\xi$ gauges\cite{tbren2}. 


\subsubsection{Decoupling scheme (DEC)}

As we will see explicitly in the next section, the $\mDR$ scheme
removes the large logarithmic terms, but there are still
non-logarithmic finite terms present, which can give non-decoupling
effects.  It has been recently proposed\cite{Haber}
that those finite terms can be removed by hand, forcing the decoupling
to happen. 
This decoupling (DEC) scheme is defined as:
\begin{subequations}
\begin{align}
\delta Z_{\mathcal H_1}^{\rm{DEC}}&=-\left[\re \Sigma_{HH}'(p^2)\right]_{\alpha=0, \,p^2=0}\ ,\\
\delta Z_{\mathcal H_2}^{\rm{DEC}}&=-\left[\re \Sigma_{hh}'(p^2)\right]_{\alpha=0, \, p^2=0}\ ,\\
\delta\tb^{\rm{DEC}}&=\frac12\left(\delta Z_{\mathcal H_2}^{\rm{DEC}}-\delta Z_{\mathcal H_1}^{\rm{DEC}}\right)\ .
\end{align}
\end{subequations}

The convenience of this scheme in the context of effective field theories
has been discussed in \citere{Haber}. The advantage of this scheme is that, by
construction, it implements the proper matching between the high energy
theory and the intermediate energy effective theory.  However, we prefer
here not to use an effective field theory approach where the heavy degrees  
are explicitly integrated out (like the possible use of a derived one-loop
effective potential), because we do not want to assume in the present
computation any specific intermediate low energy effective theory, but we wish
simply to ensure that the final low energy effective theory where all the
non-SM particles are decoupled is indeed the SM as expected. Consequently, in
our  analysis we perform the one-loop computation in the full
high energy theory including explicitly the heavy particles with several
different mass scales involved (using an appropriate renormalization
scheme), and use these masses as input parameters that 
will be varied in the posterior numerical analysis within a wide range from
high to low energies. Correspondingly, the disadvantage of the DEC scheme is
that by assuming the MSSM as the explicit intermediate low energy effective
theory, any dependence on the heavy neutrinos/sneutrinos is by construction
fully removed already at the intermediate (SUSY) energy scales.


\subsubsection{Higgs mass scheme (HM)}

Another possibility is to demand that some physical quantity, e.g., the mass $m_H$, is given at one loop level by its tree level expression:
\begin{equation}
 m_{H,\text{1 loop}}^2=m_{H,\text{tree}}^2+\hat\Sigma_{HH}(p^2=m_{H,\text{tree}}^2)=m_{H,\text{tree}}^2\ . 
 \end{equation}
This condition defines the \textit{Higgs mass} (HM) scheme and fixes, from \refeq{deltamH}, the $\tb$ counterterm to:
\begin{align}
\delta\tb^{\rm HM}=& \frac1{\MZ^2\sin2\beta\sin2(\alpha+\beta)}
\bigg\{ \delta M_A^2\sin^2(\alpha-\beta)+\delta \MZ^2\cos^2(\alpha+\beta) -\Sigma_{HH}(m_{H,\text{tree}}^2)\\
&-\frac e{2\MW\sw}\Big(\delta T_H\cos(\alpha-\beta)\big(1+\sin^2(\alpha-\beta)\big)+\delta T_h\sin(\alpha-\beta)\cos^2(\alpha-\beta)\Big)\bigg\}\ .\non
\end{align}

The HM scheme, as any other scheme that is defined in terms of physical masses, provides manifestly a gauge-independent definition of $\tb$ \cite{tbren2}. However, it is not numerically stable either, as has been shown in \cite{tbren2}, so the convergence of the perturbative expansion is again not ensured.


\section{Analytic results and analysis of the relevant terms}
\label{Analytic}

In this section we discuss the calculation of the higher-order
corrections to the light Higgs boson mass, and in particular discuss
analytically the decoupling behavior of the various schemes in the
case of three generations of (s)neutrinos. Going from the one
generation to the three generations case, due to the appearance of
relevant generation mixing, the corresponding radiative corrections cannot be
trivialy extrapolated and they must be explicitly and separately computed. 

We have used the Feynman diagrammatic (FD) approach to calculate the
one-loop corrections from the neutrino/sneutrino sector to the MSSM
Higgs boson masses. The full one-loop neutrino and sneutrino corrections
to the self-energies, $\Sigma^{\nu/\tilde{\nu}}_{hh}$ and
$\Sigma^{\nu/\tilde{\nu}}_{HH}$, entering the computation have been
evaluated with the help of \textit{FeynArts}~\cite{FeynArts1,FeynArts2,FeynArts3,FeynArts4,FeynArts5,FeynArts6} and
\textit{FormCalc}~\cite{FormCalc}.  The relevant Feynman rules for the
present computation with three generations of Majorana neutrinos and
sneutrinos have been  derived from the Lagrangians of section
\ref{sec:nN} and expressed in terms of the physical basis. The results
are collected in \refapp{sec:fmr} (to our knowledge, they  are
not available in the previous literature).
These Feynman rules have also
been inserted into a new model file which is available upon request. 

\begin{figure}[hbtp]
\centering
\includegraphics[trim = 20mm 140mm 20mm 5mm, clip, width=15cm]{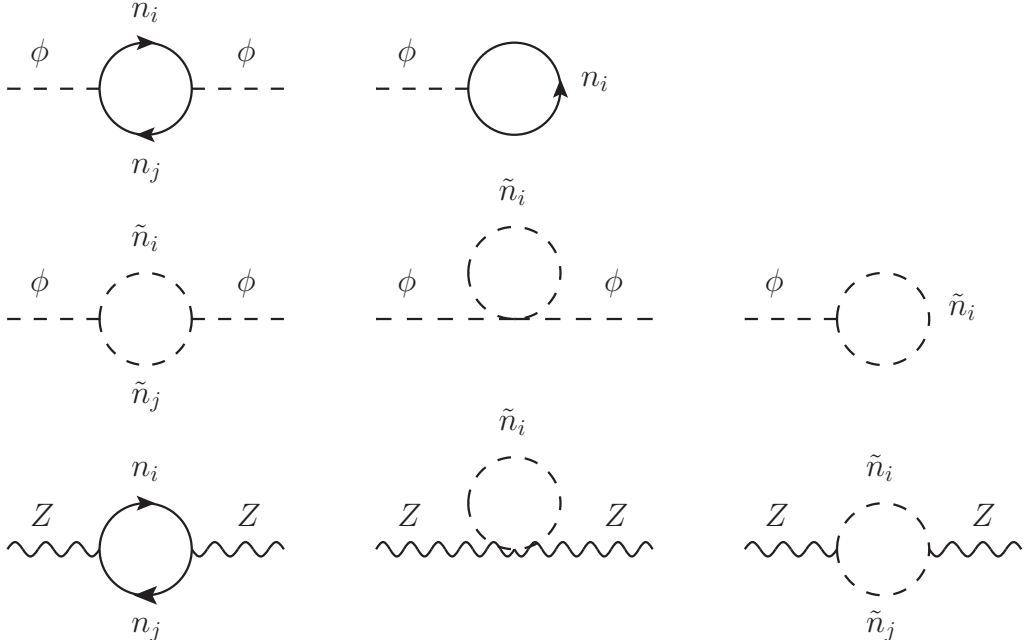}
\caption{Generic one-loop Feynman-diagrams contributing to the computation of the one-loop new corrections to the Higgs boson mass form neutrinos and sneutrinos. Here $\phi=h, H, A$.}
\label{diagrams}
\end{figure}

The generic one-loop Feynman diagrams that enter in the computation of the renormalized self-energies are collected in \reffi{diagrams}. They include the two-point (one-point) diagrams in the Higgs self-energies (tadpoles), and the two-point diagrams in the $Z$ boson self-energy. Here the notation is: $\phi$ refers to all physical neutral Higgs bosons, $h$, $H$, and $A$; $n_{i}\,\,\left(\text{where}\,\,i=1,\ldots,6\right)$ refers to all physical neutrinos; $\tilde{n}_{i}\,\,\left(\text{where}\,\,i=1,\ldots,12\right)$ refers to all physical sneutrinos; and $Z$ refers to the $Z$ gauge boson.\\

Following a similar analysis here as the one performed in \cite{Ana} for the one generation case, it is illustrative to expand the renormalized self-energy in powers \footnote{Notice that only even powers of $m_D$ are present in this expansion\cite{Ana}.} of $m_D$:
\begin{equation}
\hat\Sigma^{\nu/\tilde{\nu}}(p^2)=\Big(\hat\Sigma^{\nu/\tilde{\nu}}(p^2)\Big)_{m_D^0}+\Big(\hat\Sigma^{\nu/\tilde{\nu}}(p^2)\Big)_{m_D^2}+\Big(\hat\Sigma^{\nu/\tilde{\nu}}(p^2)\Big)_{m_D^4}+\dots
\end{equation}
where $\Big(\hat\Sigma^{\nu/\tilde{\nu}}(p^2)\Big)_{m_D^n}$ means $\mathcal{O}(m_D^n)$ terms in the expansion of $\hat\Sigma^{\nu/\tilde{\nu}}(p^2)$ in powers of $m_D$. For the present case with three generations $m_D^2$ represents shortly products of two Dirac matrices, such as, $m_D^\dagger m_D$ or $m_D m_D^T$; equivalently, $m_D^4$ refers to combinations of four matrices.

The first term in this expansion is independent of both $m_D$ and $m_M$, and represents, therefore, the pure gauge contribution (i.e. the result for $Y_\nu=0$), which is already present in the MSSM. 
On the other hand, the term proportional to $m_D^4$ is actually of order $\mathcal O(m_D^4/m_M^2)$ (see \cite{Ana} for details), so it is suppressed by the Majorana mass;  higher order terms in this expansion are also suppressed by inverse powers of the Majorana mass.
Thus, the new relevant contributions, coming from the neutrino and sneutrino sectors are those governed by the Yukawa couplings, and can arise only from the order $\mathcal O(m_D^2)$ terms. 
Thus we have: 
\begin{subequations}
\begin{align}
\Big(\hat\Sigma^{\nu/\tilde{\nu}}(p^2)\Big)_{\rm{full}}& = 
\Big(\hat\Sigma^{\nu/\tilde{\nu}}(p^2)\Big)_{\rm{gauge}} + \Big(\hat\Sigma^{\nu/\tilde{\nu}}(p^2)\Big)_{\rm{Yukawa}}\ \label{full},\\
\Big(\hat\Sigma^{\nu/\tilde{\nu}}(p^2)\Big)_{\rm{gauge}}&= \Big(\hat\Sigma^{\nu/\tilde{\nu}}(p^2)\Big)_{m_D^0}\ \label{gauge},\\
\Big(\hat\Sigma^{\nu/\tilde{\nu}}(p^2)\Big)_{\rm{Yukawa}}&= \Big(\hat\Sigma^{\nu/\tilde{\nu}}(p^2)\Big)_{m_D^2}+\mathcal O\Big(\frac{m_D^4}{m_M^2}\Big)\ .\label{yuk}
\end{align}
\end{subequations}
In the one generation case, the Dirac mass is related to the light, $m_\nu$, and heavy, $m_N$, neutrino physical masses by \cite{Ana} $m_D^2=-m_\nu m_N\approx- m_\nu m_M$. 
In the three generations case, a similar functional dependence of $m_D^2$ with the physical masses in $m_\nu^{\rm diag}$ and in $m_N^{\rm diag}$ is found, as it is explicitely manifested in the parametrization of (\ref{Casas}).
This means that the Yukawa contribution in \refeq{yuk}, being  proportional to $m_D^2$, grows with the Majorana masses, therefore leading to potential non-decoupling effects with respect to these masses. 
The question now is whether such a term is present in the renormalized self-energy and, in that case, if it is numerically relevant. 
This issue was first analyzed for the one generation case in \cite{Ana}, and recently in \cite{Haber}, showing that the presence and relevance of the $\mathcal O(m_D^2)$ term in \refeq{yuk} depends on the chosen renormalization scheme for $\tb$.\\

In order to better understand where these differences come from, it is
interesting to look first for the $\mathcal O(m_D^2)$ terms in the bare
self-energy, where the choice of the renormalization scheme does
not enter.
We will focus here on the lightest $\cp$-even Higgs boson
self-energy, but the conclusions will be the same for the full $(h,H)$
system.  
By computing the one-loop contributions from the $hh$ diagrams in 
\reffi{diagrams} we have obtained the following analytical result for the
$\mathcal O(m_D^2)$ contributions from three generations of neutrinos
and sneutrinos to the bare self-energy: 
\begin{eqnarray}
\label{haberlike}
\Big(\Sigma^{\nu/\tilde{\nu}}_{hh}(p^2)\Big)_{m_D^2}&=& \displaystyle{\frac{g^2}{64\pi^2\MW^2\sin^2\beta}\sum_{i=1}^{3} 
\Big(m_D^\dagger m_D\Big)_{ii}\Bigg\{
\Big[\Delta + 1-\log \frac{m_{M_i}^{\ 2}}{\mu_{\DR}^2}\Big]}\nonumber\\
&\times&\Big(\big(p^2-\MZ^2\big)2\cos^2\alpha
  \nonumber \\
&&-\MZ^2\sin^2\beta \big( \cos^2\alpha (4-3 \cot^2\beta)+ 
2 \sin 2\alpha \cot\beta - \sin^2 \alpha
 \big) \Big) \nonumber \\
&+&\displaystyle{  \Big[\Delta-\log\frac{m_{M_i}^2}{\mu_{\DR}^2}\Big](4m^2_{\text{SUSY}}\cos^2\alpha)}\Bigg\}\ .
\end{eqnarray}
In this expression, for shortness, we have set $a_\nu=b_\nu=\mu=0$, and
we have considered the most simple case with just one single soft mass
scale in the slepton sector, 
$m_{\tilde L_{i}}=m_{\tilde R_{j}} =: m_{\text{SUSY}}$, with
$i=1,2,3$. $\Delta$ is defined in (\ref{deltaeps}) and $\mu_{\DR}$ is
again the renormalization scale. 
The corresponding result for the $\Sigma_{HH}$ is obtained from the
above formula by replacing $\cos\alpha\rightarrow\sin\alpha,\,
\sin\alpha\rightarrow-\cos\alpha$.

First of all, it should be noted that the result in (\ref{haberlike}) is a pure ${\cal O}(Y_\nu^2)$ radiative correction with an overall factor given by:
\begin{equation}
\displaystyle{\frac{g^2}{64\pi^2\MW^2\sin^2\beta}\sum_{i=1}^{3} 
\Big(m_D^\dagger m_D\Big)_{ii}=\frac{1}{32\pi^2}\sum_{i=1}^{3}\Big(Y_\nu^\dagger Y_\nu\Big)_{ii}.}
\end{equation}    
Secondly, a good check of our computation in (\ref{haberlike})  is that by setting to zero all the entries in the 
$m_{D_{ij}}$ matrix except for one in the diagonal (for instance,  $m_{D_{11}}$) we recover the result of the one generation case, in full agreement with the expressions in the Appendix~A of \cite{Haber} (with $m_{D_{11}}= m_{D}$ and $m_{M_1}=m_M$).

The result in  (\ref{haberlike})  shows, most importantly,  that the
bare self-energy  has a non-negligible $\mathcal O(m_D^2)$ term, which
grows logarithmically with the Majorana masses. Nevertheless, as
we have already said, we will analyze whether such a term is present or
not in the  renormalized self-energy.  
If one assumes that the Yukawa contribution from neutrinos/sneutrinos to
the bare self-energy is approximated by the previous result in
(\ref{haberlike}), one arrives at the following
$\mathcal O(m_D^2)$ expressions for the $\tb$ counterterms in the various
schemes: 
\begin{eqnarray}
(\delta\tb^{\DR})_{m_D^2} &=&
\displaystyle{-\frac{g^2}{64\pi^2\MW^2\sin^2\beta}\sum_{i=1}^{3} 
\Big(m_D^\dagger m_D\Big)_{ii} 
\Big[\Delta  \Big]},\nonumber\\
(\delta\tb^{\mDR})_{m_D^2} &=&
\displaystyle{-\frac{g^2}{64\pi^2\MW^2\sin^2\beta}\sum_{i=1}^{3} 
\Big(m_D^\dagger m_D\Big)_{ii} 
\Big[\Delta -\log \frac{m_{M_i}^{\ 2}}{\mu_{\DR}^2} \Big]},\nonumber\\
(\delta\tb^{\rm OS})_{m_D^2} &=&(\delta\tb^{\rm DEC})_{m_D^2}=
(\delta\tb^{\rm HM})_{m_D^2}=\nonumber\\
& &\displaystyle{-\frac{g^2}{64\pi^2\MW^2\sin^2\beta}\sum_{i=1}^{3} 
\Big(m_D^\dagger m_D\Big)_{ii} 
\Big[\Delta + 1 -\log \frac{m_{M_i}^{\ 2}}{\mu_{\DR}^2} \Big]}.
\label{tanbetacount}
\end{eqnarray}
Then, one can easily find the relation among the corresponding renormalized $\tb$ values, at this same level of approximation. Using, for instance, the renormalized value in the OS scheme, $\tb^{\rm OS}$,  which is $\mu_{\DR}$ independent, as the reference value to be compared with in this illustrative exercise, we get:
\begin{eqnarray}
(\tb^{\rm OS})_{m_D^2}&=&(\tb^{\rm DEC})_{m_D^2}=(\tb^{\rm HM})_{m_D^2} ,\nonumber\\
(\tb^{\DR})_{m_D^2}-(\tb^{\rm OS})_{m_D^2}&=& 
\displaystyle{-\frac{g^2 \tb}{64\pi^2\MW^2\sin^2\beta}\sum_{i=1}^{3} 
\Big(m_D^\dagger m_D\Big)_{ii} 
\Big[ 1 -\log \frac{m_{M_i}^{\ 2}}{\mu_{\DR}^2} \Big]},\nonumber\\
(\tb^{\mDR})_{m_D^2}-(\tb^{\rm OS})_{m_D^2}&=& 
\displaystyle{-\frac{g^2 \tb}{64\pi^2\MW^2\sin^2\beta}\sum_{i=1}^{3} 
\Big(m_D^\dagger m_D\Big)_{ii} 
\Big[ 1 \Big]}.
\label{tanbetar}
\end{eqnarray}  
Finally, using the computed expressions at $\mathcal O(m_D^2)$ of the
bare self-energy and the counterterms one obtains the
renormalized self-energy at this same order. In the case of the $\DR$
scheme we get: 
\begin{eqnarray}
\label{sigmarenDR1}
\Big(\hat{\Sigma}^{\nu/\tilde{\nu}\,\,{\DR}}_{hh}(p^2)\Big)_{m_D^2}&=& \displaystyle{\frac{g^2}{64\pi^2\MW^2\sin^2\beta}\sum_{i=1}^{3} 
\Big(m_D^\dagger m_D\Big)_{ii}
\Big[ 1-\log \frac{m_{M_i}^{\ 2}}{\mu_{\DR}^2}\Big]}\nonumber\\
&\times&
\Big[  -2 M_A^2 \cos^2(\al-\be)\cos^2\be +2 p^2 \cos^2\al \nonumber \\
&&- \MZ^2 \sin\be\sin(\al+\be)\left(2 \left(1+\cos^2\be\right)\cos\al-\sin2\beta\sin\al \right)  \Big],
\end{eqnarray}
which can be rewritten in terms of $m_{h, {\rm tree}}$ simply as:
\begin{eqnarray}
\label{sigmarenDR2}
\Big(\hat{\Sigma}^{\nu/\tilde{\nu}\,\,{\DR}}_{hh}(p^2)\Big)_{m_D^2}&=& \displaystyle{\frac{g^2}{64\pi^2\MW^2\sin^2\beta}\sum_{i=1}^{3} 
\Big(m_D^\dagger m_D\Big)_{ii}
\Big[ 1-\log \frac{m_{M_i}^{\ 2}}{\mu_{\DR}^2}\Big]}\nonumber\\
&\times&
\Big[
\big(p^2-m_{h,\text{tree}}^2\big)2\cos^2\alpha-\MZ^2\sin2\beta\sin2(\alpha+\beta)  \Big].  
\end{eqnarray}
Notice that there are no terms proportional to $m_{\rm SUSY}^2$ in~\refeq{sigmarenDR2}, since they are cancelled by the $\delta T_h,\delta T_H,\delta M_A^2$ and $\delta M_Z^2$ counterterms.
We have numerically studied the accuracy of these approximate  $\mathcal
O(m_D^2)$ results, both for the renormalized self-energy and the finite
contribution in the bare self-energy, and
compared with their corresponding full results. 
We have found that they constitute extremely good approximations, leading to
relative differences below $10^{-4}$ w.r.t.\ the full expressions for all the explored parameter space (including for non-zero values of $a_\nu$, $b_\nu$ and $\mu$). 
  
It is also straight forward to check that by setting properly the $m_D$
matrix entries in (\ref{sigmarenDR1}) and (\ref{sigmarenDR2}) we recover again
the proper results for the one generation case, in agreement with \cite{Ana}
and \cite{Haber}. 

Similarly, one can derive the corresponding  $\mathcal O(m_D^2)$ expressions in the other considered schemes.  In the 
$\mDR$ we get:
\begin{eqnarray}
\Big(\hat{\Sigma}^{\nu/\tilde{\nu}\,\,{\mDR}}_{hh}(p^2)\Big)_{m_D^2}&=& \displaystyle{\frac{g^2}{64\pi^2\MW^2\sin^2\beta}\sum_{i=1}^{3} 
\Big(m_D^\dagger m_D\Big)_{ii}
\Big[ 1 \Big]}\nonumber\\
&\times&
\Big[
\big(p^2-m_{h,\text{tree}}^2\big)2\cos^2\alpha-\MZ^2\sin2\beta\sin2(\alpha+\beta)  \Big].  
\end{eqnarray}
And in the OS, DEC and HM we get the expected decoupling behavior at this order, in agreement with the results for the one generation case in \cite{Ana} and \cite{Haber}:
\begin{eqnarray}
\label{sigmarenmDR}
\Big(\hat{\Sigma}^{\nu/\tilde{\nu}\,\,{\rm OS}}_{hh}(p^2)\Big)_{m_D^2}&=&
 \Big(\hat{\Sigma}^{\nu/\tilde{\nu}\,\,{\rm DEC}}_{hh}(p^2)\Big)_{m_D^2} = 
 \Big(\hat{\Sigma}^{\nu/\tilde{\nu}\,\,{\rm HM}}_{hh}(p^2)\Big)_{m_D^2}
 \nonumber\\
&=& \displaystyle{\frac{g^2}{64\pi^2\MW^2\sin^2\beta}\sum_{i=1}^{3} 
\Big(m_D^\dagger m_D\Big)_{ii}
\Big[ 0 \Big]}\nonumber\\
&\times&
\Big[
\big(p^2-m_{h,\text{tree}}^2\big)2\cos^2\alpha-\MZ^2\sin2\beta\sin2(\alpha+\beta)  \Big].  
\end{eqnarray}

In summary, in this section we have analyzed the relevant differences
among the various schemes for $\tb$ and the wave function
renormalizations, and these differences have been understood in terms of
$\mathcal O(m_D^2)$ contributions to the self-energies. Once we have set
clearly these differences, it is a simple exercise to find the
prediction in one scheme and then extract from it the prediction in
another scheme. 

\begin{figure}[hbtp]
\centering
\includegraphics[width=0.52\textwidth]{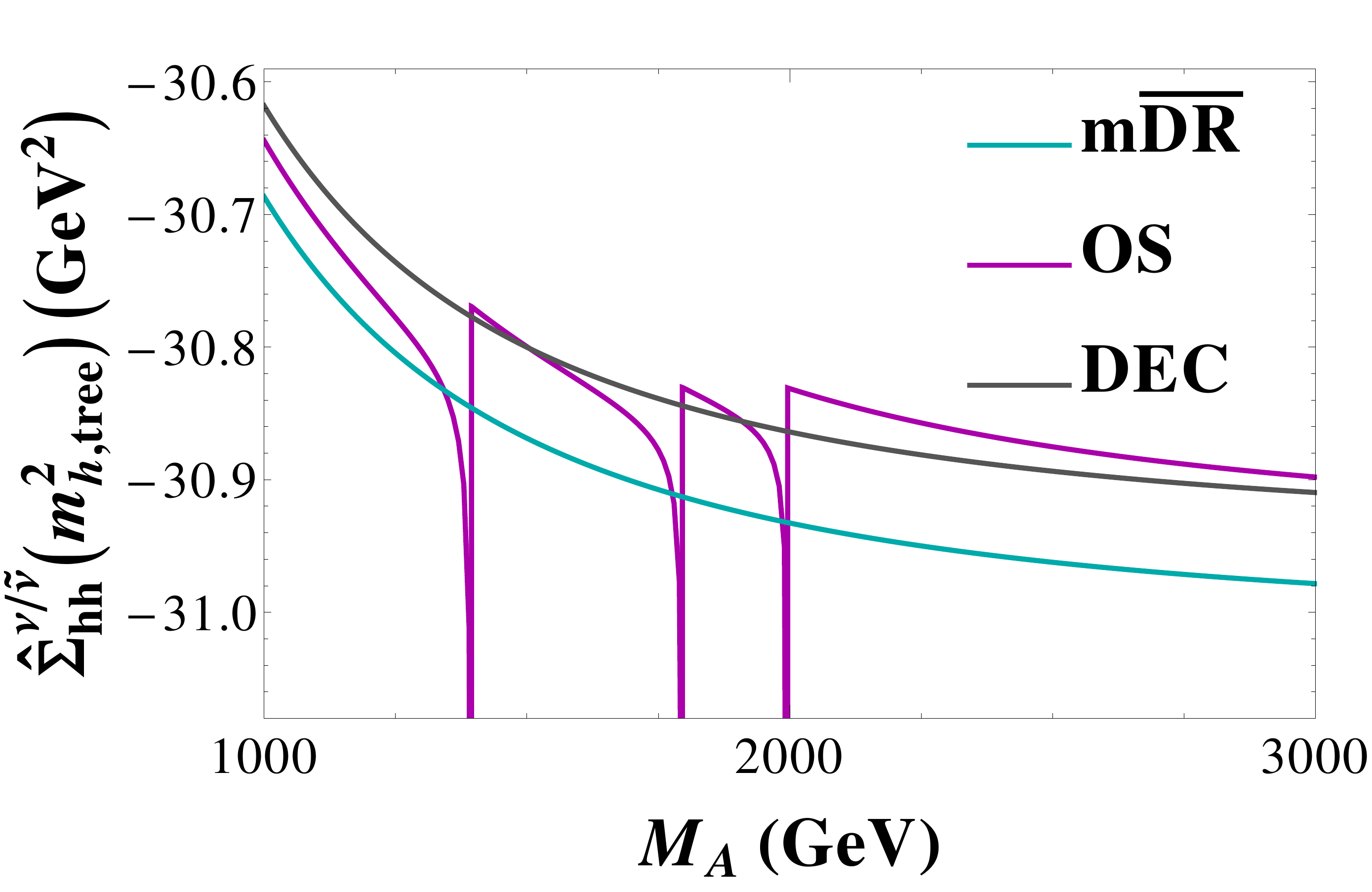} 
\includegraphics[width=0.47\textwidth]{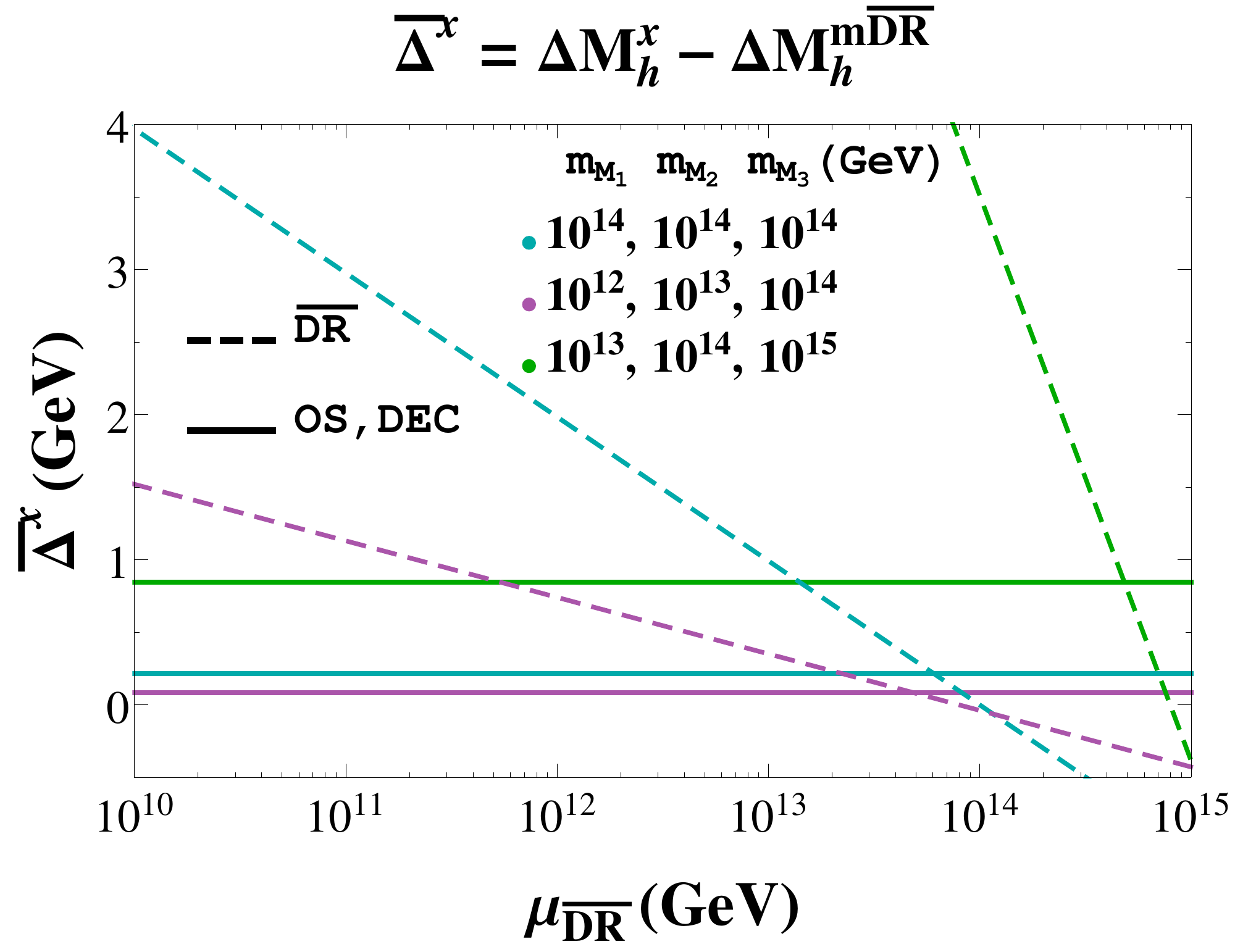}
\caption{ Comparison among the various schemes. The plot in the left
  shows the renormalized self-energies evaluated at  $p^2=m_{h,{\rm
      tree}}^2$ in the OS, DEC and $\mDR$ as functions of $M_A$, for
  $m_{M_{1,2,3}}=10^{12}~{\rm GeV}$, $m_{\nu_1}=0.01~{\rm eV}$ and
  $m_{\tilde L_{1,2,3}}=(700,900,1000)~{\rm GeV}$.  
The plot in the right shows the predictions of the mass differences
${\overline \Delta}^{\rm x} = \Delta M_h^{\rm x} - \Delta M_h^{\mDR}$, 
for ${\rm x} = {\DR}$ (dashed
lines), OS, DEC (solid lines) , as functions of $\mu_{\DR}$, and for
several choices of the Majorana masses, 
$(m_{M_1},m_{M_2},m_{M_3})$(GeV): $(10^{14},10^{14},10^{14})$ (in light
blue); $(10^{12},10^{13},10^{14})$ (in purple) and 
$(10^{13},10^{14},10^{15})$ (in green). The rest of input parameters are
fixed as in (\ref{values}). $\Delta M_h$ is defined
in~\refeq{delta_mass}.}
\label{compareschemes}
\end{figure}

\medskip
We illustrate numerically the most relevant differences among the
various schemes in \reffi{compareschemes}. The plot on the left displays
the renormalized self-energies in three schemes that are $\mu_{\DR}$
independent: OS, DEC and $\mDR$. In all the cases we plot the full
one-loop result from neutrinos and sneutrinos evaluated at the tree
Higgs mass,  
$p^2=m_{h,{\rm tree}}^2$, as a function of $M_A$. In this example 
the instabilities that are found in the OS scheme are clearly
visible, in comparison with
the stability of the $\mDR$ and DEC schemes. 
These ``dips'' are due to thresholds encountered in the loop diagrams
and, as can be seen in \reffi{compareschemes}, appear at $\MA$ values
approximately twice each one of the soft SUSY-breaking parameters 
$m_{\tilde L_i}$. We have checked that these ``dips'' are indeed very narrow
and profound. For aritrary close values to threshold they go
to $-\infty$ due to the fact that the imaginary part of the standard one-loop $B_0$~function\cite{Denner}
is not differentiable at threshold. These instabilities occur as long
as width effects are not taken into account.
We also see that,
for the input values in this plot, the numerical values for the
renormalized self-energies of the OS, DEC and $\mDR$ are quite close to
each other. In particular, in the region out of the dips, the OS and DEC
values are practically identical. We have also checked that the numerical
results in the HM scheme (not shown) also manifest instabilities and,
furthermore, they turn out to be substantially different than in the
other  $\mu_{\DR}$ independent schemes. This difference of the HM has
been studied in \cite{Haber} in the one generation case and it has been
understood in terms of the substantially different contributions in the
pure gauge part, i.e  of $\mathcal O(m_D^0)$,  which are numerically
relevant. For instance, comparing the HM with the DEC approximate
results for the mass correction  in \cite{Haber}, the first one is a
factor of  
$(\cos^2 2 \beta)^{-1}$ larger than the last one (for $\tb=2$,
e.g., this yields a factor of 2.8).
We have found agreement with this numerical factor in our numerical
results for $\hat{\Sigma}^{\nu/\tilde{\nu}\,\,{\rm HM}}_{hh}(m_{h,\text{tree}}^2)$,
in the region out of the instabilities.

The right plot of \reffi{compareschemes} compares the predictions
for the Higgs mass correction among the different renormalization
schemes in various examples with different choices for the Majorana
masses and their hierarchies. Again the full one-loop renormalized
self-energies are considered and the simple formula for the Higgs mass
correction in (\ref{delta_mass}) is used. In this plot we have chosen
the $\mDR$ as the reference scheme to be compared with, such that 
${\overline \Delta}^{\rm x} = \De M_h^{\rm x} - \De M_h^{\mDR}$
represents the difference in 
the prediction of the mass correction in the scheme $x$ respect to the
prediction in the $\mDR$ scheme. Firstly, we have found again that the
results of the OS and the DEC schemes are practically
indistinguishable. We also see that for the input values explored in
this plot, the predictions in these OS and DEC schemes differ from the
predictions in the 
$\mDR$ scheme in $1 \gev$ at most, and this largest difference is for the
case when the heaviest Majorana mass is at the largest considered value
of $10^{15} \gev$. The comparison with the $\DR$ scheme, whose result is
$\mu_{\DR}$ dependent, shows that, in order to get a prediction close to
the other schemes, within say a 1 GeV interval, a value of $\mu_{\DR}$
at  the near proximity of the highest  Majorana mass should be chosen.


\section{Numerical analysis of $\mathbf {\Delta M_{h}}$}
\label{Numerical}

In this final section we show some numerical results for the
one-loop corrections to the light Higgs boson mass, $\Delta M_h$
(via \refeq{delta_mass}).
Using the DEC scheme, the OS scheme or another scheme that decouples the heavy mass
scales completely, would yield small effects (except where the
numerical instabilities occur as demonstrated in \refse{sec:tb-ren}). 
Since every scheme, however, 
has its advantages and disadvantages as discussed in
\refse{sec:tb-ren} we choose here to use the $\mDR$ scheme. The
numerical results in other schemes can be 
inferred from these by using the results in the previous section. 
While by definition not showing full decoupling, 
the $\mDR$ combines
several of the desired properties: stability, perturbativity and
gauge invariance at the one-loop level. Besides, this scheme is safe of
large logarithms introduced by the large Majorana scales. The fact that
the non-logarithmic finite terms are not removed in this scheme,
translates into a finite contribution of  $\mathcal O(m_D^2)$ which will
leave a non-vanishing radiative contribution from the neutrinos and
sneutrinos into the Higgs mass correction.
Furthermore, we are interested in different scenarios where the Majorana masses can range from the extreme large values of order $10^{14}-10^{15}$~GeV down to low values of order $10^3$~GeV and, correspondingly, we will explore these scenarios keeping explicitly the contributions from the $\nu/\Snu$ particles.
Consequently, the
numerical analysis is performed as a function of all relevant
parameters that will be varied in a wide range: the masses of the light neutrinos, the masses of the
heavy Majorana neutrinos and the mixing provided by the $R$ matrix in
the case of three generations, as well as the MSSM parameters. Unless
stated otherwise, we set the parameters to the following reference
values: 
\begin{equation}
\label{values}
\begin{array}{rcl}
m_{M_1} = m_{M_2} = m_{M_3} \equiv m_{M} = 10^{14} \text{ GeV}\,, &\quad& m_{\nu_1} = 0.1\text{ eV}\,,\\
m_{\tilde{L}_1}=m_{\tilde{L}_2}=m_{\tilde{L}_3}\equiv m_{\tilde{L}} = 2 \text{ TeV} \,,&\quad& M_{A} = 500 \text{ GeV}\,,\\
m_{\tilde{R}_1}=m_{\tilde{R}_2}=m_{\tilde{R}_3}\equiv m_{\tilde{R}} = 2 \text{ TeV} \,,&\quad& \mu = 500 \text{ GeV}\,,\\
a_{\nu} = 2 \text{ TeV} \,,&\quad& \tb = 2\,,\\  
b_{\nu} = 2 \text{ TeV}\,,&\quad& R=
\one. 
\end{array}
\end{equation}
The masses of the other two light neutrinos are obtained from
$m_{\nu_1}$ and the mass differences given in \refeq{input_values}, 
implying that these light neutrinos of our reference case are
quasi-degenerate.

We assume that the other MSSM parameters, 
in particular from the top/scalar top sector,
which do not affect our results, give a corrected Higgs mass of
$\Mh \sim 125 \gev$. Here it should be noted that in the
non-(s)neutrino part of the calculation a \DRbar\ renormalization of
$\tb$ and the wave function of the two Higgs doublets has been used
(with $\mu_{\DR} = \mt$). 
The choice of a different renormalization scale
in the estimate of $\Mh$ within the MSSM has been discussed at length in the
literature (see for instance \citeres{tbren1,tbren2}), but it is not relevant
for the present work given the fact that we are using this $\Mh$ as a given
value (fixed here to $125 \gev$) and we are estimating just the shift 
$\Delta \Mh$ with respect 
to this value due to the new sectors $\nu/\Snu$ (given by
\refeq{delta_mass}).

Two different scenarios for the mass hierarchy of the light neutrinos
can be set, the normal hierarchy (NH) case and the inverted hierarchy (IH)
case: 
\begin{itemize}
\item{Normal hierarchy (NH):}\\
$\nu_1$ is the lightest neutrino, and its mass will be our input value. The mass of the other two neutrinos are fixed by the experimental mass differences:
\begin{eqnarray}
m_{\nu_2}^{{\rm NH}}&=&\sqrt{m_{\nu_1}^2+\Delta m_{21}^2}\,, \nonumber\\
m_{\nu_3}^{{\rm NH}}&=&\sqrt{m_{\nu_1}^2+\Delta m_{21}^2+\Delta m_{32}^2}\,.
\end{eqnarray}
\item{Inverted hierarchy (IH):}\\
$\nu_3$ is the lightest neutrino, and its mass will be our input value. The mass of the other two neutrinos again, are fixed by the experimental mass differences:
\begin{eqnarray}
m_{\nu_1}^{{\rm IH}}&=&\sqrt{m_{\nu_3}^2-\Delta m_{21}^2-\Delta m_{32}^2}\,, \nonumber\\
m_{\nu_2}^{{\rm IH}}&=&\sqrt{m_{\nu_3}^2-\Delta m_{32}^2}\,.
\end{eqnarray}
\end{itemize}
with $\Delta m_{21}^2$ and $\Delta m_{32}^2$ are given in section \ref{sec:nN}. 
The default choice used below is the NH case, and the IH case will be
especially indicated.

Notice that we are using the Casas-Ibarra parametrization \refeq{Casas}
that provides a prediction of the full $3\times 3$ $v_2 Y_\nu$
(i.e. $m_D$) matrix in terms of the input parameters of the light
sector, $m_{\nu_{i}}$ and $\theta_{ij}$, and of the heavy sector
$m_{M_{i}}$ and $\theta_{i}$, and the last two can take in principle any
value.  Therefore the size of the Yukawa couplings
that we are generating is  related directly to these parameters, and
in consequence they can be large and even non-perturbative. In order to
ensure that $Y_\nu$ is inside the perturbative region, for every
set of input parameters we first check that all of the entries of the
Yukawa matrix fulfill a perturbative condition that we set here to 
  \begin{equation}
   \dfrac{\vert (Y_\nu)_{ij}\vert^2}{4\pi}<1.5,
 \end{equation}  
 otherwise, the point in the parameter space is rejected.
 

\subsection{Relation with the one-generation case}

\begin{figure}[hbtp]
\centering
\includegraphics[width=0.49\textwidth]{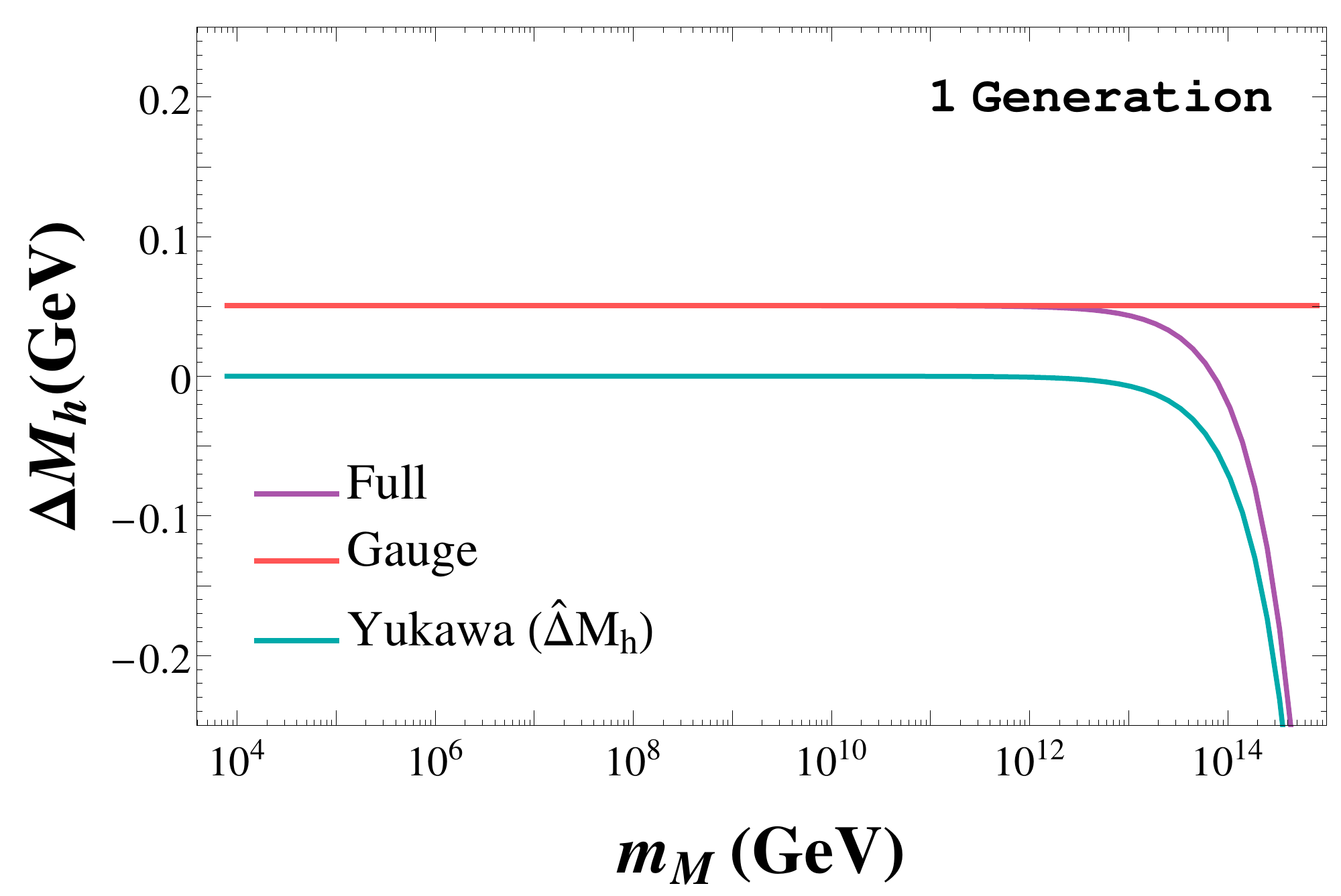} 
\includegraphics[width=0.49\textwidth]{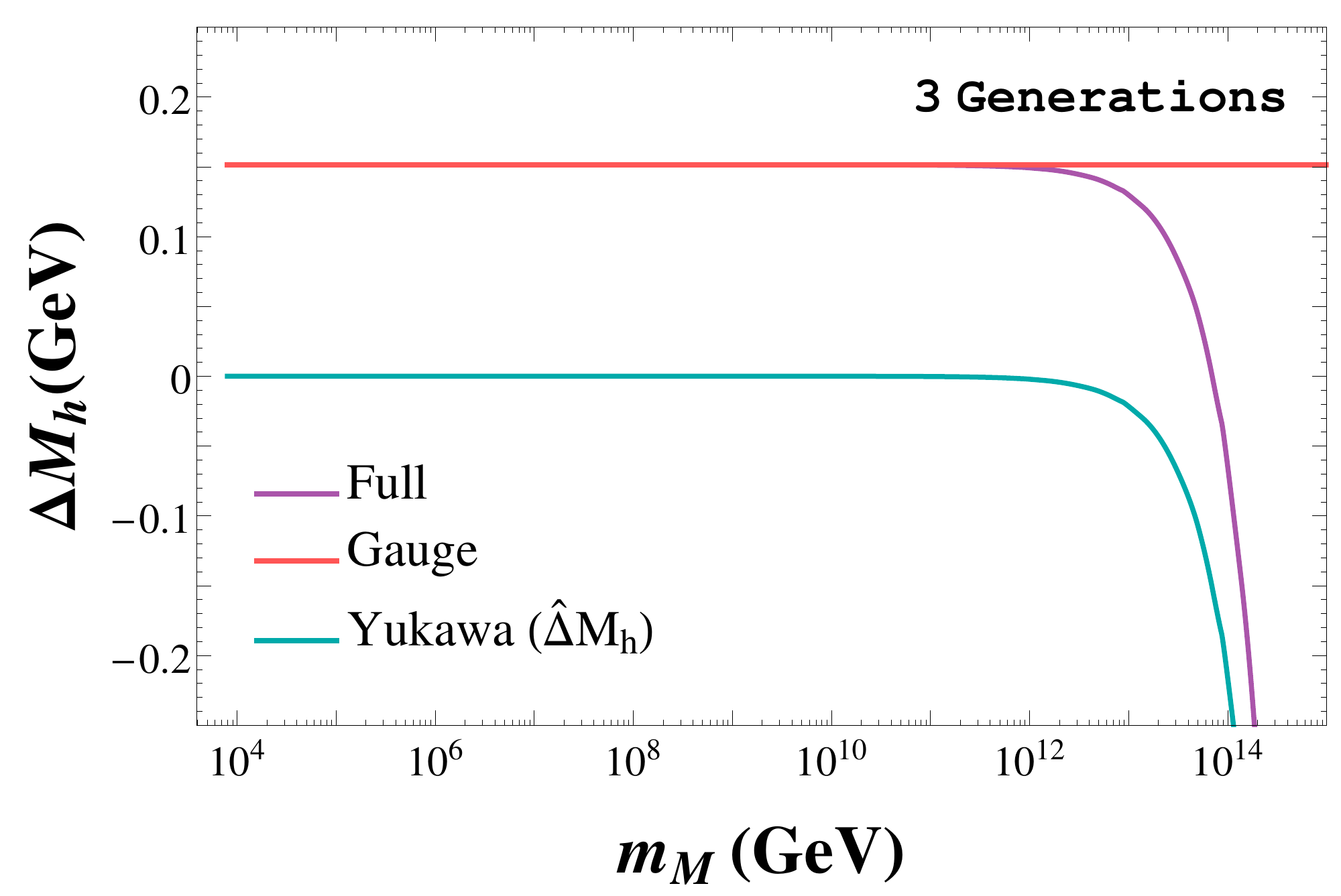}
\caption{Left panel: $\Delta M_h$ as a function of $m_M$ for the one
  generation case. Right panel: $\Delta M_h$ as a function of the scale $m_M$
  for the degenerate three generations case. The rest of the model
  parameters are set as in~\refeq{values}.} 
\label{1G3G}
\end{figure}

As a first check of our three generations code, we have reproduced with
this code the same behavior of the Higgs mass correction, $\Delta M_h$,
with the Majorana mass  as in the one generation case \cite{Ana}.  The
connection with the one generation case is done by setting the
corresponding absent  entries in the Dirac mass  matrix to zero. For
this analysis, the mass of the light and  heavy Majorana neutrinos have
been set to $0.1$ eV and $10^{14} \gev$ respectively. The result for the
one-generation case delivered in such a way is shown in the left plot of
\reffi{1G3G}. In the right plot it is shown the behavior of the three
generations case with three equaly heavy neutrino masses,
i.e. $m_{M_i}=m_M$.  As expected, we obtain that the  Higgs mass
corrections  in the three generations case are three times the ones of
the one generation case.  Notice that we have separated the
contributions to the full mass correction coming from the gauge and the
Yukawa parts, according to (\ref{full}): 
\begin{equation}
\Delta M_h=\left(\Delta M_h\right)_{\text{gauge}}+\left(\Delta M_h\right)_{\text{Yukawa}}\,,
\label{Dmhfull}
\end{equation}
where $\left(\Delta M_h\right)_{\text{gauge}}$ corresponds to setting all the Yukawa couplings to zero and $\left(\Delta M_h\right)_{\text{Yukawa}}$ is the remaining contribution. Within our approximation of \refeq{delta_mass}, they are related with the renormalized self energy as follows:
\begin{subequations}
\begin{align}
\left(\Delta M_h\right)_{\text{gauge}}&=-\dfrac{\hat\Sigma^{\nu/\tilde\nu}_{\text{gauge}}(M_{h}^{2})}{2 M_{h}} 
\label{Dmhgauge},\\
\left(\Delta M_h\right)_{\text{Yukawa}}&=-\dfrac{\hat\Sigma^{\nu/\tilde\nu}_{\text{Yukawa}}(M_{h}^{2})}{2 M_{h}}\equiv \hat\Delta M_h 
\label{Dmhyuk}\,.
\end{align}
\end{subequations}
It should also be noted that, similarly to the one generation case, the full mass correction changes from positive values in the low $m_M$ region to negative values in the region of large $m_M \gsim 10 ^{14} \gev$.   In particular, for the reference values in (\ref{values}), it is $\Delta M_h=-0.25 \gev$. 

%
As mentioned before, the gauge part of the Higgs mass correction represents the common part with the MSSM, therefore in the following, we will focus  the discussion mainly in the Yukawa part which is the new contribution, denoted here and from now on shortly as $\hat\Delta M_h$.

\subsection{Sensitivity of the Higgs mass correction to the relevant SUSY parameters}
We next study the effects on $\Delta M_h$ of the other parameters
entering the calculation: $\tb$, $M_A$, $m_{\tilde{L}_i}$,
$m_{\tilde{R}_i}$, $a_{\nu}$, $b_{\nu}$ and $\mu$. In order to
explore these behaviours of $\Delta M_h$ with the relevant MSSM
parameters in presence of three Majorana neutrinos and their SUSY
partners, we run with one of the parameters while the others are set to
the reference values given in~\refeq{values}. 

\begin{figure}[hbtp]
\centering
\includegraphics[width=0.49\textwidth]{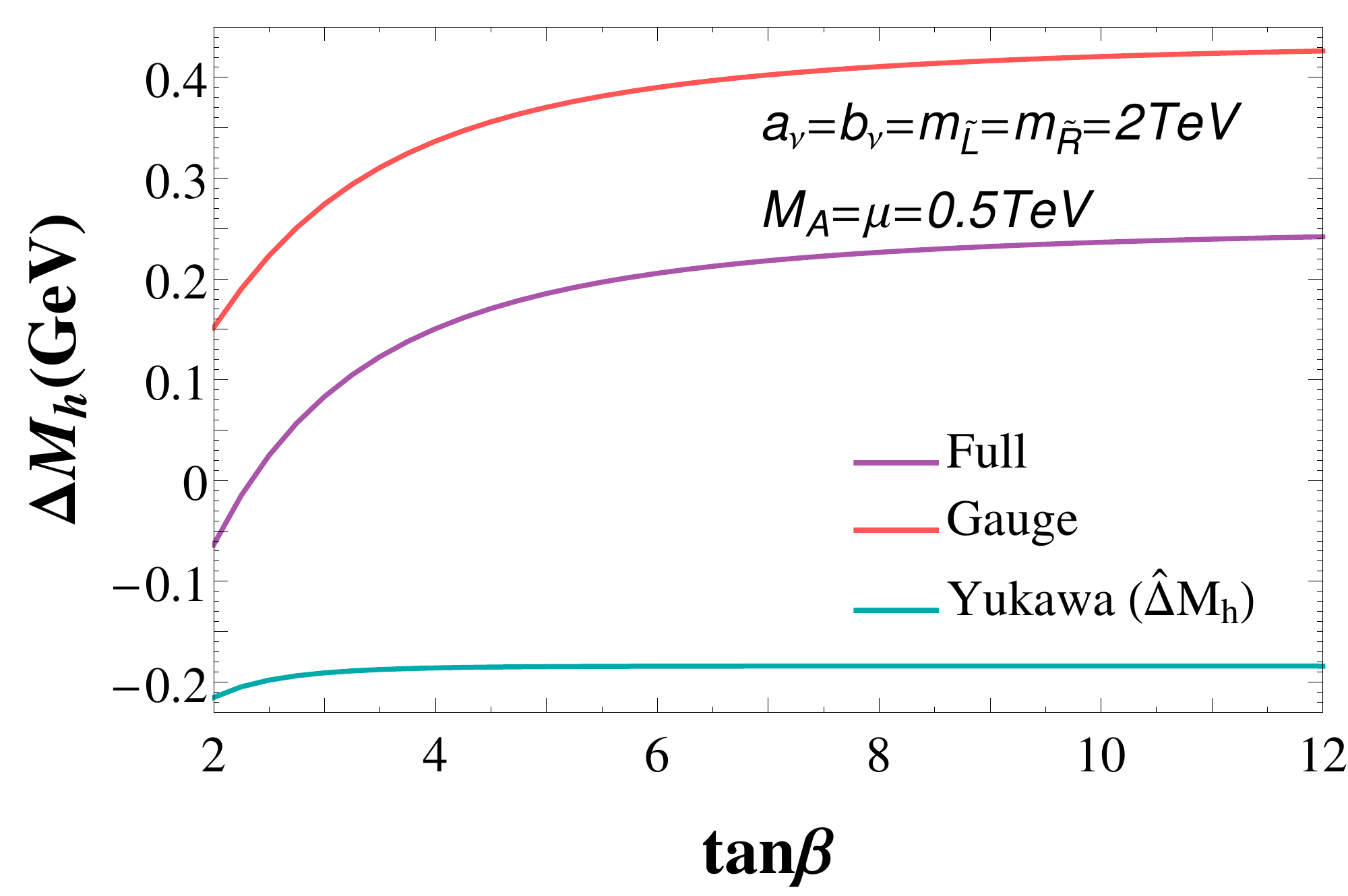}   
\includegraphics[width=0.49\textwidth]{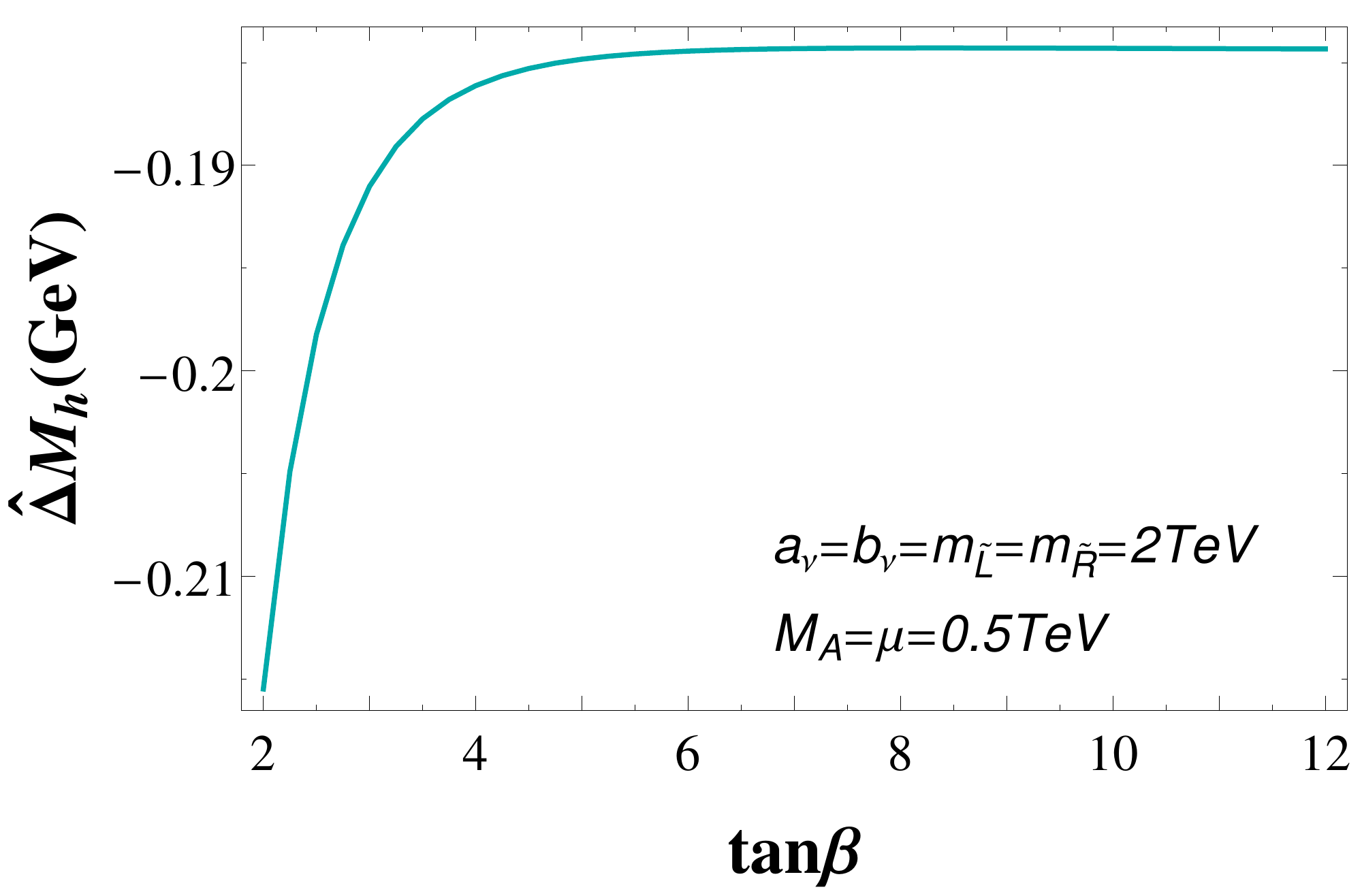}
\caption{Left panel: The full, gauge and Yukawa contributions to $\Delta M_h$ defined in \refeq{Dmhfull}, \refeq{Dmhgauge} and \refeq{Dmhyuk}, respectively, are plotted as  functions of $\tb$. Right panel:  Zoom of $\hat\Delta M_h$ as a function of $\tb$.}
\label{tanb}
\end{figure}

\begin{figure}[hbtp]
\centering
\includegraphics[width=0.49\textwidth]{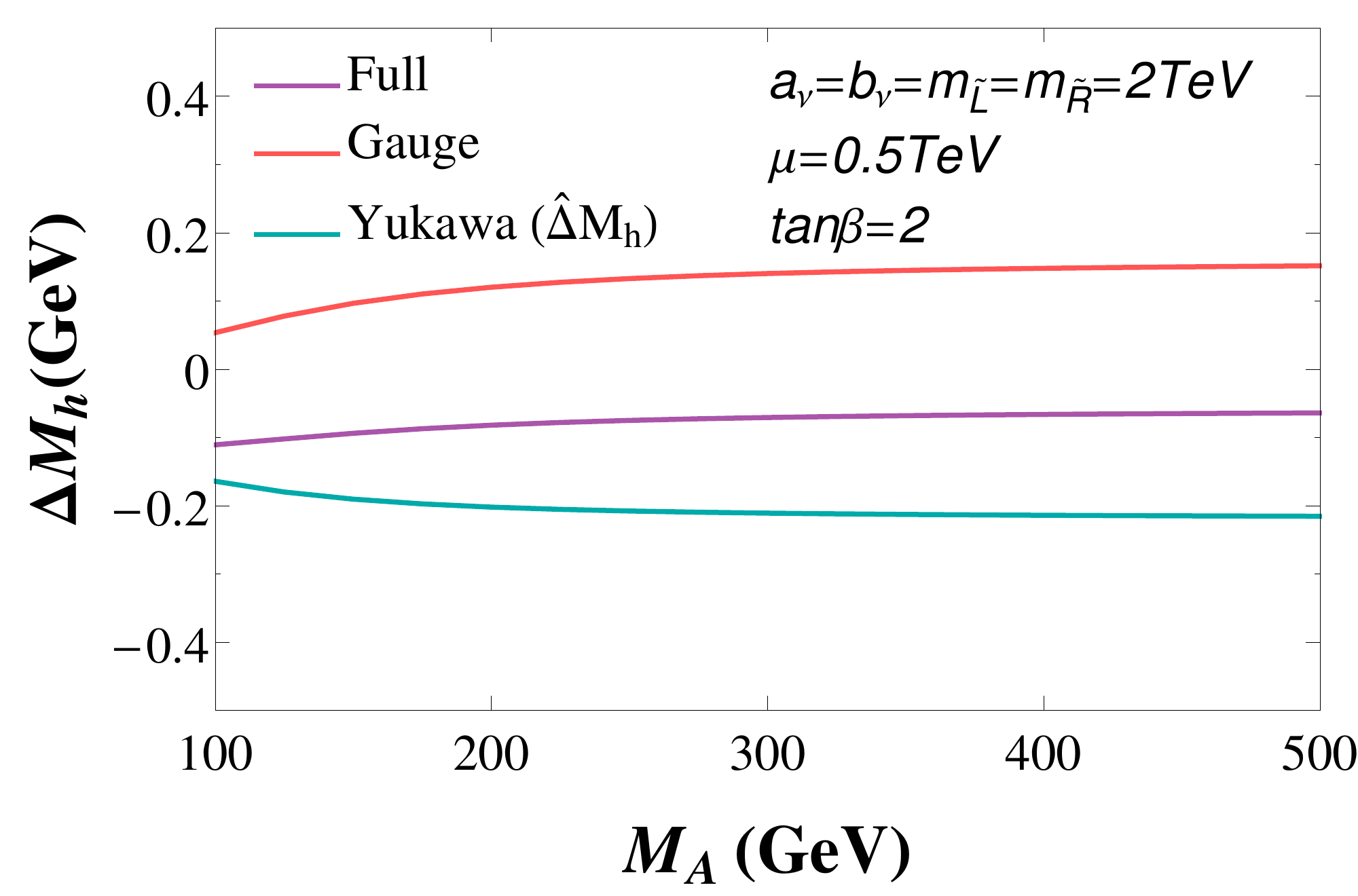}   
\includegraphics[width=0.49\textwidth]{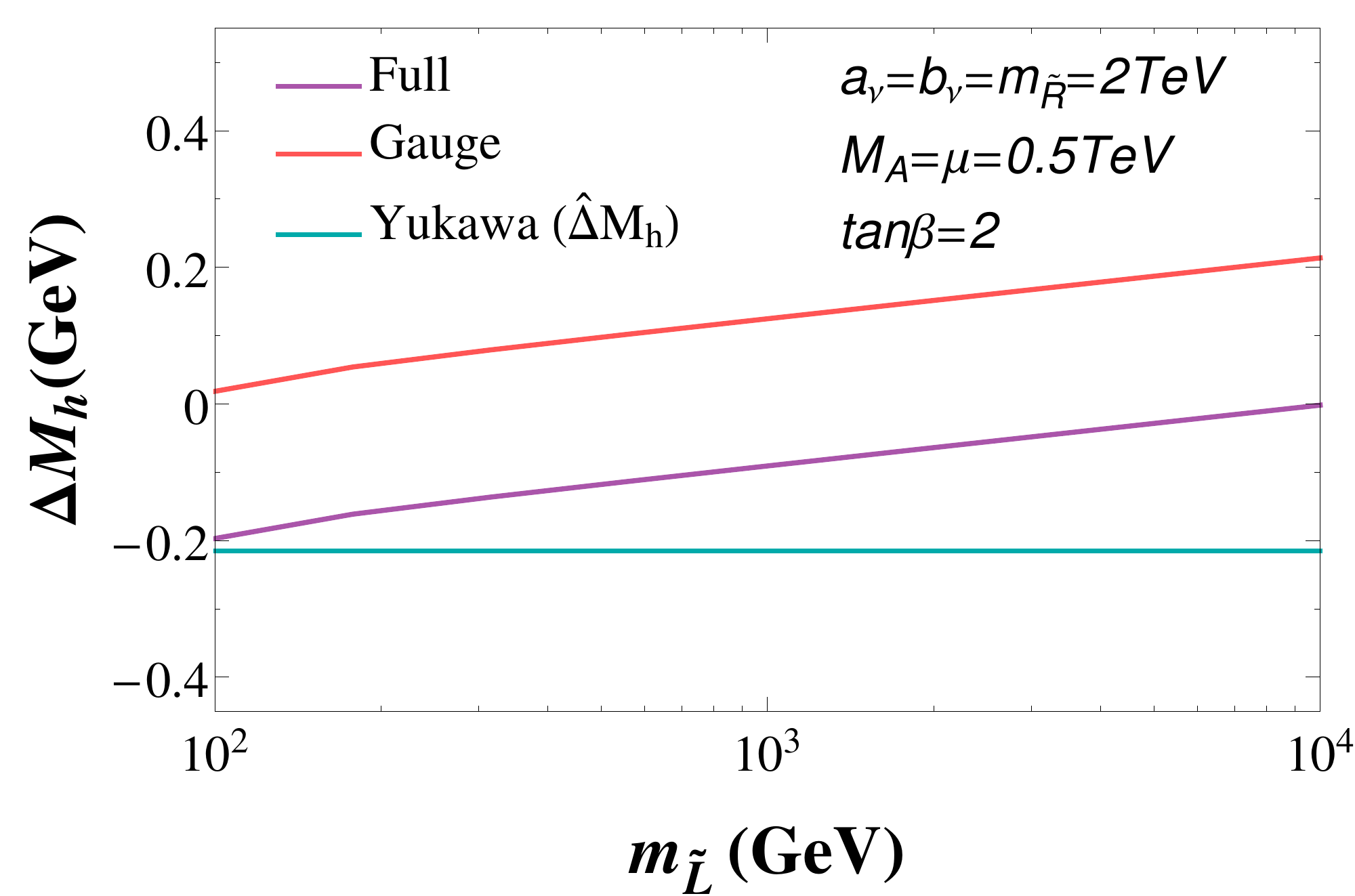}
\label{mL}
\includegraphics[width=0.49\textwidth]{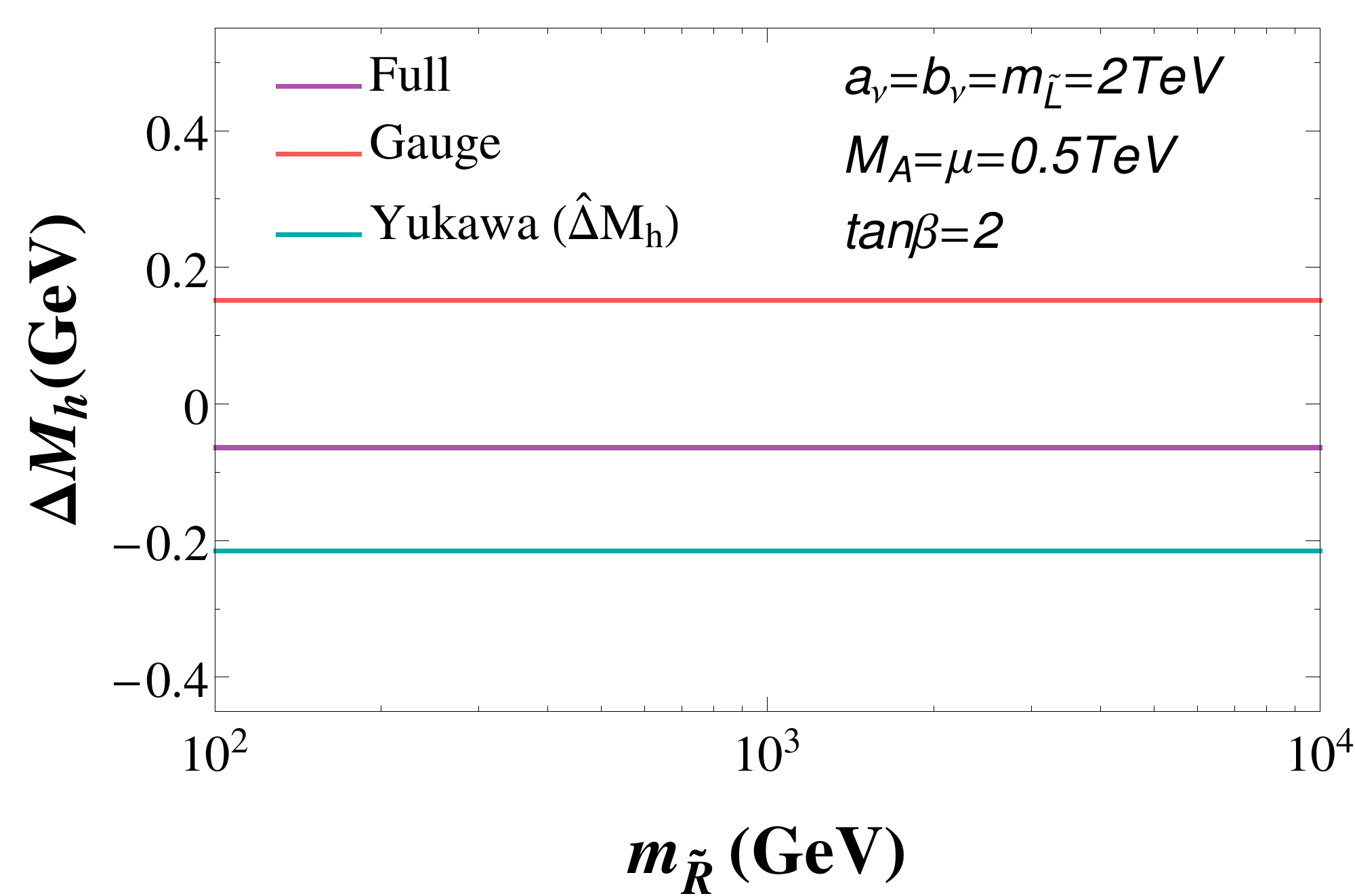}
\includegraphics[width=0.49\textwidth]{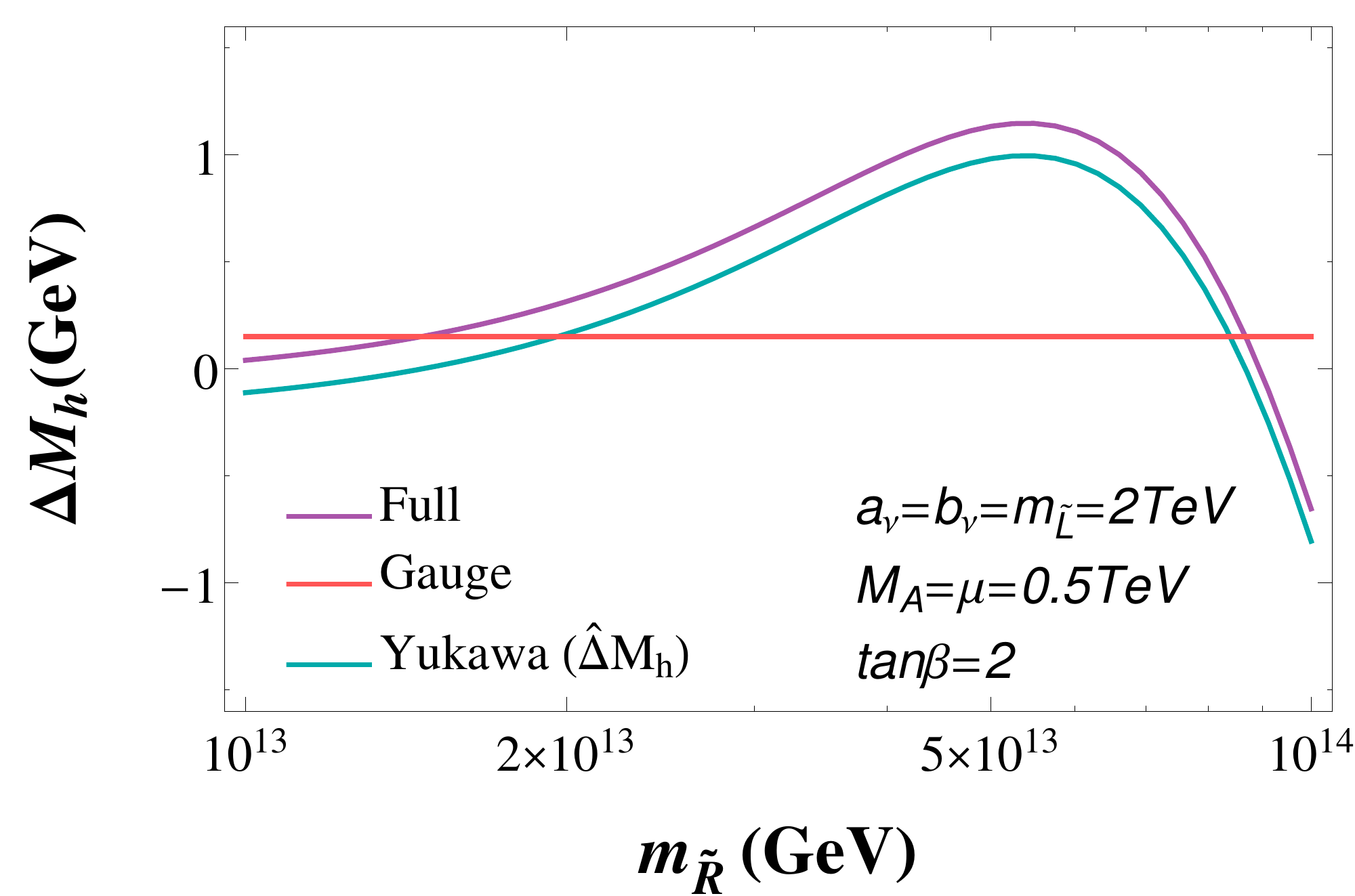}
\caption{Left upper panel: $\Delta M_h$ as a function of $M_A$. Upper right panel: $\Delta M_h$ as a function of $m_{\tilde{L}}$. 
Lower left panel: $\Delta M_h$ as a function of $m_{\tilde{R}}$ for low mass values $10^2 \text{ GeV} <m_{\tilde{R}}< 10^4 \text{ GeV}$. Lower right panel: $\Delta M_h$ as a function of $m_{\tilde{R}}$ for high mass values $10^{13} \text{ GeV} <m_{\tilde{R}}< 10^{14} \text{ GeV}$.}
\label{softpara}
\end{figure}

The behaviour of the one-loop corrections to the lightest Higgs boson
mass in the $\text{m}\overline{\text{DR}}$ scheme with  these relevant
parameters  are shown in figs \ref{tanb} and  \ref{softpara}. 

We start with the analysis of the behaviour with $\tb$, which is
shown in \reffi{tanb}. In the 
left plot the behavior of the full mass correction as well as the gauge
and Yukawa parts are shown. In the right plot we focus on the
Yukawa contribution to the mass correction. The biggest negative
correction $\hat\Delta M_h$ is obtained for  the lowest considered value
of $\tb = 2$; so in the following, motivating $\tb=2$ as our reference
value.  The numerical results for other choices of $\tb$ in the
remaining plots of this work can be easily inferred from this plot on
the right. 

The dependence on the pseudoscalar Higgs boson mass is analyzed in
\reffi{softpara}. For $M_A$ larger than 200 GeV the behavior with $M_A$
is nearly flat.  
The dependence on the soft SUSY-breaking mass of the
``left handed'' $SU(2)$ doublet, $m_{\tilde{L}}$ is also flat as shown
in \reffi{softpara}. The behavior with the soft mass of the ``right
handed'' sector $m_{\tilde{R}}$ in a range similar to the other soft
SUSY-breaking parameters is shown in \reffi{softpara}. In addition,
also values of $m_{\tilde{R}}$ closer to $m_{M_i}$ are explored in this
figure. The correction to the Higgs boson $h$ mass stays flat with
$m_{\tilde{R}}$ up to about $m_{\tilde{R}}\sim 10^{13} \gev$. Above this
mass scale the correction grows rapidly, reaching $\Delta M_h\sim -1$
GeV at $m_{\tilde{R}}\sim 10^{14} \gev$, in agreement with the results
found for the one generation case in~\cite{Ana}.  

We have also checked that the behavior of $\Delta M_{h}$ with the
remaining parameters, $a_\nu$, $b_\nu$ and $\mu$, in the intervals
$-1000\text{ GeV}<a_\nu<1000\text{ GeV}$, $100\text{
  GeV}<b_\nu<10^4\text{ GeV}$ and $-1000\text{ GeV}<\mu<1000\text{ GeV}$
are also flat as in the case of the low mass values of $m_{\tilde{R}}$. 
The behaviors of $\hat\Delta M_{h}$ with all these parameters agree as well with the results obtained in the one-generation case \cite{Ana}.

\subsection{Sensitivity of the Higgs mass corrections to the light neutrinos}

In this section we analyze the sensitivity of the mass correction to the
mass hierarchy of the light neutrinos.  
Here we investigate the two cases of NH and IH, where
the values of the rest of the parameters are fixed to the ones of our
reference scenario given in \eqref{values}. 

\begin{figure}[hbtp]
\centering
\includegraphics[width=0.56\textwidth]{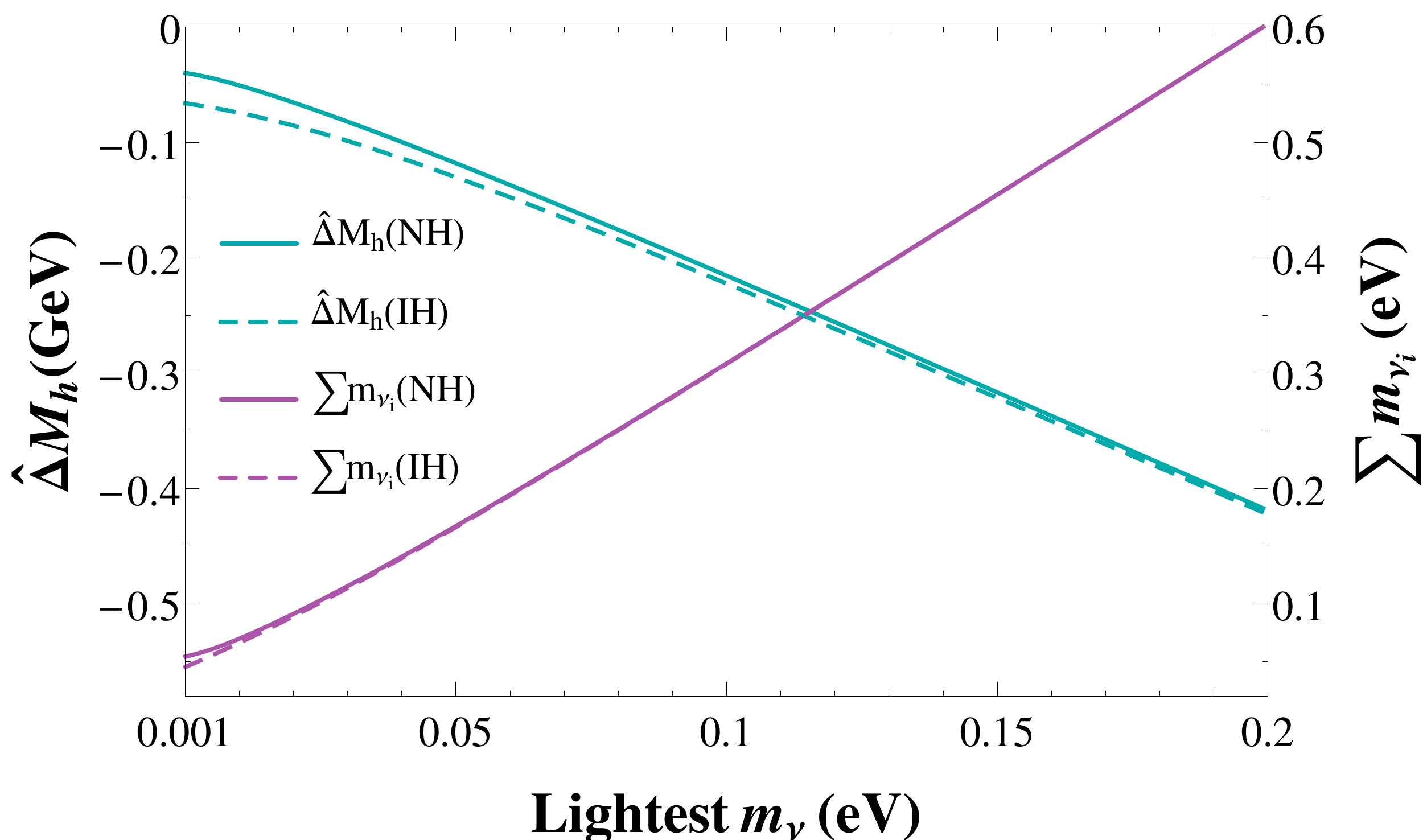}
\caption{$\hat\Delta M_h$ (blue) and $\Sigma\,m_{\nu_i}$ (purple) as a function of the lightest neutrino mass, $m_{\nu_1}$ for a normal hierarchy (solid) and $m_{\nu_3}$ for an inverted hierarchy (dashed). The rest of the model parameters are set as in~\refeq{values}.}
\label{lightneutrinos}
\end{figure}

\reffi{lightneutrinos} shows the behavior of the Yukawa part of the mass correction with the mass of the lightest neutrino, $\nu_1$ and $\nu_3$ for the NH (solid lines) and IH (dashed lines), respectively.
We show the Yukawa contribution to the mass correction (vertical left
axis) as well as the sum of the three neutrino masses (vertical right
axis) for each value as a function of the lightest neutrino mass.  
We conclude that, even though the numerical result of $\hat\Delta M_h$ for both hierarchies are quite similar, the Higgs mass corrections found in the IH case are slightly bigger than the ones of the NH case.

\subsection{Sensitivity of the Higgs mass corrections to the heavy neutrino masses}

In this section the behaviors of the mass correction with the masses of the heavy Majorana neutrinos as well as with the $R$ matrix are analyzed. As mentioned before, the $R$ matrix of \refeq{Casas} parametrizes the mixing in the heavy neutrino sector.

\begin{figure}[hbtp]
\centering
\includegraphics[width=0.49\textwidth]{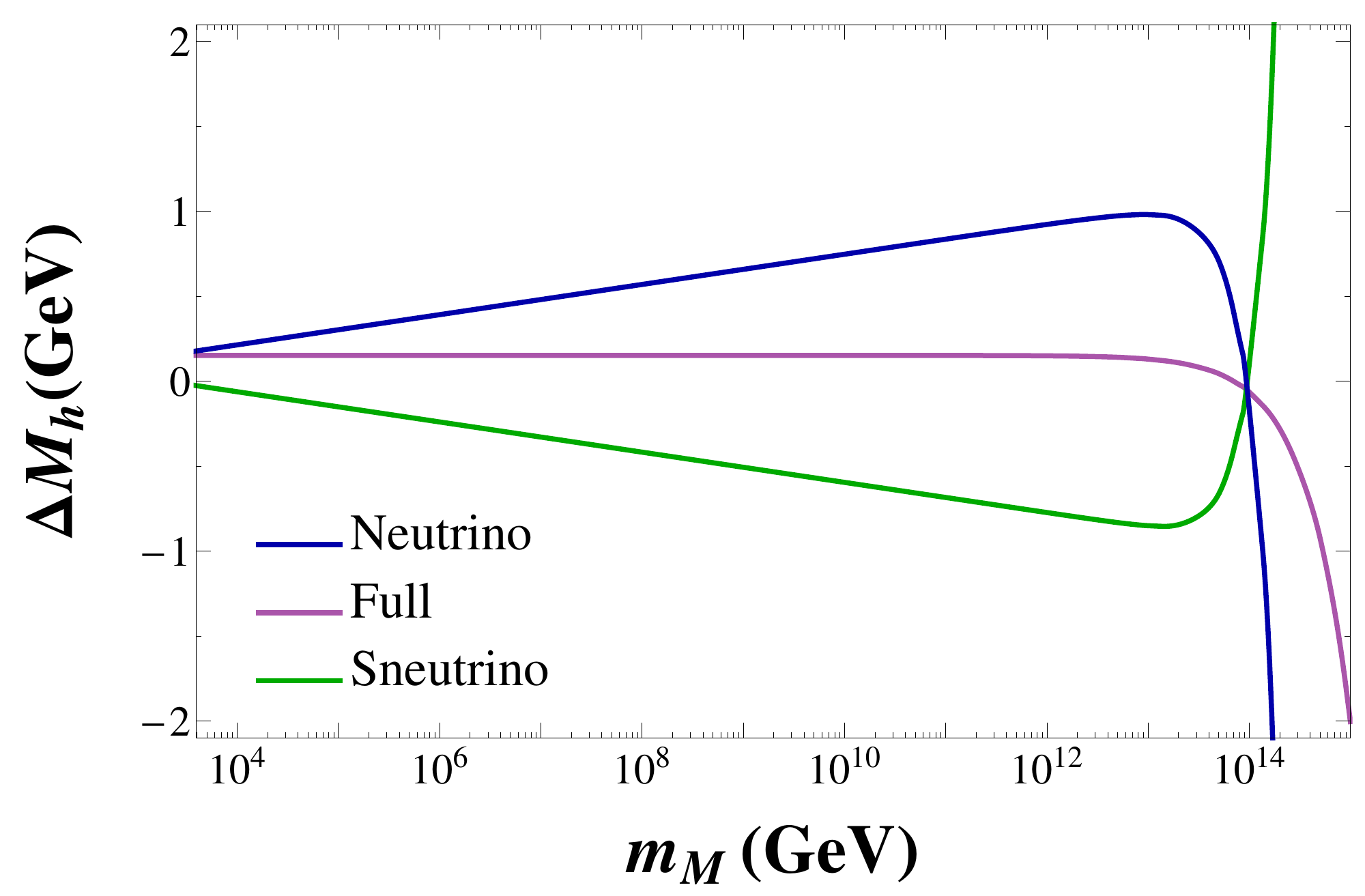}
\includegraphics[width=0.49\textwidth]{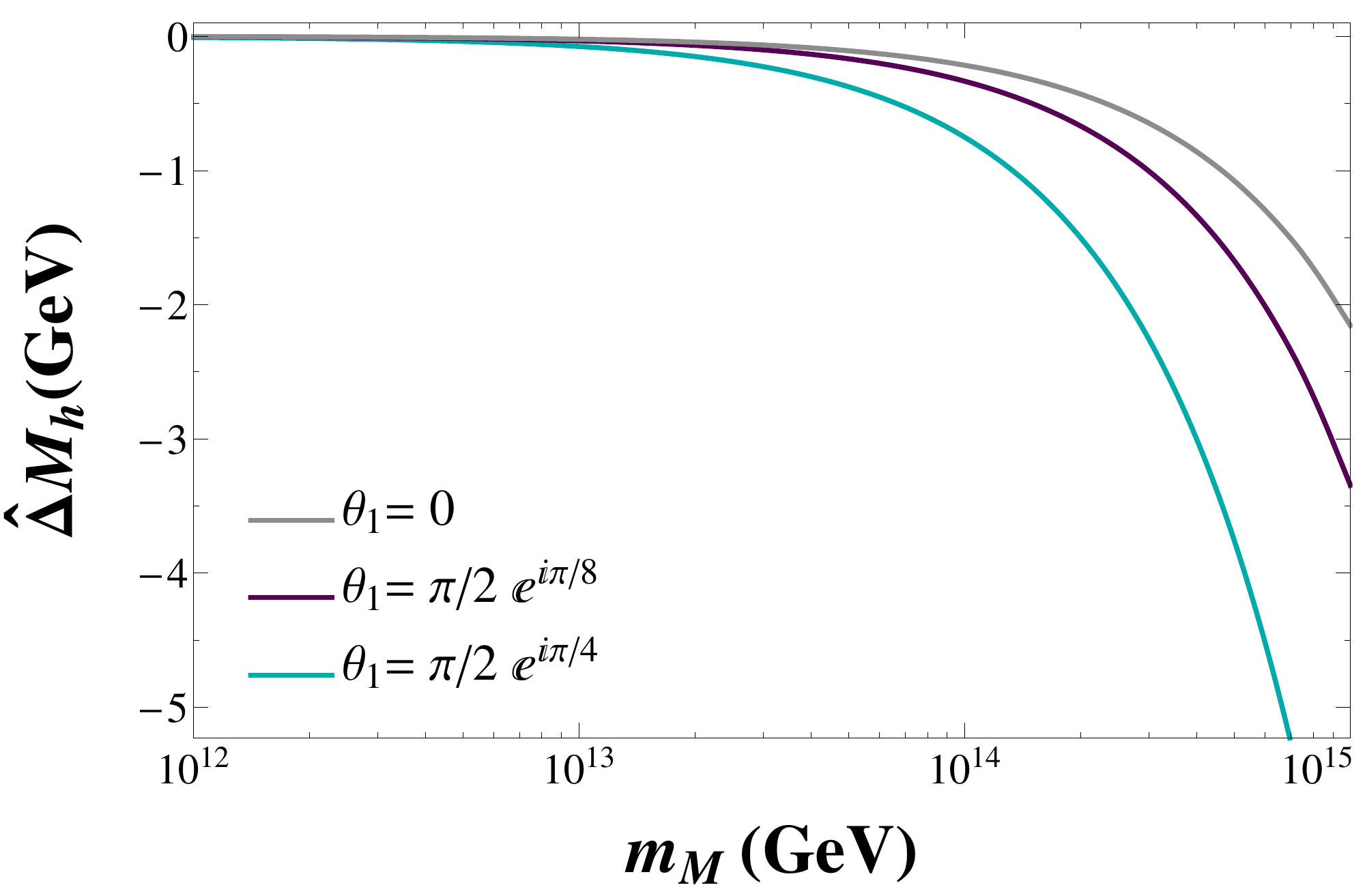}
\caption{Left panel: $\Delta M_h$ as a function of $m_M$ for degenerate
  heavy neutrinos and $R=1$. The corrections from neutrinos,
  sneutrinos and the total are displayed separately. Right panel:
  $\hat\Delta M_h$ as a function of $m_{M}$ for
  $\theta_1=\left(0,\pi/2 e^{i\pi/8},\pi/2 e^{i\pi/4}\right)$ and
  $\theta_2=\theta_3=0$. The rest of the model parameters are set as
  in~\refeq{values}.} 
\label{majorana_plots}
\end{figure}

First of all, we show the results for the degenerate heavy neutrino
scenario where the three heavy Majorana neutrinos have all the same
mass, i.e. $m_{M_1}=m_{M_2}=m_{M_3}=m_{M}$.  The mass of the lightest
neutrino as well as the SUSY parameters are set to the reference values
given in \refeq{values}. In the left panel of
\reffi{majorana_plots} we show the behaviour of the full mass
correction $\Delta M_h$ with the common Majorana  $m_M$. We have
separated the contribution to the mass correction coming from the
neutrino and sneutrino sectors in order to show the 
remarkable cancellation between the two parts that it is happening. It
can also be seen that the behavior of the total $\Delta M_h$ with $m_M$
at very large $m_M\gtrsim 10^{14} \gev$ is dominated by
the neutrino contributions. 
It should be noted that the same result as in \reffi{majorana_plots} is obtained for any other real $R$ matrix different from the reference value $R=\one$. This independence on the particular real $R$ value can be understood from the fact that, as we have mentioned before, $\hat\Delta M_{h}\varpropto m_{D}^{\dagger}m_{D}$, and with the definition of $m_{D}$ given in \eqref{Casas}, we find:
\begin{eqnarray}
\label{mddagamd}
m_{D}^{\dagger}m_{D}&=&\sqrt{m_N^\text{diag}} \, R^{*} \, \sqrt{m_{\nu}^\text{diag}} \, U_\text{PMNS}^T \, U_\text{PMNS}^* \, \sqrt{m_{\nu}^\text{diag}} \, R^{T} \sqrt{m_N^\text{diag}}\nonumber\\
&=& \sqrt{m_N^\text{diag}} \, R^{*} \, m_{\nu}^\text{diag} \, R^{T} \sqrt{m_N^\text{diag}} \, .
\end{eqnarray}
As the three light neutrinos in \refeq{values} are quasi degenerate, $m_{\nu}^\text{diag}\thickapprox m_{\nu_{1}} \one$, and since here $R$ is a real and orthogonal matrix, then \eqref{mddagamd} becomes independent on $R$, i.e.:
\begin{equation}
\label{mddagamdR}
m_{D}^{\dagger}m_{D}\thickapprox m_{\nu_{1}} \sqrt{m_N^\text{diag}} \, R \, R^{T} \sqrt{m_N^\text{diag}} =  m_{\nu_{1}} m_N^{\text{diag}}\, .
\end{equation}
In contrast, when a complex $R$ matrix is implemented, the result in
\refeq{mddagamdR} is no longer true and $\hat\Delta M_{h}$ grows with
the size of both Re$(\theta_i)$ and Im$(\theta_i)$, as can be seen in
the right panel of \reffi{majorana_plots}. There we plot
$\hat\Delta M_{h}$ for three different values $\theta_1=\left(0,\pi/2
e^{i\pi/8},\pi/2 e^{i\pi/4}\right)$ while the other two angles,
i.e. $\theta_2$ and $\theta_3$, are set to zero. We have checked that
similar growing behaviors with the other complex $\theta_{2,3}$ angles
are found.    

\begin{figure}[hbtp]
\centering
\includegraphics[width=0.49\textwidth]{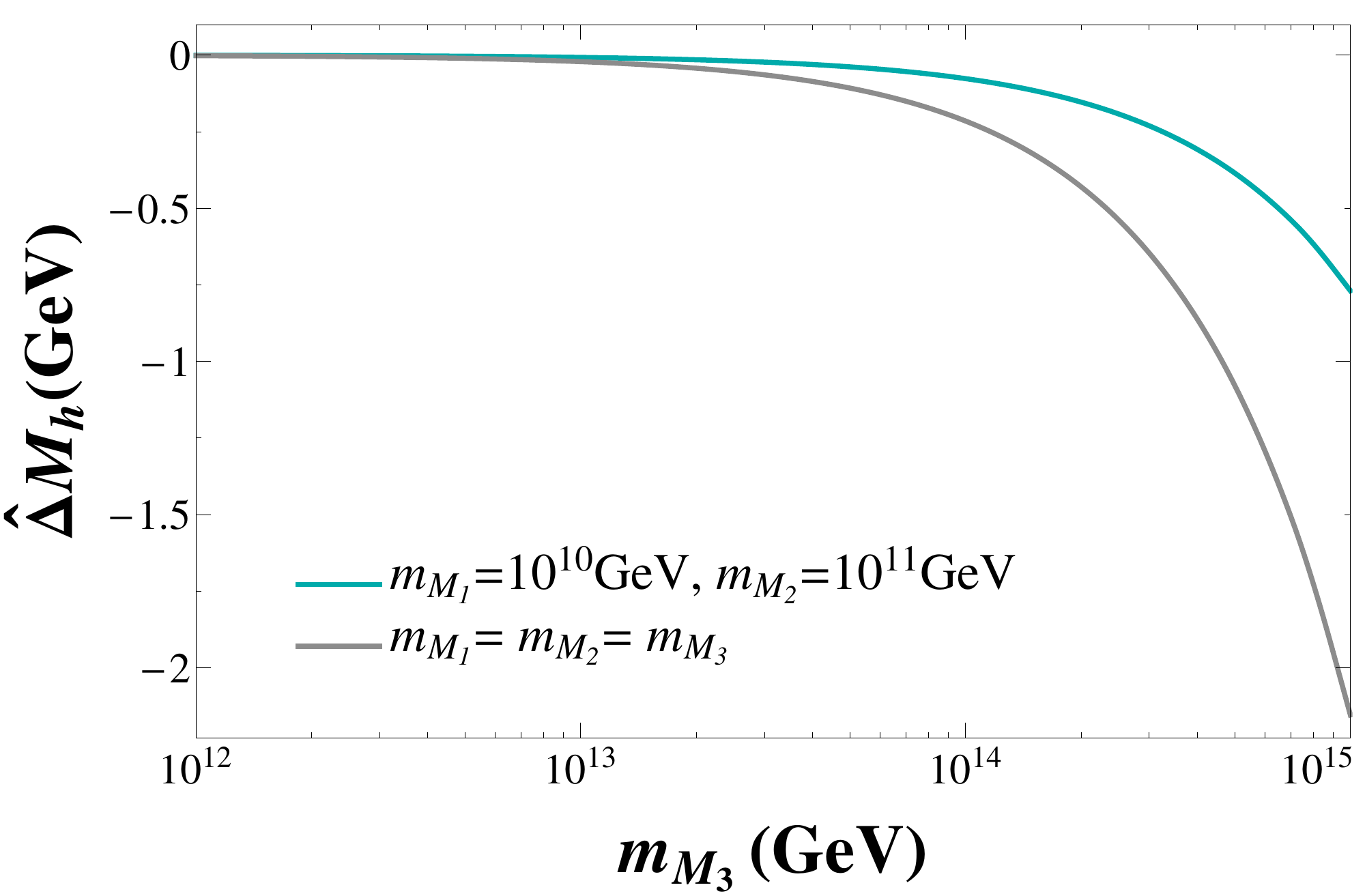}
\caption{$\hat\Delta M_h$ as a function of $m_{M_3}$ for hierarchical (blue) and degenerate (grey) heavy neutrinos. The rest of input parameters are set to the reference values in (\ref{values}).}
\label{majorana_hier_plots}
\end{figure}

Next we study the case where there is a hierarchy between the three heavy Majorana neutrino masses. 
First we consider the simplest case of $R=\one$ and analyze the behavior with the heaviest Majorana mass, chosen here to be $m_{M_3}$, while the other two masses are fixed to $m_{M_1}=10^{10} \gev$ and $m_{M_2}=10^{11} \gev$. \reffi{majorana_hier_plots} compares the behavior of $\hat\Delta M_h$ with $m_{M_3}$ in both degenerate and hierarchical cases. This figure shows that the size of the correction $\hat\Delta M_h$ in the hierarchical case is dominated by the heaviest Majorana mass, $m_{M_3}$ in this example.  Furthermore, the obtained Higgs mass correction for a given $m_{M_3}$ value is smaller than the corresponding mass correction  in the degenerate case with the common $m_M$ fixed to this same value, i.e for $m_M=m_{M_3}$ .

\begin{figure}[hbtp]
\centering
\includegraphics[width=0.49\textwidth]{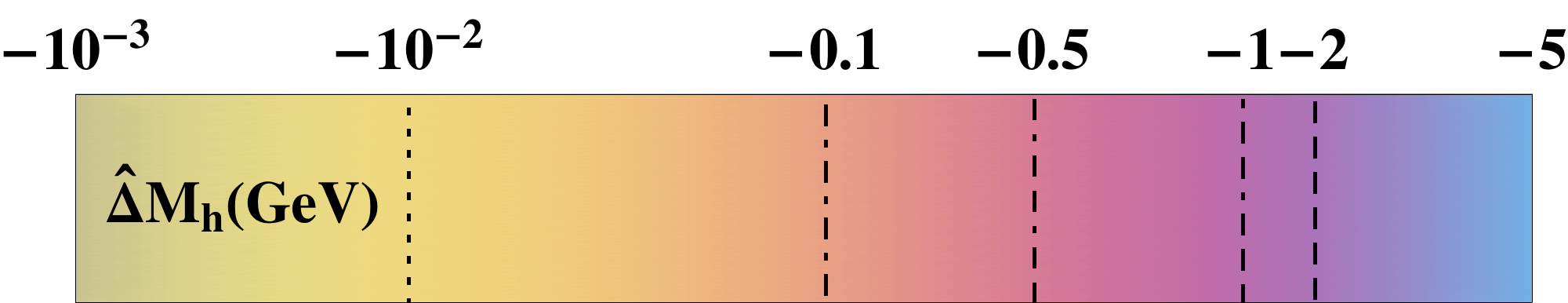}\\
\includegraphics[width=0.39\textwidth]{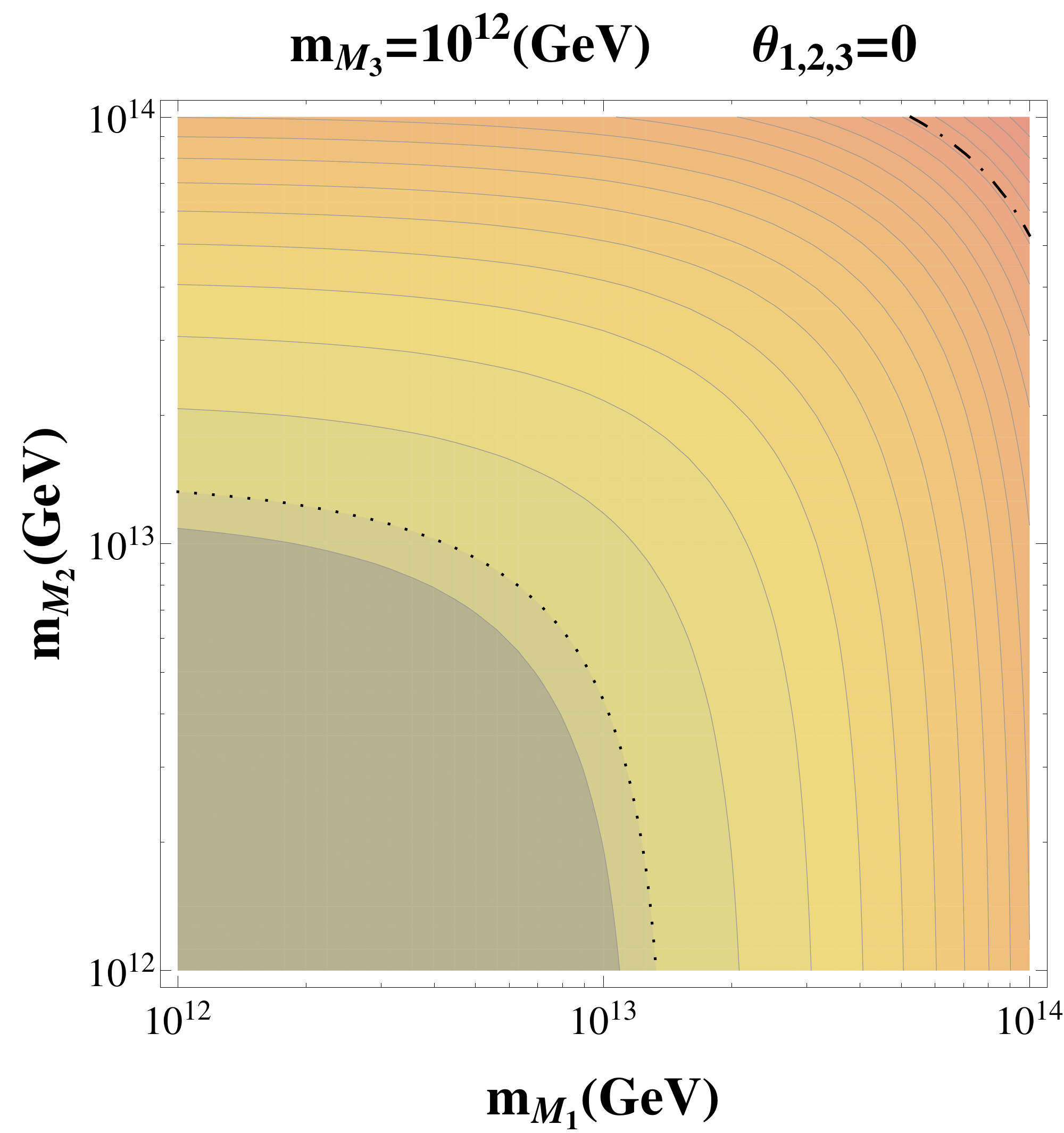}
\includegraphics[width=0.39\textwidth]{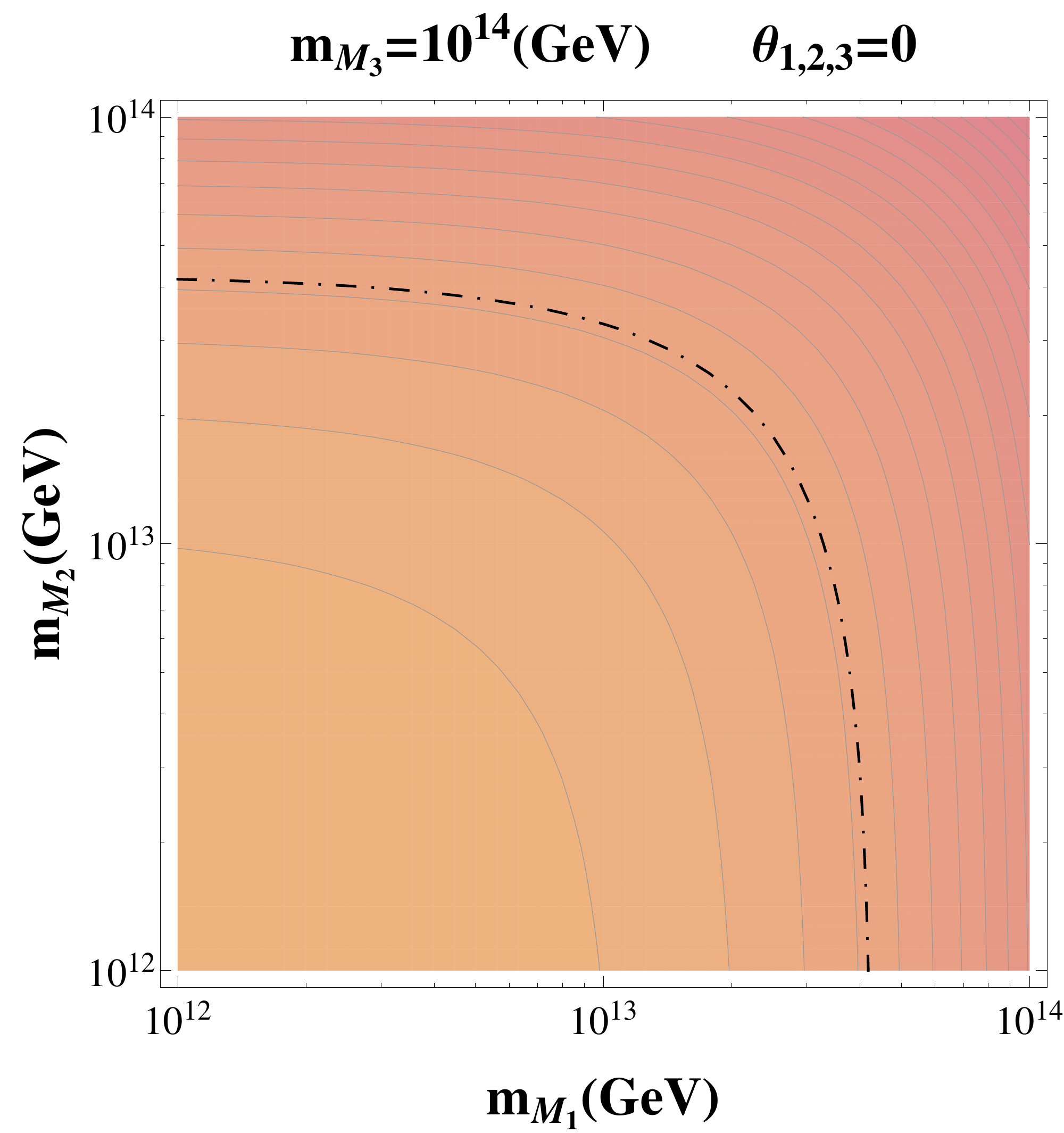}
\caption{$\hat\Delta M_h$ as a function of $m_{M_1}$ and $m_{M_2}$ in
  the hierarchical heavy neutrinos scenario. Left panel: with
  $m_{M_3}=10^{12} \gev$. Right panel: with $m_{M_3}=10^{14} \gev$.  The
  rest of input parameters are set to the reference values in
  (\ref{values}).
} 
\label{contour}
\end{figure}

In order to perform a complete analysis with hierarchical heavy neutrinos, we have scanned the Majorana masses $m_{M_1}$ and $m_{M_2}$ in the range $10^{12}\leqslant m_{M_{1,2}}\leqslant 10^{14} \gev$ for two different values of $m_{M_3}$. As a result, we have obtained the two contour plots that are shown in \reffi{contour}. Due to the fact that we are assuming in practice that the light neutrinos are quasi degenerate and that there is no mixing among the heavy Majorana neutrinos 
($R=\one$), the behavior of the Higgs mass correction is symmetric in all the three Majorana masses and consequently, the biggest correction is obtained when the three masses are equal and set to the highest value, i.e. $10^{14} \gev$ in these plots. We have checked that once the value of  $m_{M_3}$ lies below $10^{12} \gev$ there is no appreciable sensitivity to that mass, so the result will be the same as in the left panel of \reffi{contour}. Similarly to the previous degenerate case, there is not sensitivity to the choice of the real $R$ matrix in the hierarchical case either, as can be understood from the result in (\ref{mddagamdR}) that also holds here. Therefore, the results in \reffi{contour} are valid for all values of real $R$.
 
\begin{figure}[htb!]
\centering
\includegraphics[width=0.39\textwidth]{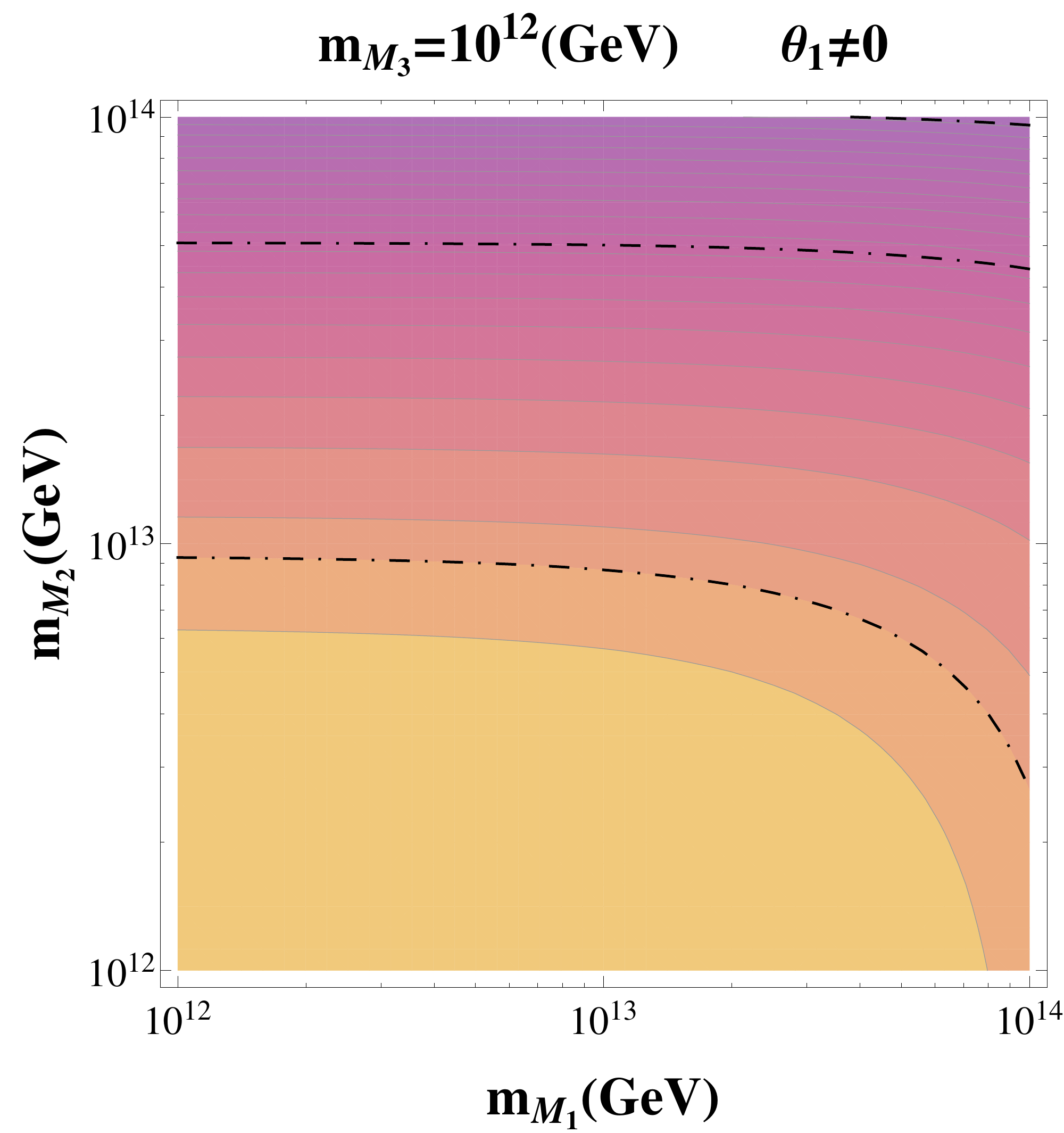}
\includegraphics[width=0.39\textwidth]{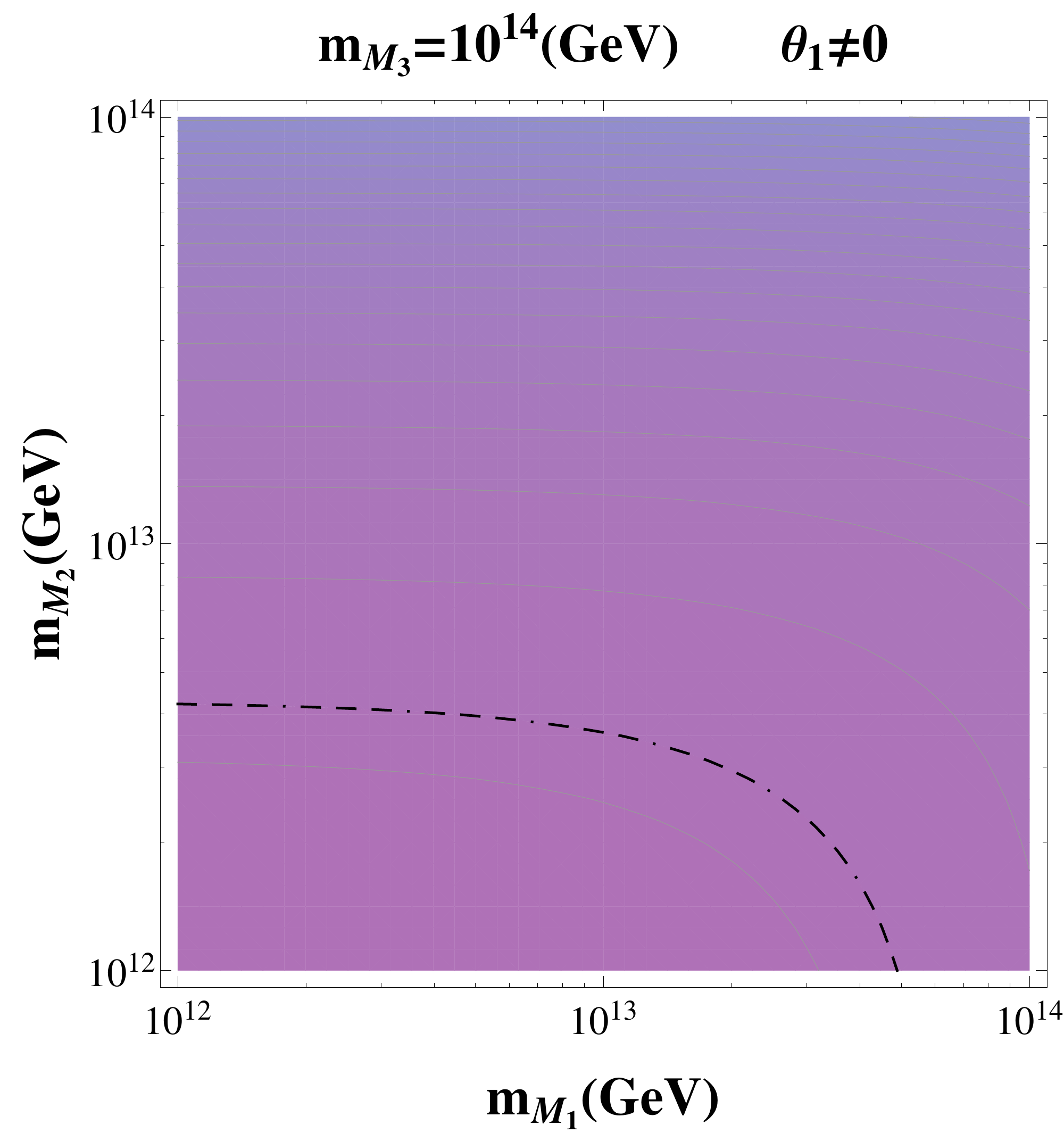}
\includegraphics[width=0.39\textwidth]{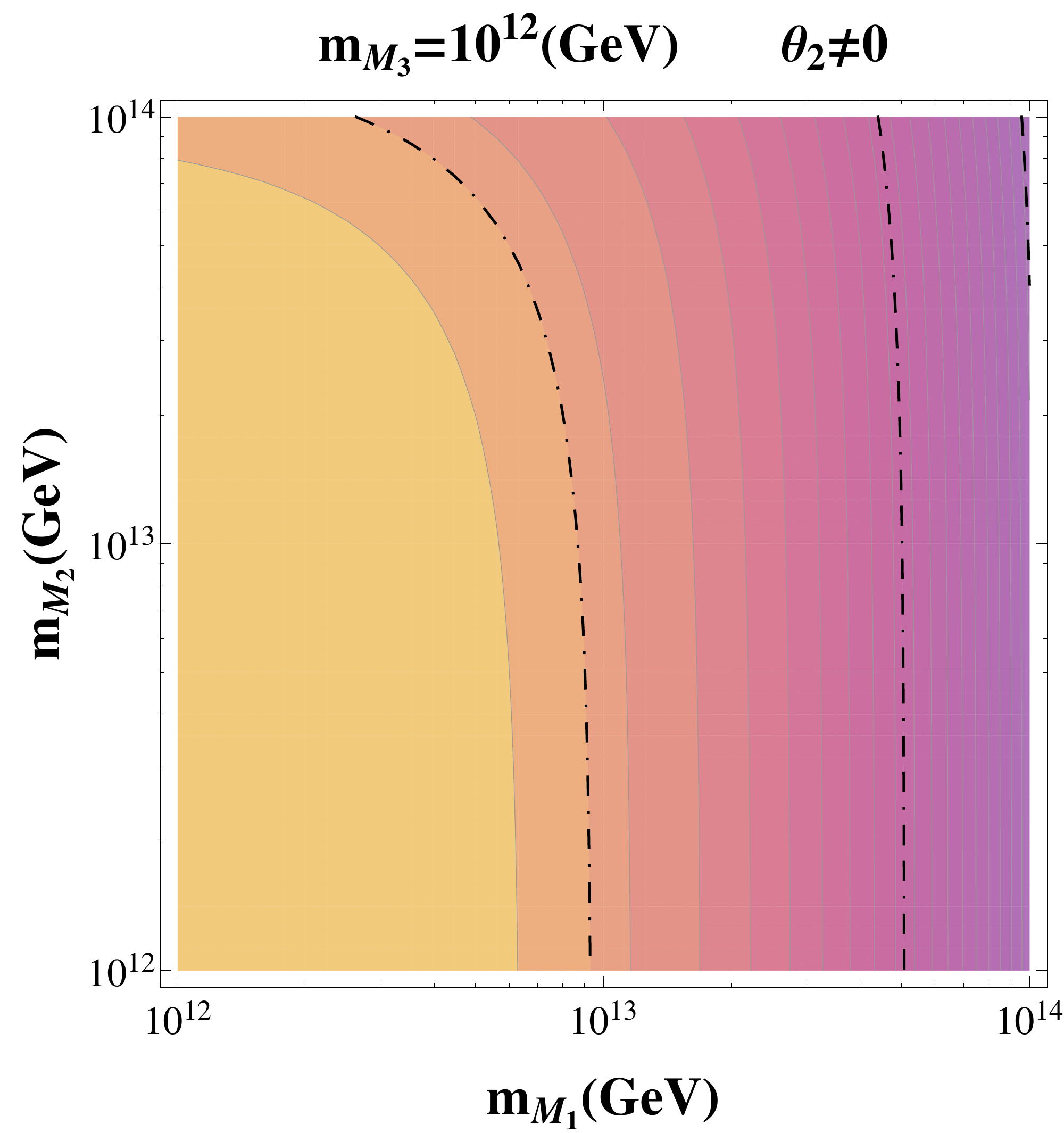}
\includegraphics[width=0.39\textwidth]{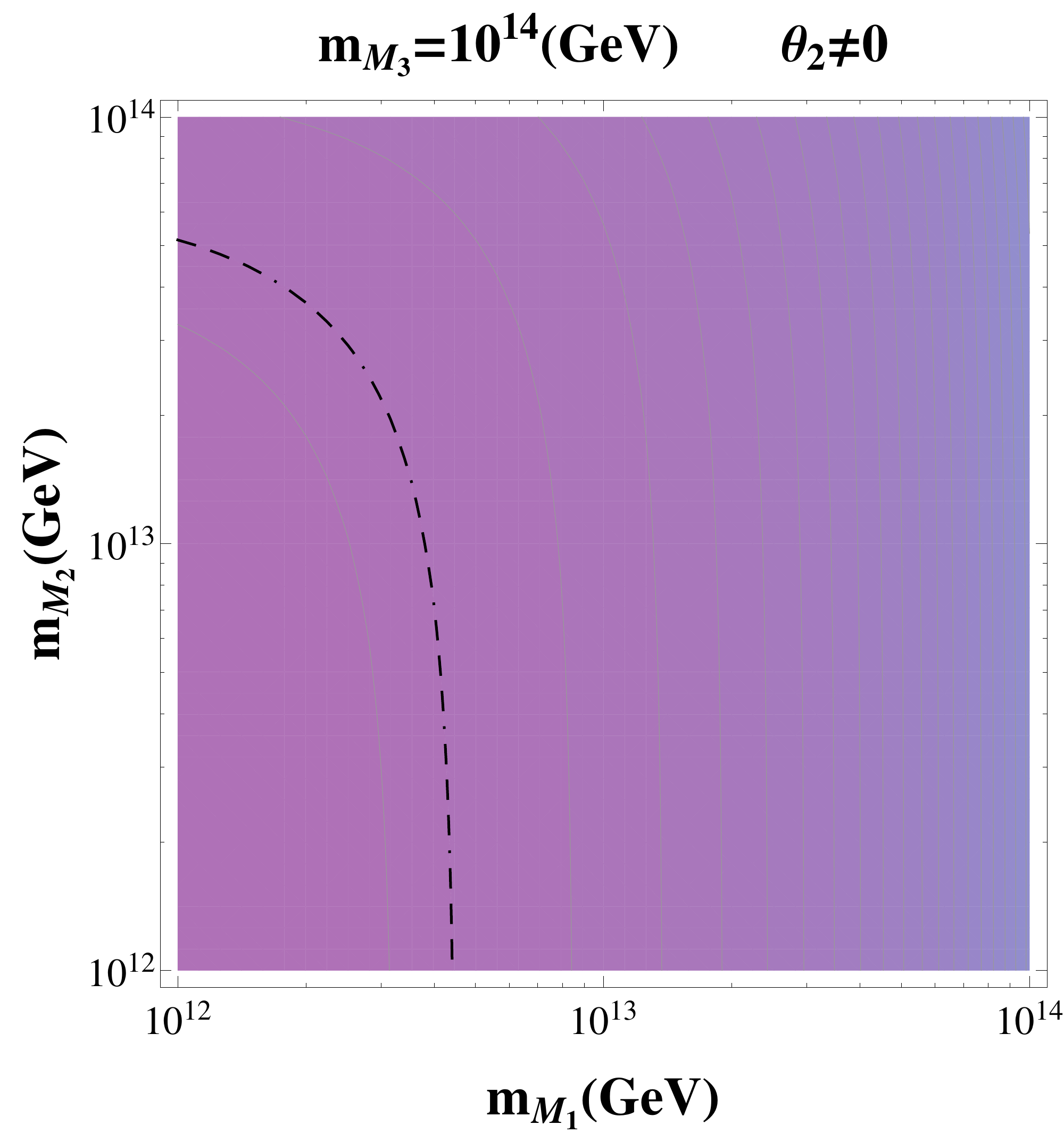}
\includegraphics[width=0.39\textwidth]{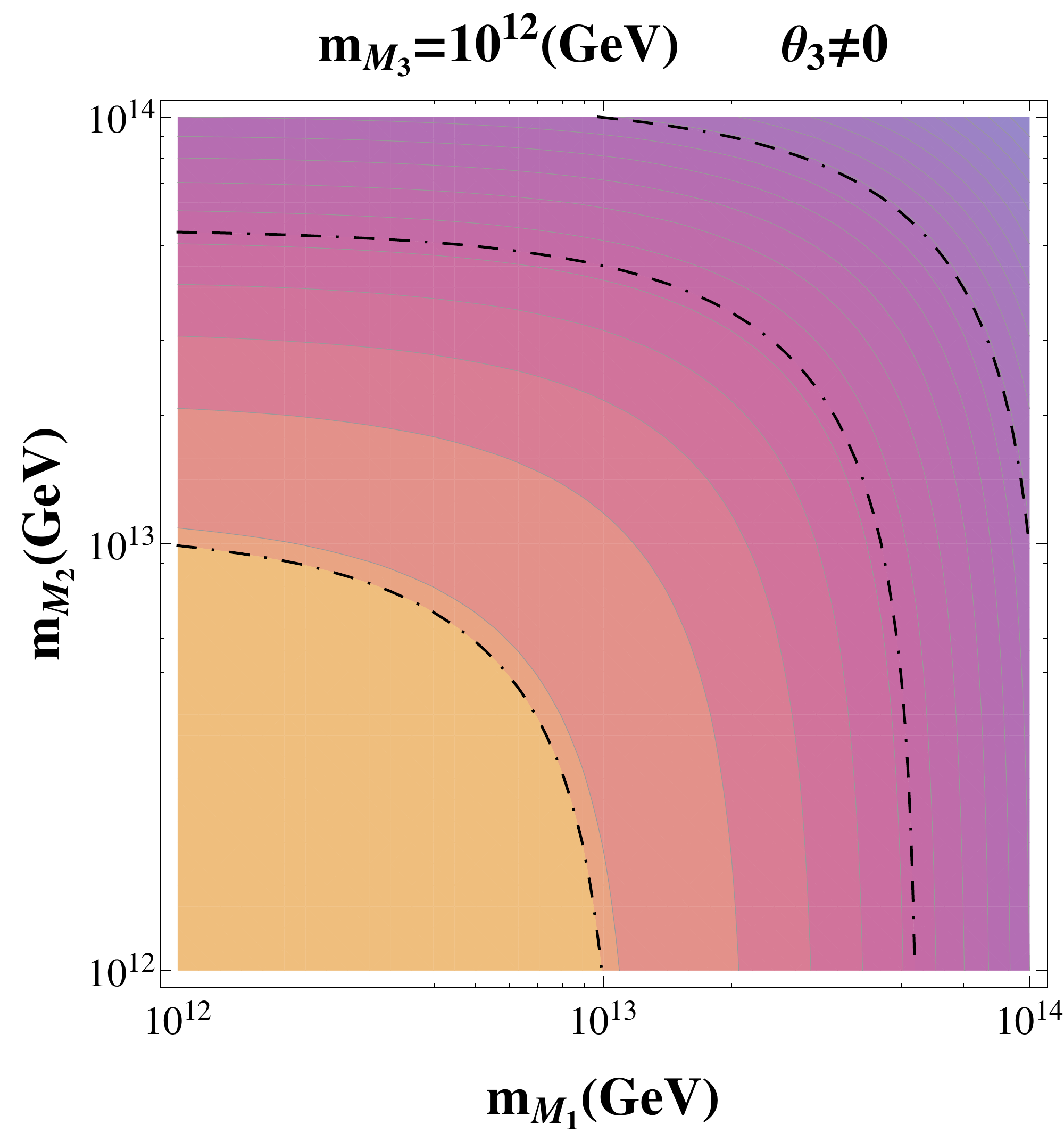}
\includegraphics[width=0.39\textwidth]{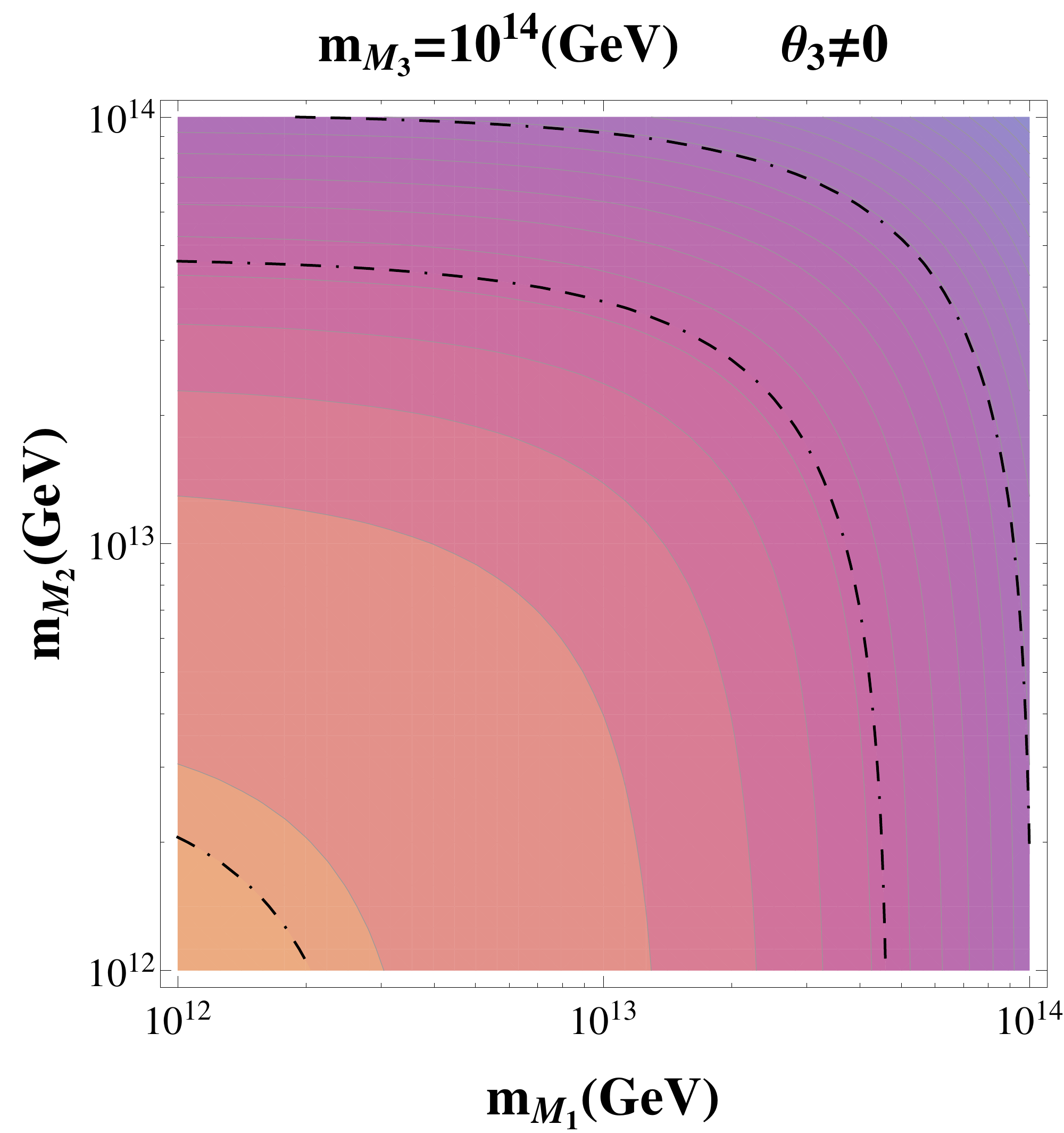}
\includegraphics[width=0.49\textwidth]{figs/Ruler}
\caption{$\hat\Delta M_h$ contour lines in the ($m_{M_1}$, $m_{M_2}$)
  plane for two values of $m_{M_3}$ and for a single complex non
  vanishing $\theta_i$ angle. Left panels: with $m_{M_3}=10^{12}
  \gev$. Right panels: with $m_{M_3}=10^{14} \gev$. Top panels:
  $\theta_{1}=3\pi/4\,e^{i\pi/4}$, $\theta_{2}=\theta_{3}=0$. Middle
  panels: $\theta_{2}=3\pi/4\,e^{i\pi/4}$,
  $\theta_{1}=\theta_{3}=0$. Bottom panels:
  $\theta_{3}=3\pi/4\,e^{i\pi/4}$, $\theta_{1}=\theta_{2}=0$. The rest
  of input parameters are set to the reference values in
  (\ref{values}).} 
\label{contour_tetas}
\vspace{-8em}
\end{figure}

\begin{figure}[hbtp]
\centering
\includegraphics[width=0.39\textwidth]{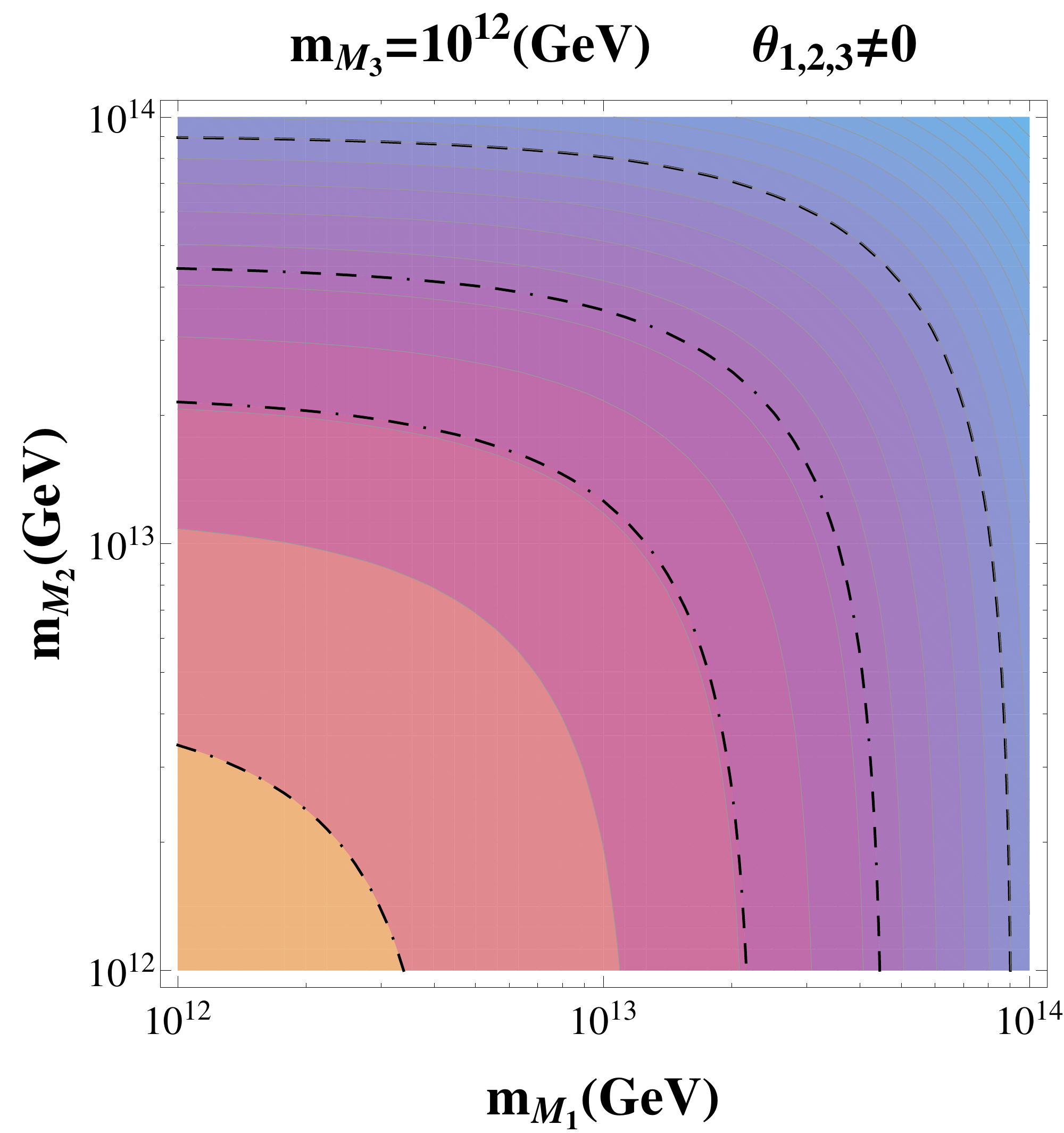}
\includegraphics[width=0.39\textwidth]{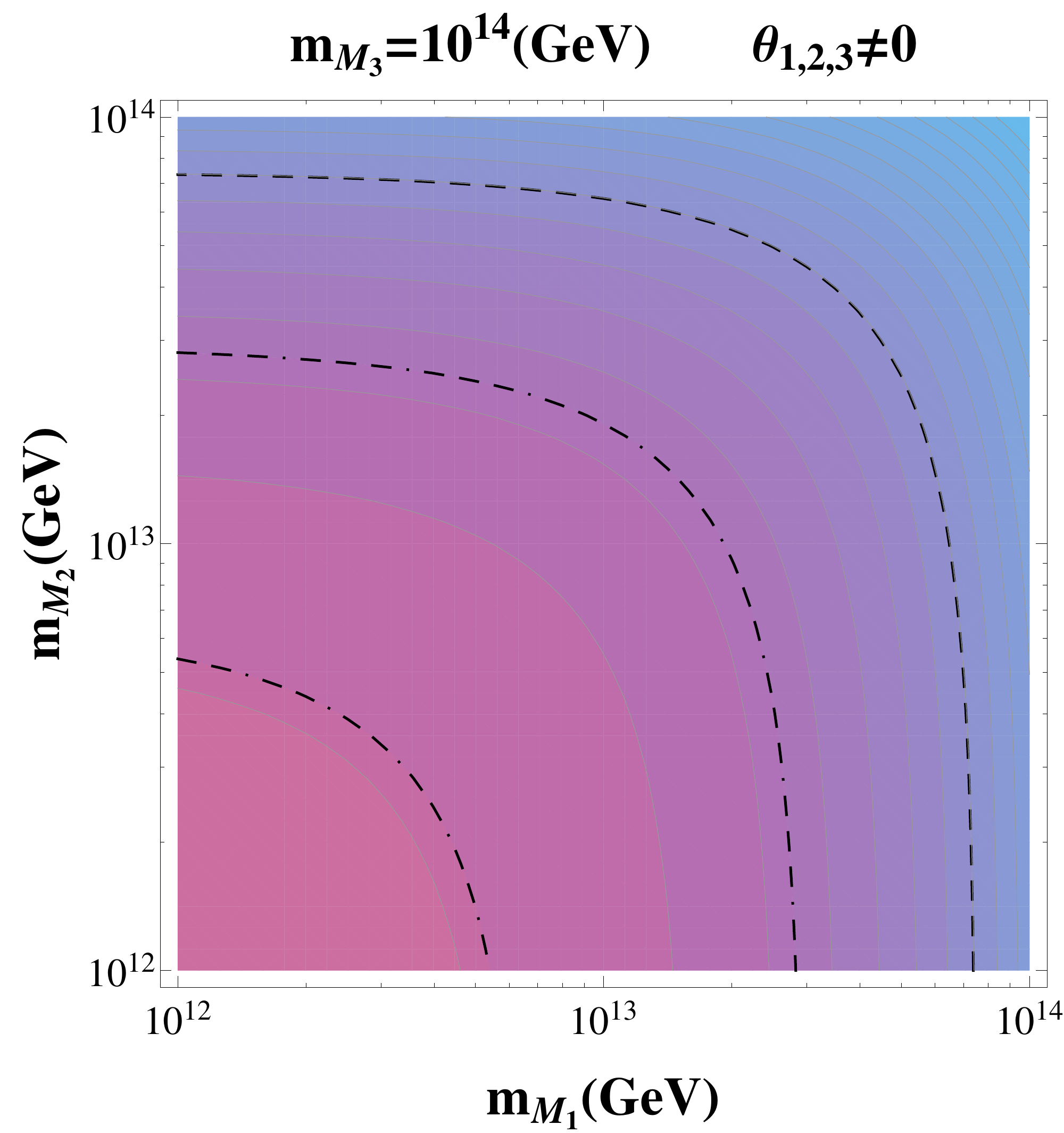}
\includegraphics[width=0.49\textwidth]{figs/Ruler}
\caption{$\hat\Delta M_h$ contour lines in the ($m_{M_1}$, $m_{M_2}$) plane for two values of $m_{M_3}$ and for three 
non vanishing complex $\theta_i$ angles: $\theta_{1}=3\pi/8\,e^{i\pi/4}$, $\theta_{2}=\pi/2\,e^{i\pi/5}$ and $\theta_{3}=3\pi/4\,e^{i\pi/7}$.  Left panel: $m_{M_3}=10^{12} \gev$. Right panel: $m_{M_3}=10^{14} \gev$. The rest of input parameters are set to the reference values in (\ref{values}).}
\label{contour_3tetas}
\end{figure}

Finally we analyze the imprints of the mixing of the hierarchical heavy neutrinos in $\hat \Delta M_{h}$ when a complex $R$ matrix is implemented. \reffi{contour_tetas} shows  the $\hat\Delta M_{h}$ contours in the general case of three Majorana masses, $m_{M_{1,2,3}}$, and  when one of the three $\theta_{i}$ angles is fixed to $3\pi/4\,e^{i\pi/4}$ while the other two are set to zero. As before, the biggest correction is obtained when all the three Majorana masses are degenerate and have their biggest considered value of $10^{14} \gev$. The symmetry shown in \reffi{contour_tetas} with respect to the three masses 
$m_{M_{1,2,3}}$ is a consequence of the quasi degeneracy assumed of the
three light neutrinos.  When the three $\theta_i$ angles are non zero
and complex, $\hat\Delta M_{h}$ becomes considerably larger than in the
real case, as can be seen in \reffi{contour_3tetas} where we have chosen
as an illustrative example, $\theta_{1}=3\pi/8\,e^{i\pi/4}$,
$\theta_{2}=\pi/2\,e^{i\pi/5}$ and $\theta_{3}=3\pi/4\,e^{i\pi/7}$. The
larger the arguments of the angles $\theta_{i}$ are, the larger
$\hat\Delta M_{h}$ becomes. 
However, the size of these $\theta_i$, as well as the size of the
$m_{M_i}$, are constrained by perturbativity of the Yukawa
coupling. In this context it should be remembered that large corrections for $\hat\De\Mh$ would not be reliable within the approximation used here of~\refeq{delta_mass}. 

\section{Conclusions}
\label{Conclusions}

In this paper we have presented the full one-loop radiative corrections
to the renormalized $\cp$-even Higgs boson self-energies and to the
lightest Higgs boson mass, $\Mh$, from the three-generations in the
neutrino-sneutrino sector within the context of the MSSM-seesaw. 
This work extends and completes the previous calculation in the
simplified one-generation case~\cite{Ana}. 
The most interesting features in this MSSM-seesaw are that the neutrinos,
contrary to other fermions, are assumed to be
Majorana particles, and that the origin for the light neutrino masses, 
again in contrast to the other fermions, are generated by means 
of the seesaw mechanism with the addition of heavy right handed neutrinos 
with large Majorana masses.

As a by-product, we have included here the complete set of Feynman
rules in this MSSM-seesaw for the three-generation (s)neutrino case
relevant for this work (again extending and completing \citere{Ana}). 
This includes the vertices for the interactions of the neutrinos and
sneutrinos with the Higgs sector and with the $Z$~boson.
These Feynman rules have been presented
in terms of all the physical masses and mixing angles of the 
particles involved, 
in particular in the mass eigen basis of the light and heavy Majorana
neutrinos, as well as their light and heavy  SUSY partners.

Our computation is a complete one-loop calculation in the Feynman
diagrammatic approch without any simplifying assumptions. The
corresponding analytical results are also presented in  
terms of the physical neutrinos, sneutrinos,~$Z$, and Higgs bosons masses. 

In particular we have discussed the renormalization of $\tb$ and
the wave function of the two Higgs doublets in the case of three
generations of (s)neutrinos.
As was discussed previously in the literature (in the one-generation
case), the dependence of the
prediction of $\Delta\Mh$ on the Majorana mass scales depends strongly on the
choice of the $\tb$ renormalization. Various schemes have been analyzed,
where each scheme exhibits advantages and disadvantages. 
Especially, the ``modified \DRbar'' scheme ($\mDR$) was contrasted to
other schemes, like the ``more physical'' OS and HM  schemes and the
``decoupling'' scheme (DEC). 
The latter one leads, hence its name, to a full decoupling of the heavy
Majorana mass scales in $\hat\Delta \Mh$, which we confirm here for
the three generations case.  
Regarding the comparison with the  ``more physical'' schemes, like OS
and HM, we have seen that they can lead to  
potentially unstable numerical behavior in certain regions of the
MSSM-seesaw parameter space. Therefore the convergence of the
perturbative expansion may not be ensured in the presence of
heavy scales. We have also
found that the use of the ``more traditional'' 
${\DR}$ scheme  is not convenient either, since there is an extremely
high sensitivity to the choice of the  $\mu_{\DR}$ scale. When this
$\mu_{\DR}$ scale is set to the high Majorana scale, then the large
logarithmic contributions disappear and one reaches a more stable
result. The absence of large logarithmic contributions, $\log(m_{M_i}/
\mu_{\DR})$, is automatically implemented in the  $\mDR$ scheme. 
This $\mDR$ scheme, by construction not exhibiting complete
decoupling behavior, leads
to numerically stable predictions for $\Delta\Mh$, gauge invariant to one-loop, while exhibiting a residual
dependence of $\Delta\Mh$ on the heavy Majorana mass scales. 
The analytic structure of those terms in the $\mDR$ scheme as
well as in the OS and the \DRbar\ scheme
have been derived and fully analyzed here for the
three-generations case.

Finally, in order to cover several scenarios and hierarchies,
in the numerical investigation we have analyzed the neutrino/sneutrino
corrections to the renormalized $\cp$-even Higgs self-energies and $\Delta\Mh$
with respect to all the involved masses and parameters: 
$m_{M_i}$, $\tb$, $M_A$, $m_{\tilde{L}_i} $,  $m_{\tilde{R}_i}$,  
$a_{\nu}$, $m_{\nu_i}$, $\theta_i$
and $b_{\nu}$. These analyses have been performed in the $\mDR$ scheme.
A clear prescription has also been presented to pass from this scheme to
the other introduced schemes (where, by definition, the DEC
scheme would lead to very small effects.)
We have ensured that our numerical scenarios are in
agreement with experimental data by using the Casas-Ibarra
parametrization of the neutrino sector and choosing the relevant values,
e.g.\ of neutrino mass differences, according to the most recent
experimental results. We have investigated both, the normal and the
inverted hierarchy. 

The pure gauge contributions, which are already present in the MSSM
and grow with $\tb$, can amount about $\De\Mh \sim 150 \mev$, in the
low $\tb \sim 2$ region of interest here, i.e.\ about half of the
current 
experimental uncertainty. These corrections arise from the sneutrino
sector only, and thus are independent of the assumed hierarchy in the
neutrino sector. The remaining contributions,  $\hat \Delta M_{h}$,
which are sensitive to the heavy neutrinos/sneutrinos via the Yukawa
couplings in the $\mDR$ scheme, are larger than the pure gauge
contributions in presence of 
very heavy scales, and are in contrast larger at the lower values of
$\tb$.
We have studied the size of the corrections $\Delta\Mh$ with respect to
\ the Majorana mass scales (where no dependence would have been
found in the DEC scheme). The largest corrections are found in
the degenerate case and for the largest allowed Majorana mass
values. In the present work these maximum values have been set  
to $10^{15} \gev$ in order to respect the perturbativity condition on
the Yukawa couplings. In the large region of  
$10^{14} \, {\rm GeV} \lsim m_M \lsim 10^{15} \, {\rm GeV}$  
we find negative
corrections of up to  $\De\Mh \sim{\cal O}(-5) \gev$. 
We have also found that the corrections in the three generations case are generally larger than  in the one generation case. Particularly, we have checked that for the degenerate Majorana masses scenario with no generation mixing, the corrections are indeed approximately three times larger.
Finally, the dominant corrections in this work are found to be proportional to the square of the
neutrino Dirac mass scale, specifically to  
$m_D^\dagger m_D$, 
and therefore they can be enhanced when complex
$\theta_i$ parameters are taken into account. However, the above
commented perturbativity requirements on the Yukawa of couplings will
always restrict the size of the mass correction.

\newpage

\section*{Appendix}
\appendix
\renewcommand{\sectionmark}[1]{\markright{Appendix \thesection}{} }

\section{New Feynman rules}
\label{sec:fmr}

In this Appendix we collect the Feynman rules derived from the interaction Lagrangian terms of section \ref{sec:nN} within the MSSM-seesaw that are relevant for the present work. They represent the interactions between the neutrinos and sneutrinos with the MSSM neutral Higgs bosons and between the neutrinos and sneutrinos with the Z gauge bosons. All the Feynman rules are written here in the physical basis. Here $\cw\equiv\cos\theta_W$ and we have shortened the notation as in \refeq{rotationR}, i.e. $U_{ij}\equiv U_{i,j}$, $\tilde{U}_{ij}\equiv \tilde{U}_{i,j}$. 

\subsection*{Neutrinos}

Three-point couplings of two Majorana neutrinos to one MSSM Higgs boson and of two Majorana neutrinos to the Z gauge boson. \\

\begin{table}[h!]
\begin{tabular}{ll}
\parbox[c]{1em}{\includegraphics[scale=0.75]{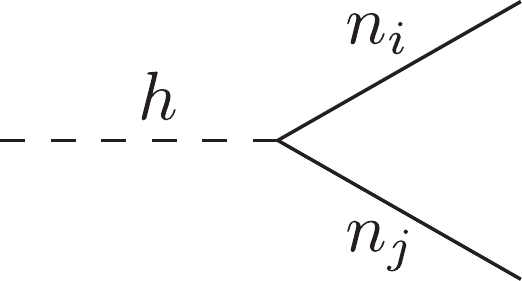}}
 &\put(100,-15){$\begin{array}{rl}
i V_{hn_in_j}^L P_L+ iV_{hn_in_j}^R P_R=&-\dfrac{ig \cos{\alpha}}{2M_W \sin{\beta}} \Big( U^*_{m+3,i}(m_D^\dagger)_{mn}U^*_{nj}P_L \, \\\\
&+ \,U_{mi}\left(m_D\right)_{mn}U_{n+3,j}P_R \Big) + \left(i\leftrightarrow j\right)
\end{array}$}\\
&\\
&\\
\parbox[c]{1em}{\includegraphics[scale=0.76]{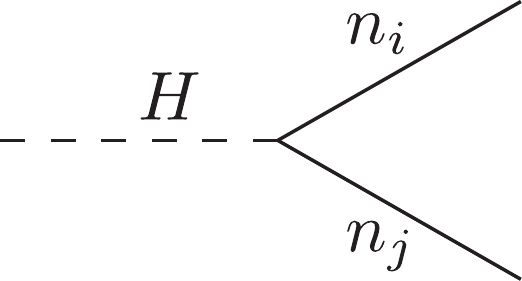}}
 &\put(100,-15){$\begin{array}{rr}
i V_{Hn_in_j}^L P_L+ iV_{Hn_in_j}^R P_R=&-\dfrac{ig \sin{\alpha}}{2M_W \sin{\beta}} \Big( U^*_{m+3,i}(m_D^\dagger)_{mn}U^*_{nj}P_L \, \\\\
&+ \,U_{mi}\left(m_D\right)_{mn}U_{n+3,j}P_R \Big) + \left(i\leftrightarrow j\right)
\end{array}$}\\
&\\
&\\
\parbox[c]{1em}{\includegraphics[scale=0.75]{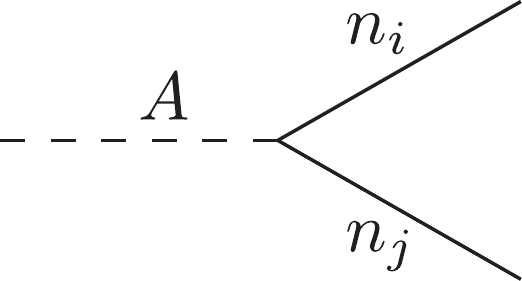}}
 &\put(100,-15){$\begin{array}{rr}
i V_{An_in_j}^L P_L+ iV_{An_in_j}^R P_R=&\dfrac{g \cos{\beta}}{2M_W \sin{\beta}} \Big( U^*_{m+3,i}(m_D^\dagger)_{mn}U^*_{nj}P_L \,\\\\
&- \,U_{mi}\left(m_D\right)_{mn}U_{n+3,j}P_R \Big) + \left(i\leftrightarrow j\right)
\end{array}$}\\
&\\
&\\
\parbox[c]{1em}{\includegraphics[scale=0.75]{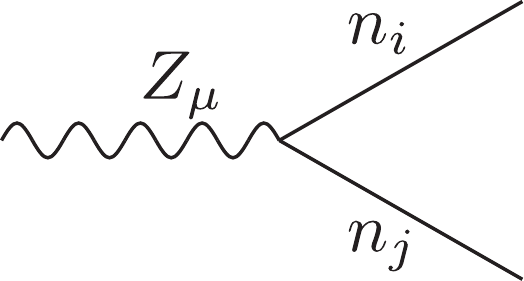}}
 &\put(100,0){$iV_{Zn_in_j}^L\gamma^\mu P_L=\dfrac{-ig}{2\cw}\Big( U_{mi}U^*_{mj} \Big) \gamma^\mu P_L + \left(i\leftrightarrow j\right)$}\\
&\\
\end{tabular}
\end{table}

\subsection*{Sneutrinos}

\begin{table}[H]
Three-point couplings of two sneutrinos to one MSSM Higgs boson and of two sneutrinos to the Z gauge boson. All the couplings not shown here vanish.\\ \\
\begin{tabular}{ll}
\parbox[c]{1em}{\includegraphics[scale=0.75]{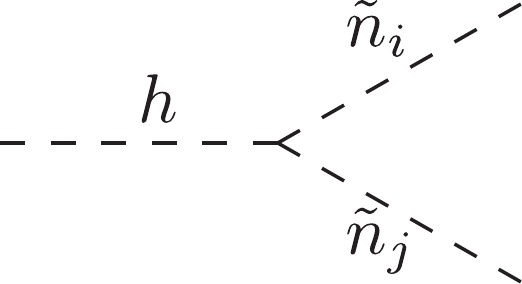}}
&\put(100,-60){$\begin{array}{l}
iV_{h\tilde{n}_i\tilde{n}_j}= \dfrac{i\cos{\alpha}}{\sqrt{2}}\left\lbrace \tilde{U}_{mi} \left( A_\nu\right)_{mn} \tilde{U}_{n+6,j} 
+ \tilde{U}_{m+3,i} (A^\dagger_\nu)_{nm} \tilde{U}_{n+9,j}\right\rbrace \\ \\

\dfrac{-ig \cos{\alpha}}{M_W \sin{\beta}} \left\lbrace \tilde{U}_{m+9,i} (m_D^\dagger)_{mn} \left( m_D\right)_{nl} \tilde{U}_{l+6,j} 
+ \tilde{U}_{mi} \left( m_D\right)_{mn} (m_D^\dagger)_{nl} \tilde{U}_{l+3,j}\right\rbrace \\ \\

\dfrac{-ig \cos{\alpha}}{2 M_W \sin{\beta}} \left\lbrace \tilde{U}_{m+3,i} (m_D^\dagger)_{nm} \left(m_M\right)_{nl} \tilde{U}_{l+6,j} 
+ \tilde{U}_{mi} \left( m_D\right)_{mn} (m^\dagger_M)_{ln} \tilde{U}_{l+9,j} \right\rbrace\\ \\ 
 
\dfrac{-i g \sin{\alpha}}{2 M_W \sin{\beta}} \left\lbrace \mu^* \tilde{U}_{mi} \left( m_D\right)_{mn} \tilde{U}_{n+6,j} 
+ \mu \tilde{U}_{m+3,i} (m_D^\dagger)_{nm} \tilde{U}_{n+9,j}\right\rbrace \\ \\

\dfrac{+i g M_Z \sin{(\alpha+\beta)}}{2 \cw} \left\lbrace \tilde{U}_{m+3,i} \tilde{U}_{mj} \right\rbrace + \left(i\leftrightarrow j\right)
\end{array}$}\\
&\\
&\\
\parbox[c]{1em}{\includegraphics[scale=0.75]{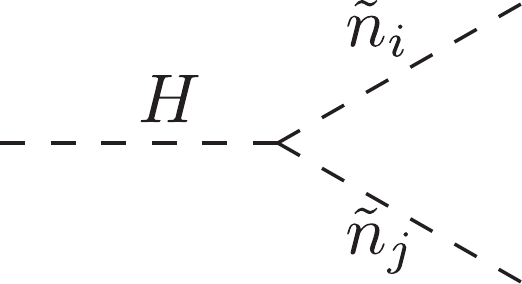}}
&\put(100,-60){$\begin{array}{l}
iV_{H\tilde{n}_i\tilde{n}_j}= \dfrac{i\sin{\alpha}}{\sqrt{2}}\left\lbrace \tilde{U}_{mi} \left( A_\nu\right)_{mn} \tilde{U}_{n+6,j} 
+ \tilde{U}_{m+3,i} (A^\dagger_\nu)_{nm} \tilde{U}_{n+9,j}\right\rbrace \\ \\

\dfrac{-ig \sin{\alpha}}{M_W \sin{\beta}} \left\lbrace \tilde{U}_{m+9,i} (m_D^\dagger)_{mn} \left( m_D\right)_{nl} \tilde{U}_{l+6,j} 
+ \tilde{U}_{mi} \left( m_D\right)_{mn} (m_D^\dagger)_{nl} \tilde{U}_{l+3,j}\right\rbrace \\ \\

\dfrac{-ig \sin{\alpha}}{2 M_W \sin{\beta}} \left\lbrace \tilde{U}_{m+3,i} (m_D^\dagger)_{nm} \left(m_M\right)_{nl} \tilde{U}_{l+6,j} 
+ \tilde{U}_{mi} \left( m_D\right)_{mn} (m^\dagger_M)_{ln} \tilde{U}_{l+9,j} \right\rbrace\\ \\ 
 
\dfrac{+i g \cos{\alpha}}{2 M_W \sin{\beta}} \left\lbrace \mu^* \tilde{U}_{mi} \left( m_D\right)_{mn} \tilde{U}_{n+6,j} 
+ \mu \tilde{U}_{m+3,i} (m_D^\dagger)_{nm} \tilde{U}_{n+9,j}\right\rbrace \\ \\

\dfrac{-i g M_Z \cos{(\alpha+\beta)}}{2 \cw} \left\lbrace \tilde{U}_{m+3,i} \tilde{U}_{mj} \right\rbrace + \left(i\leftrightarrow j\right)
\end{array}$}\\
&\\
&\\
\parbox[c]{1em}{\includegraphics[scale=0.75]{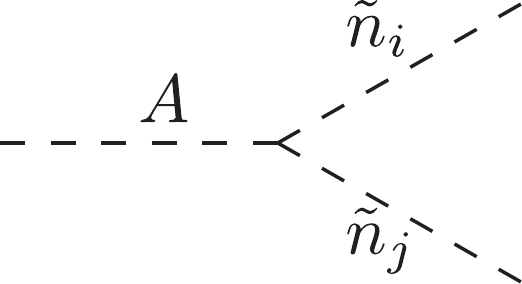}}
&\put(100,-20){$\begin{array}{l}
iV_{A\tilde{n}_i\tilde{n}_j}= \dfrac{g}{2 M_W} \left\lbrace \mu^* \tilde{U}_{mi} \left(m_D\right)_{mn} \tilde{U}_{n+6,j} - \mu \tilde{U}_{m+3,i} (m_D^\dagger)_{nm} \tilde{U}_{n+9,j} \right\rbrace\\ \\

\dfrac{-g\cos{\beta}}{2 M_W \sin{\beta}}\left\lbrace\tilde{U}_{m+3,i} (m_D^\dagger)_{nm} \left(m_M\right)_{nl} \tilde{U}_{l+6,j} - \tilde{U}_{mi} \left(m_D\right)_{mn} (m_M^\dagger)_{ln} \tilde{U}_{l+9,j}\right\rbrace \\ \\

 \dfrac{+\cos{\beta}}{3\sqrt{2} }\left\lbrace\tilde{U}_{mi} \left(A_\nu\right)_{mn} \tilde{U}_{n+6,j}-\tilde{U}_{m+3,i} \left(A_\nu^\dagger\right)_{nm} \tilde{U}_{n+9,j}\right\rbrace  + \left(i\leftrightarrow j\right)
\end{array}$}\\
&\\
\parbox[c]{1em}{\includegraphics[scale=0.75]{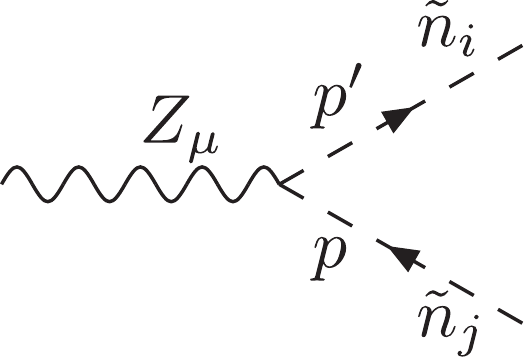}}
&\put(100,0){$\begin{array}{l}
iV_{Z\tilde{n}_i\tilde{n}_j}=\dfrac{ig}{2\cw}\left\lbrace \tilde{U}_{m+3,i}\tilde{U}_{mj}- \tilde{U}_{m+3,j}\tilde{U}_{mi}\right\rbrace\left( p+p^{\prime}\right)^{\mu}
\end{array}$}\\
\end{tabular}
\end{table}
\newpage
Four-point couplings of two neutrinos to two MSSM Higgs bosons and two neutrinos to two Z gauge bosons. All the couplings not shown vanish.

\begin{table}[H]
\begin{tabular}{ll}
\parbox[c]{1em}{\includegraphics[scale=0.75]{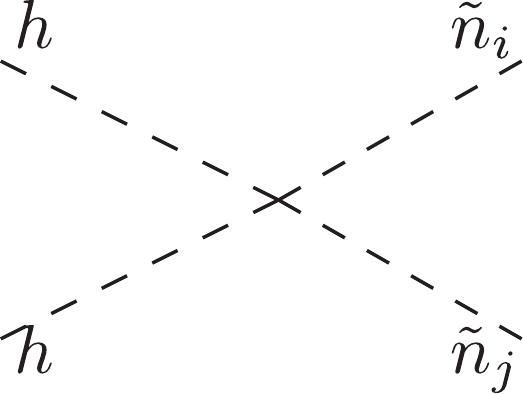}}
&\put(100,0){$\begin{array}{l}
iV_{hh\tilde{n}_i\tilde{n}_j}= -\dfrac{ig^2 \cos^2{\alpha}}{2 M_W^2 \sin^2{\beta}}\left\lbrace  \tilde{U}_{m+9,i} (m_D^\dagger)_{mn}\left(m_D\right)_{nl} \tilde{U}_{l+6,j} \, \right.\\ \\

\left. +\, \tilde{U}_{mi}\left(m_D\right)_{mn} (m_D^\dagger)_{nl} \tilde{U}_{l+3,j}\right\rbrace + \dfrac{ig^2\cos{2\alpha}}{4 \cw^2}\left\lbrace  \tilde{U}_{m+3,i}  \tilde{U}_{mj}\right\rbrace + \left(i\leftrightarrow j\right)
\end{array}$}\\
&\\
&\\
\parbox[c]{1em}{\includegraphics[scale=0.75]{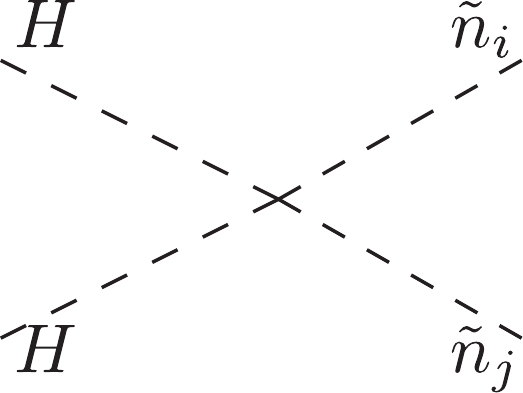}}
&\put(100,0){$\begin{array}{l}
iV_{HH\tilde{n}_i\tilde{n}_j}= -\dfrac{ig^2 \sin^2{\alpha}}{2 M_W^2 \sin^2{\beta}}\left\lbrace  \tilde{U}_{m+9,i} (m_D^\dagger)_{mn}\left(m_D\right)_{nl} \tilde{U}_{l+6,j} \, \right. \\ \\

\left. +\, \tilde{U}_{mi}\left(m_D\right)_{mn} (m_D^\dagger)_{nl} \tilde{U}_{l+3,j}\right\rbrace - \dfrac{ig^2\cos{2\alpha}}{4 \cw^2}\left\lbrace  \tilde{U}_{m+3,i}  \tilde{U}_{mj}\right\rbrace + \left(i\leftrightarrow j\right)
\end{array}$}\\
&\\
&\\
\parbox[c]{1em}{\includegraphics[scale=0.75]{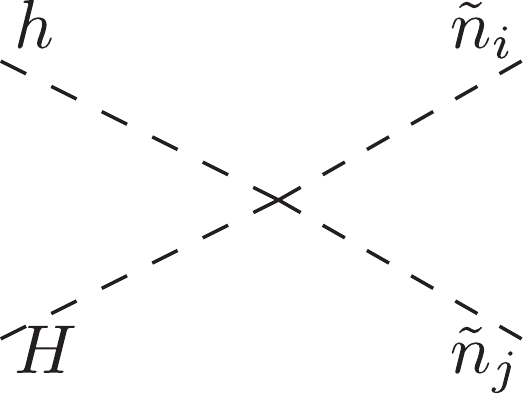}}
&\put(100,0){$\begin{array}{l}
iV_{hH\tilde{n}_i\tilde{n}_j}= -\dfrac{ig^2 \sin{2\alpha}}{4 M_W^2 \sin^2{\beta}}\left\lbrace  \tilde{U}_{m+9,i} (m_D^\dagger)_{mn}\left(m_D\right)_{nl} \tilde{U}_{l+6,j} \, \right. \\ \\

\left. +\,  \tilde{U}_{mi}\left(m_D\right)_{mn} (m_D^\dagger)_{nl} \tilde{U}_{l+3,j}\right\rbrace + \dfrac{ig^2\sin{2\alpha}}{4 \cw^2}\left\lbrace  \tilde{U}_{m+3,i}  \tilde{U}_{mj}\right\rbrace + \left(i\leftrightarrow j\right)
\end{array}$}\\
&\\
&\\
\parbox[c]{1em}{\includegraphics[scale=0.75]{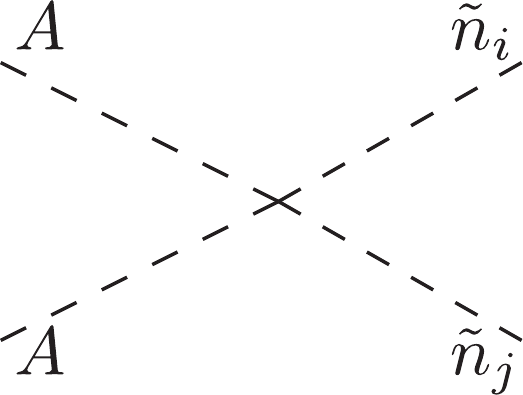}}
&\put(100,0){$\begin{array}{l}
iV_{AA\tilde{n}_i\tilde{n}_j}= -\dfrac{ig^2 \cot^2{\beta}}{2 M_W^2}\left\lbrace  \tilde{U}_{m+9,i} (m_D^\dagger)_{mn}\left(m_D\right)_{nl} \tilde{U}_{l+6,j} \,\right.\\ \\

\left. +\,\tilde{U}_{mi}\left(m_D\right)_{mn} (m_D^\dagger)_{nl} \tilde{U}_{l+3,j}\right\rbrace + \dfrac{ig^2\cos{2\beta}}{4 \cw^2}\left\lbrace  \tilde{U}_{m+3,i}  \tilde{U}_{mj}\right\rbrace + \left(i\leftrightarrow j\right)
\end{array}$}\\
&\\
&\\
\parbox[c]{1em}{\includegraphics[scale=0.75]{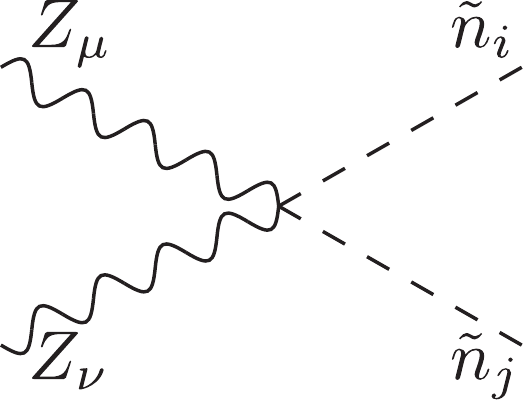}}
&\put(100,0){$\begin{array}{l}
iV_{ZZ\tilde{n}_i\tilde{n}_j}=\dfrac{ig^2}{2\cw^2}\Big( \tilde{U}_{m+3,i}\tilde{U}_{mj}\Big) g^{\mu\nu} + \left(i\leftrightarrow j\right)
\end{array}$}\\
&\\
\end{tabular}
\end{table}


\subsection*{Conflict of Interests}

The authors declare that there is no conflict of interests regarding the publication of this paper.


\subsection*{Acknowledgements}

We thank H.~Haber for helpful and clarifying discussions.
The work of S.H.\ was supported 
by the Spanish MICINN's Consolider-Ingenio 2010
Program under grant MultiDark CSD2009-00064. 
The work of M.H., J.H.\ and X.M.\ was partially supported by the
European Union FP7 ITN INVISIBLES (Marie Curie Actions, PITN- GA-2011-
289442), by the CICYT through the project FPA2012-31880,  
by the Spanish Consolider-Ingenio 2010 Programme CPAN (CSD2007-00042) 
and by the Spanish MINECO's ``Centro de Excelencia Severo Ochoa''
Programme under grant SEV-2012-0249.
X.~M. is also supported through the FPU grant AP-2012-6708.
J.~H. acknowledges financial support by the European Union through the FP7 Marie Curie Actions CIG NeuProbes (PCIG11- GA-2012-321582).

\end{document}